%% file: ms-method_20120615_a.tex
\documentclass[apj]{emulateapj-rtx4}
\usepackage{float}
\usepackage{rotating} 
\usepackage{times}
\usepackage{lscape}
\usepackage{subfigure}
\usepackage{amsmath}
\usepackage{amsfonts}
\usepackage{amssymb}
\usepackage{epsfig}
\usepackage{hyperref}
\usepackage{graphicx}
\usepackage{natbib}

\usepackage{bm}
\usepackage{color}
\usepackage{hyperref}

\usepackage{amsmath}
\usepackage{amsfonts}
\usepackage{amssymb}
\usepackage{epsfig}
\usepackage{hyperref}
\usepackage{graphicx}

\newcommand{\herschel}{{\it Herschel}}
\newcommand{\atlas}{H-ATLAS}

\newcommand{\spitzer}{{\it Spitzer}}

\newcommand{\um}{\,$\mu$m}

\begin{document}

\title{{\it Spitzer} IRAC identification of {\it Herschel}-ATLAS SPIRE sources}

\author{ 
Sam Kim\altaffilmark{1}, 
Julie~L.~Wardlow\altaffilmark{1}, 
Asantha~Cooray\altaffilmark{1}, 
S.~Fleuren\altaffilmark{2}, 
W.~Sutherland\altaffilmark{2},
A.~A.~Khostovan\altaffilmark{1}, 
R.~Auld\altaffilmark{3}, 
M.~Baes\altaffilmark{4}, 
R.~S.~Bussmann\altaffilmark{5}, 
S.~Buttiglione\altaffilmark{6}, 
A.~Cava\altaffilmark{7}, 
D.~Clements\altaffilmark{8}, 
A.~Dariush\altaffilmark{8}, 
G.~De Zotti\altaffilmark{6,9},
L.~Dunne\altaffilmark{10}, 
S.~Dye\altaffilmark{11}, 
S.~Eales\altaffilmark{3}, 
J.~Fritz\altaffilmark{4}, 
R.~Hopwood\altaffilmark{8}, 
E.~Ibar\altaffilmark{12}, 
R.~Ivison\altaffilmark{12}, 
M.~Jarvis\altaffilmark{14,15}, 
S.~Maddox\altaffilmark{10}, 
M.~J. Micha{\l}owski\altaffilmark{13},
E.~Pascale\altaffilmark{3}, 
M.~Pohlen\altaffilmark{3}, 
E.~Rigby\altaffilmark{10}, 
D.~Scott\altaffilmark{16}, 
D.~J.~B. Smith\altaffilmark{14}, 
P.~Temi\altaffilmark{17},  
and P.~van der Werf\altaffilmark{18}
}

\altaffiltext{1}{Department of Physics and Astronomy, University of California, Irvine, CA 92697}
\altaffiltext{2}{School of Mathematical Sciences, Queen Mary, University of London, Mile End Road, London, E1 4NS, UK}
\altaffiltext{3}{School of Physics and Astronomy, Cardiff University, Cardiff, CF24 3AA, UK}
\altaffiltext{4}{Sterrenkundig Observatorium, Universiteit Gent,
 Krijgslaan 281 S9, B-9000 Gent, Belgium}
 \altaffiltext{5}{Submillimeter Array Fellow, Harvard-Smithsonian Center
for Astrophysics, 160 Concord Ave., Cambridge, MA 02138 }
\altaffiltext{6}{INAF-Osservatorio Astronomico di Padova, Vicolo Osservatorio 5, I-35122 Padova, Italy }
\altaffiltext{7}{Departamento de Astrof\'{\i}sica, Facultad de CC. F\'{\i}sicas, Universidad Complutense de Madrid, E-28040 Madrid, Spain}
\altaffiltext{8}{Physics Department, Imperial College London, South Kensington Campus, London SW7 2AZ, UK}
\altaffiltext{9}{SISSA, Via Bonomea 265, I-34136 Trieste, Italy}
\altaffiltext{10}{Department of Physics and Astronomy, University of Canterbury, Private Bag 4800, Christchurch, 8041, New Zealand }
\altaffiltext{11}{School of Physics and Astronomy, University of Nottingham, University Park, Nottingham, NG7 2RD, UK }
\altaffiltext{12}{UK Astronomy Technology Centre, Royal Observatory, Blackford Hill, Edinburgh EH9 3HJ, UK}
\altaffiltext{13}{Scottish Universities Physics Alliance, Institute for Astronomy,
University of Edinburgh, Royal Observatory, Edinburgh, EH9 3HJ, UK}
\altaffiltext{14}{Centre for Astrophysics Research, Science \& Technology Research Institute, University of Hertfordshire, Hatfield, Herts, AL10 9AB, UK }
\altaffiltext{15}{Physics Department, University of the Western Cape, Cape Town, 7535, South Africa }
\altaffiltext{16}{Department of Physics and  Astronomy, University of British Columbia, 6224 Agricultural Road, Vancouver, BC V6T1Z1, Canada}
\altaffiltext{17}{Astrophysics Branch, NASA/Ames Research Center, MS 245-6, Moffett Field, CA 94035, USA}
\altaffiltext{18}{Leiden Observatory, Leiden University, P.O. Box 9513, 2300 RA Leiden, The Netherlands}


\begin{abstract} 
  We use \spitzer-IRAC data to identify near-infrared
  counterparts to submillimeter galaxies detected with \herschel-SPIRE
  at 250\um\ in the {\it Herschel} Astrophysical Terahertz Large
  Area Survey (H-ATLAS).  Using a likelihood ratio analysis  we identify 146 reliable IRAC counterparts to 123
  SPIRE sources out of the 159 in the survey area.  We
  find that, compared to the field population, the SPIRE counterparts
  occupy a distinct region of 3.6 and 4.5\um\ color-magnitude
  space, and we use this property to identify  further 23
  counterparts to 13 SPIRE sources. The IRAC identification rate of
  86\% is significantly higher than those that have been demonstrated with
  wide-field ground-based optical and near-IR imaging of \herschel\
  fields.  We estimate a false identification rate of 3.6\%,
  corresponding to 4 to 5 sources.    Among the 73 counterparts that are undetected in SDSS, 57
      have both 3.6 and 4.5\um\ coverage.    Of these,   43
    have $[3.6] - [4.5]> 0$ indicating that they are likely to be
    at $z\gtrsim$ 1.4. Thus, $\sim$ 40\% of identified SPIRE galaxies
      are likely to be high redshift ($z\gtrsim$ 1.4) sources.
We discuss the statistical properties of the IRAC-identified SPIRE galaxy sample including
far-IR luminosities, dust temperatures, star-formation rates, and stellar masses.
The majority of our detected galaxies have 10$^{10}$ to 10$^{11}$ $L_{\sun}$ total IR luminosities
and are not intense starbursting galaxies as those found at $z \sim 2$, but 
they have a factor of 2 to 3 above average specific star-formation rates compared to near-IR selected galaxy samples.
\end{abstract}


\section{Introduction}

The extragalactic background at far-infrared (IR) and submillimeter
(sub-mm) wavelengths is well-constrained from total intensity
measurements (Puget et al.\,1996; Fixsen et al.\,1998; Dwek et
al.\,1998). However, the properties of the discrete galaxies that make
up this background are still largely unknown. 
These sub-mm galaxies are
expected to capture the dusty star-formation out to redshifts of 4 and
beyond 
and are now understood to be an integral component of galaxy
formation and evolution (Hughes et al.\,1998; Eales et al.\,1999; Blain et al.\,2002; Chapman et al.\,2003; Austermann et al.\,2010).

A challenge for studies of sub-mm galaxies is that wide-field surveys
at these wavelengths have poor spatial resolution, caused
by the limited apertures of single dish sub-mm survey
telescopes (Smail, Ivison \& Blain 1997; Scott et al.\,2002; Coppin et al.\,2006). Typical resolutions are of the order of tens of arcseconds,
making the identification of multiwavelength counterparts challenging,
particularly in the optical where sub-mm galaxies are usually faint due
to high dust extinction. Furthermore, the surface density of faint
optical galaxies is such that several potential counterparts may be
  situated within the positional error of each sub-mm source. However,
the identification of counterparts to sub-mm galaxies is critical for
both photometric and spectroscopic studies, which yield redshifts,
spectral energy distributions (SEDs) and morphological information. The
measurements of these quantities for statistically significant samples
of sub-mm galaxies are required to compare their properties with
  theoretical predictions, and to fully understand the role of the
  sub-mm bright phase in galaxy evolution.

The most reliable way to pinpoint the positions of sub-mm galaxies is
through high-resolution sub-mm interferometry, which directly traces
the dust emission at wavelengths comparable to the original selection
function (e.g. Downes et al. 1999; Gear et al. 2000; Iono et al. 2006;
Younger et al.  2007).  The sensitivity of early generations of sub-mm
interferometric instrumentation limited anlayses to samples of no more
than a few sources (e.g. Downes et al. 1999; Gear et al. 2000; Iono et
al. 2006; Younger et al.  2007, 2008, 2009; Ivison et al. 2008; Cowie
et al. 2009; Aravena et al. 2010a; Hatsukade et al. 2010; Tamura et
al. 2010; Chen et al. 2011 ), and although rapid progress is being made
(e.g. Smolcic 2012a,b) such observations are still unfeasible for large
surveys of sources.  In addition, extended sources (e.g. Matsuda et
al. 2007), or sources that are a blend of multiple components
(e.g. Wang et al. 2011; Smolcic 2012b), can be difficult to detect.

An alternative technique to identify sub-mm galaxies is to take
advantage of the far-IR-radio correlation (e.g. Condon et al.\ 1998;
Garrett 2002) and the relatively low surface density of radio sources
by searching for counterparts in deep interferometric radio data
(e.g. Ivison et al.\ 1998, 2002; Borys et al.\ 2004; Dye et al.\ 2009;
Dunlop et al.\ 2010; Biggs et al.\ 2011). Similarly, the dusty, active
sub-mm galaxies are typically bright at mid-IR wavelengths, a property
that has also been utilized to identify counterparts (e.g. Ivison et
al.\ 2004; Pope et al.\ 2006; Dye et al.\ 2008, 2009; Clements et
al.\ 2008; Dunlop et al.\ 2010; Biggs et al.\ 2011). Deep radio and mid-IR
(typically {\it Spitzer} MIPS 24\um but also {\it Spitzer} IRAC 3.6\um; Biggs et
al.\ 2011) searches typically identify counterparts to 60--80\% of sub-mm
galaxies, and nearest-neighbor positional matching is then used to
investigate the properties of these sources at other wavelengths. The
remaining 20--40\% of sources are thought to be at high-redshifts ($z
\ga$ 3) or be dominated by cold dust ($T_D\la 25$~K at $z\sim 2$; Chapman 
et al.\ 2005).

The {\it Herschel} Astrophysical Terahertz Large Area Survey (H-ATLAS;
Eales et al.\,2010) is the largest open-time key project being
undertaken by the {\it Herschel} Space Observatory (Pilbratt et
al.\,2010)\footnote{{\it Herschel} is an ESA space observatory with
  science instruments provided by European-led Principal Investigator
  consortia and with important participation from NASA.}. The total
planned survey area is 550 deg$^2$ within which it is expected to
detect $>$ 300,000 bright sub-mm galaxies. During the Science
Demonstration Phase (SDP), H-ATLAS observed 14.4 deg$^2$ in the Galaxy
And Mass Assembly (GAMA; Driver et al.\,2011) 9~hour field, to
5$\sigma$ depths of 35 to  132 mJy in five bands from 100 to 500\um,
using PACS (100 and 160\micron; Poglitsch et al.\ 2010) and SPIRE (250,
350 and 500\micron; Griffin et al.\,2010) in parallel mode.  The
lack of deep radio and 24\um\ data across the field restricts
  counterpart identification to optical (e.g.\ Sloan Digital Sky
Survey [SDSS]; York et al.\,2000) or near-IR data (e.g.\ VISTA
Kilo-Degree Infrared Galaxy Survey [VIKING]; Sutherland et al.\,in
prep, UKIRT Infrared Deep Sky Survey [UKIDSS]; Lawrence et al.\,2007).

Smith et al.\ (2011) identified SDSS and GAMA galaxies as the
  counterparts to 37\% of the SPIRE sources in the H-ATLAS SDP
field.  In the wider area of the whole GAMA-9~hour field Fleuren
et al.\,(2012) increased the fraction of SPIRE sources with reliable
identifications to 51\% by using $K_{\rm s}$-band imaging from the
VIKING survey.  In the GAMA-15~hour field 50\% of the SPIRE sources
  have reliable counterparts identified in Wide-Field Infrared Survey Explorer
  (WISE) data at 3.4\,\micron\ (Bond et al.\ 2012).

In this paper we cross-identify H-ATLAS SPIRE sources with \spitzer\
Infrared Array Camera (IRAC; Fazio et al.\ 2004a) galaxies, which are
  selected from 3.6 and 4.5-\micron\ observations of 0.4 deg$^2$ of
the H-ATLAS SDP field.  The IRAC data are advantageous for
  counterpart identification because of their mid-IR wavelengths and
  depth ($\sim$3 magnitudes deeper than WISE). Thus, counterpart
  identification is less biased towards galaxies with
  relatively low redshifts or dust obscurations than the existing analyses.

The paper is organized as follows: the data analysis and sample
selection are presented in Section~\ref{sec:data}. In
Section~\ref{sec:ids} we describe the counterpart identification method
and in Section~\ref{sec:discussion} we discuss our results, including
the identification rate and the colors  and properties of SPIRE
counterparts. 
 Our conclusions are presented in
Section~\ref{sec:conclusion}. A table of the identified
  counterparts, including magnitudes and fluxes is presented in the
Appendix.  We use J2000 coordinates and $\Lambda$CDM cosmology with
$\Omega_m$= 0.27, $\Omega_\Lambda$= 0.73 and $H_0$ = 70\,${\rm
  km\,s^{-1}Mpc^{-1}}$ throughout. All photometry is on the AB
magnitude system where $1\,\mu{\rm Jy}$= 23.9 mag; IRAC 3.6 and
4.5\um\ AB magnitudes are designated $[3.6]$ and $[4.5]$,
respectively.

\section{Sample selection and data analysis}
\label{sec:data}

The \atlas\ fields were chosen to minimize contamination from Galactic
cirrus and to maximize the overlap with existing and planned wide-area
imaging and spectroscopic surveys.  The SDP field, which we study
here, spans $\sim14~{\rm deg}^2$ in the GAMA-9~hour field and has
existing SDSS, VIKING, and UKIDSS data.  It is observed with PACS at
100 and 160\um\ and SPIRE at 250, 350 and 500\um.  The \atlas\ PACS
and SPIRE map-making processes are described in Ibar et al.\,(2010)
and Pascale et al.\,(2011), respectively, and details of the source
extraction procedures are given in Rigby et al.\,(2011). A public
catalog of SDP sources is available from the \atlas\
website\footnote{http://www.h-atlas.org}. The data reach 5$\sigma$
point-source depths of 132, 126, 32, 36 and 45\,mJy, with beam sizes
of 8.7, 13.1, 18.1, 25.2 and 36.9\arcsec\ (FWHM) at 100, 160, 250,
350, and 500\um, respectively.  Although PACS 100 and 160\um\ data
have significantly higher angular resolution than SPIRE data we do not
use PACS data for counterpart identification because only five of the
159 SPIRE sources studied here are detected by PACS. In all five cases
the position of the identified IRAC counterpart
(table~\ref{tab:photom}) is consistent with the PACS
emission. Furthermore, the counterparts to all of these five sources
are low-redshift late-type galaxies, which is consistent with
expectations for galaxies that are detectable with PACS.

\spitzer\ IRAC 3.6 and 4.5\um\ staring mode observations of
  0.4~deg$^2$, in three regions of the \atlas\ SDP field were carried
out during the warm mission (\spitzer\ program 548; PI: A. Cooray).
Two of the areas targeted contain bright SPIRE SDP sources that are
now known to be gravitationally lensed (ID81 and ID130; Negrello et
al.\,2010; Hopwood et al.\,2011) and the third area was chosen as a
test field.  In staring mode simultaneous observations at 3.6
  and 4.5\um\ are offset from each other by 6.8$^{\prime}$.  Due to
the offset, 0.22 deg$^2$ of the targeted area has imaging data at both
3.6 and 4.5\um, and the remaining 0.18 deg$^2$ is split between 3.6
and 4.5\um\ coverage. There are 159 SPIRE sources in the total IRAC
footprint.  Of these, 101 are observed at both 3.6 and
  4.5\,\micron, and 30 (28) have only 3.6\um\ (4.5\um) data.  The total exposure time is 1080 seconds per pixel in
each filter.  Data reduction and mosaicking is performed on the
Corrected Basic Calibrated Data (cBCD) with {\sc Mopex} (MOsaicker and
Point source EXtractor; Makovoz \& Marleau 2005).  The final mosaicked
images have 0.6\arcsec$\times$ 0.6\arcsec\ pixels and the FWHM of
the IRAC point spread function (PSF) at 3.6 and 4.5\um\ is 1.66 and
1.72\arcsec, respectively.

Our source detection and extraction procedure is designed to
simultaneously select
sources at 3.6 and 4.5\,\um, and provide equivalent photometry in
both bands.  The source detection is performed on a mean combined 3.6
and 4.5\um\ image. The 3.6 and 4.5\um\ image combination is carried out
with {\sc Montage}\footnote{http://montage.ipac.caltech.edu}, which
models background flux by co-adding and re-projecting background
corrected images into a mosaic.  Source detection is performed with
{\sc SExtractor} (Bertin \& Arnouts 1996) and detected sources are
required to comprise at least three contiguous pixels with fluxes at
least 1.5$\sigma$ above the local background.  There are 27170 unique
3.6 and 4.5\um\ sources that meet this requirement and that are
  detected at $\ge$ 5$\sigma$.  Photometry is measured in 3.8\arcsec\
diameter apertures with the {\sc apphot} task in {\sc iraf}.  The advantage of {\sc apphot} is that the photometry is
measured in fixed apertures at specified source positions.  The
measured aperture photometry is corrected to ``total'' fluxes, assuming
point-source profiles, and using the calibration derived by the SWIRE
team for IRAC data with multiplicative correction factors of 1.36
and 1.40 at 3.6 and 4.5\um, respectively (Surace et al.\,2005).

The completeness of the IRAC catalog is established by inserting
artificial galaxies with a maximum half-light radius of 1.5\arcsec\ in
the IRAC images.  Our source extraction and photometric procedure is
repeated and a source is considered to be recovered if the extracted
position and magnitude are within 1.5\arcsec\ and 5$\sigma$,
  respectively, of the input values.  The 50\% completeness limits
are 22.5 mag (3.63 $\mu$Jy) and 22.2 mag (4.79 $\mu$Jy) at 3.6 and
4.5\um\, respectively.

The far-IR emission detected by SPIRE traces emission from cold
dust. Therefore, point sources in blank-field surveys, such as
H-ATLAS, are primarily expected to be external galaxies and not stars
in the Milky Way.  Indeed, of the $\sim$ 6600 SPIRE sources in
$\sim$ 14 deg$^2$ in the H-ATLAS SDP field only 78 are Galactic stars,
and two are candidate debris disks (Thompson et al.\ 2010).  Since
  stars dominate the IRAC catalog at the brightest fluxes and have
  magnitude distributions that are different to galaxies, it is
  necessary to remove them prior to counterpart identification
  (e.g. Smith et al.\ 2011).  Unfortunately the angular resolution of
  IRAC (FWHM~$\sim$ 1.7$\arcsec$) is insufficient to morphologically
  distinguish between stars and high-redshift galaxies. Furthermore, while
  stars can be identified in $[3.6] - [4.5]$ color-magnitude space
  (e.g. Eisenhardt et al.\,2004) we are limited to just single IRAC
  band data for $\sim$ 45\% of the total IRAC catalog.

Therefore,  instead of using IRAC data alone to reliably identify stars, we make
  use of the stellar classification in SDSS. For this we match the
  IRAC catalog to SDSS DR7 (Abazajian, et al.\ 2009) using positional
  information.  $\sim40\%$ of IRAC  sources  are matched   with SDSS catalog  within search radius 1.5\arcsec. 
   Out of total 27170 IRAC sources, there are 4239 (16\%) sources that are classified
  as stars by SDSS. Once these are removed, the final IRAC catalog
  has 22931 galaxies. This catalog is used for the SPIRE
    source identification and subsequent analysis.   We note that Fazio et al (2004b) showed that
 faint stars are a only minor contribution to the IRAC population,
 with  $<10^{3}$~mag$^{-1}$ deg$^{-2}$ for magnitudes fainter than 16
 mag at both 3.6 and 4.5\um. Therefore, stars that are
 too faint to be detected as such by SDSS are not expected to affect our
 statistical analysis. 

In addition, SDSS QSO population statistics (Schneider et
  al. 2007) indicate that only 2--3 QSOs are expected in the IRAC area ($<1$ in
  the counterpart search area around SPIRE centroids), and
  therefore the effect of QSOs on our statistics is also expected to
  be negligible. 
We conclude that the counterpart identification statistics are unlikely to be affected by  unresolved source  contamination in our input catalog.

\begin{figure}
  \includegraphics[width=9cm]{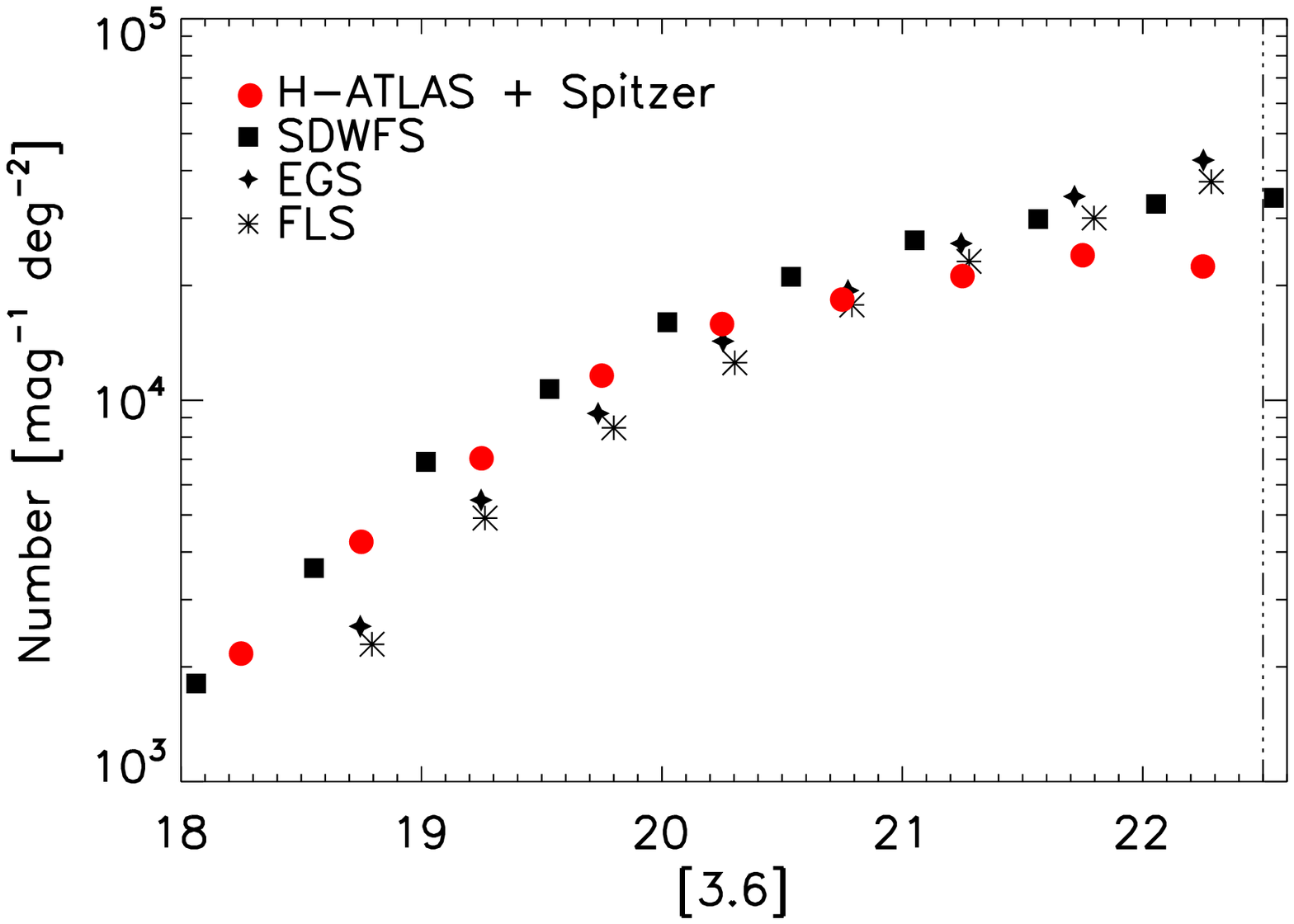}
  \includegraphics[width=9cm]{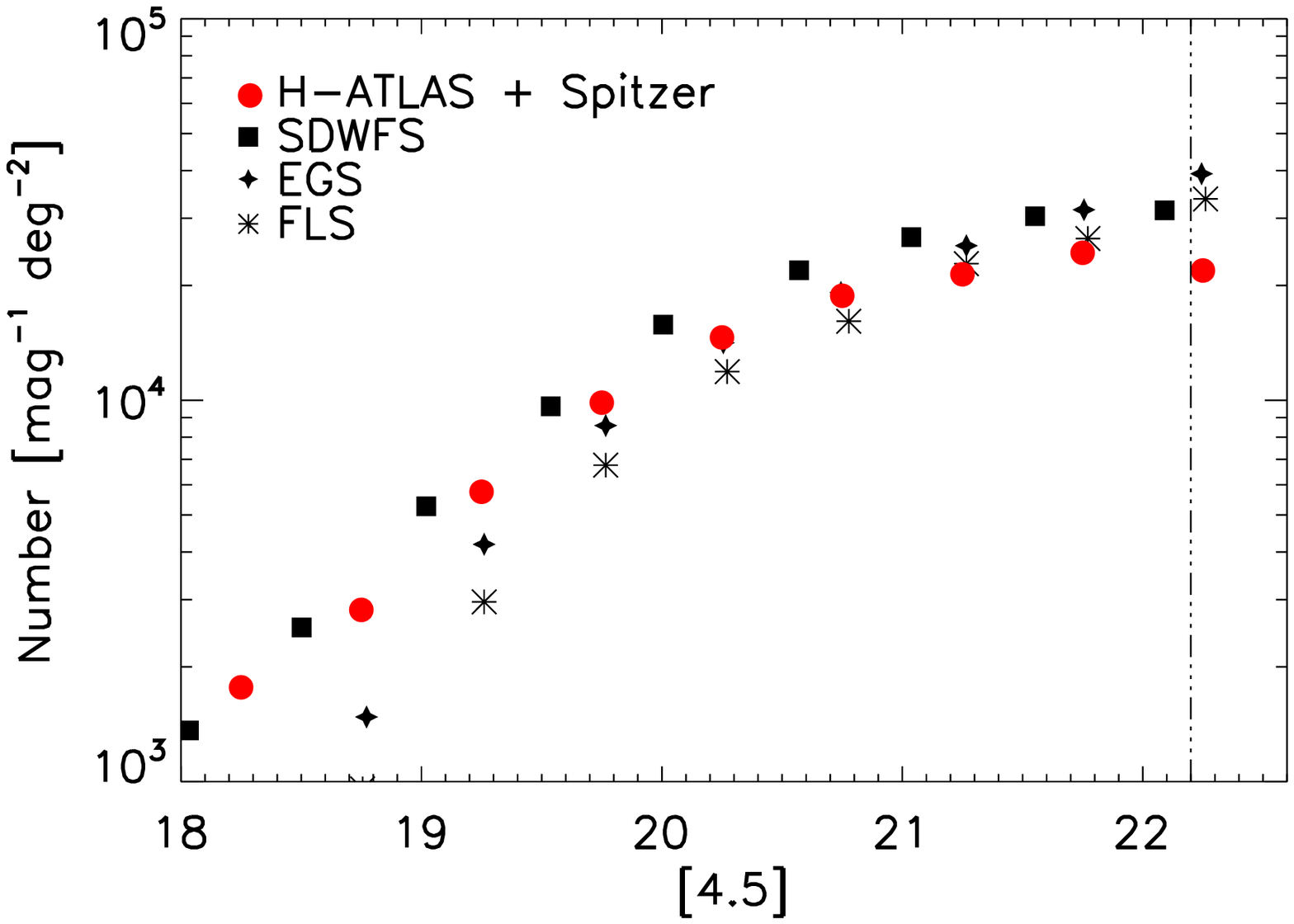}
  \caption{
Galaxy number counts from H-ATLAS/{\it Spitzer}  area compared with previous IRAC deep 
and wide fields: SDWFS (Ashby et al. 2009),
First Look Survey (FLS; Fazio et al. 2004b) and  EGS (Barmby et al. 2008). 
H-ATLAS galaxy counts are more consistent  with SDWFS.
The 50\% IRAC completeness limits of our imaging at 22.5 and 22.2 mag at 3.6 and 4.5\um, respectively, 
are indicated by vertical dash-dot-dot lines.   5$\sigma$ detection limit of SDWFS
with  3\arcsec\ aperture is  22.8 and  22.3 mag at  3.6 and 4.5\um, respectively.    }
  \label{fig:count}
\end{figure}

In figure~\ref{fig:count}   we compare the IRAC galaxy counts  in H-ATLAS/{\it Spitzer} area with some of the previous  IRAC deep and wide fields.  
H-ATLAS galaxy counts are more consistent with  SDWFS (Ashby et
al. 2009),  although the variations between fields are likely to be
  due to cosmic aariance.

\section{{\it Spitzer} IRAC  identification of SPIRE sources } 
\label{sec:ids}

The size of the SPIRE beam is 18.1\arcsec\ (FWHM) at 250\um, and 25.2
and 36.9\arcsec\ at 350 and 500\um, respectively. 
 The spatial density
of IRAC-bright galaxies is high enough that multiple sources may lie within  the SPIRE beam. 
Typically, this will include the true SPIRE counterpart and
  unassociated  foreground and background galaxies. However, the
  surface density of 250-\um\ bright sources is sufficient that a single SPIRE source 
  may  be composed of emission from  multiple galaxies.   
Therefore, choosing the nearest object as the counterpart
of a given SPIRE source can be misleading. Instead, we perform a
likelihood ratio (LR) analysis (Section~\ref{sec:lr}; Sutherland \&
Saunders 1992; Ciliegi et al.\,2003; Brusa et al.\,2007; Smith et
al.\,2011; Fleuren et al.\ 2012), which uses positional information and the magnitude
distribution of counterparts and background sources, to identify SPIRE
sources in the IRAC catalog.  Furthermore, the LR analysis
shows that galaxies associated with SPIRE sources occupy a
distinct region of IRAC color-flux space (figure~\ref{fig:color}),
which we use in
Section~\ref{sec:color} to identify additional
counterparts.

We note that other statistical matching techniques have been used in
astronomical implementations. For example, the corrected Poissonian
probability ($p$-statistic; Downes et al.\ 1986) uses the surface
density to calculate the probability of a source with the observed
magnitude being detected within the radius investigated (e.g. Ivison et
al. 2002, 2007; Pope et al. 2006; Chapin et al. 2009).  However, this
technique is most appropriate for catalogues in which the surface
density is low (e.g. radio data) and favors counterparts that are
brighter than the background population. Our IRAC data do not have
low-surface density, and we do not wish to assume that the {\it
Herschel} counterparts are typically  the brightest IRAC sources.  Bayesian techniques
(e.g. Budavari \& Szalay 2008), which use a priori knowledge of the
counterpart population to guide the identification process, could also
be employed (e.g. Brand et al. 2006; Gorjian et al.\ 2008). However, we
do not wish to bias our results by assuming a priori knowledge of the
SPIRE population at IRAC wavelengths. The LR analysis is advantageous
because it computes the intrinsic IRAC magnitude distribution of the
SPIRE sources from the data provided (see e.g. Fleuren et al.\ 2012).

\subsection{The Likelihood Ratio method} 
\label{sec:lr}

\begin{figure}
  \includegraphics[width=9cm]{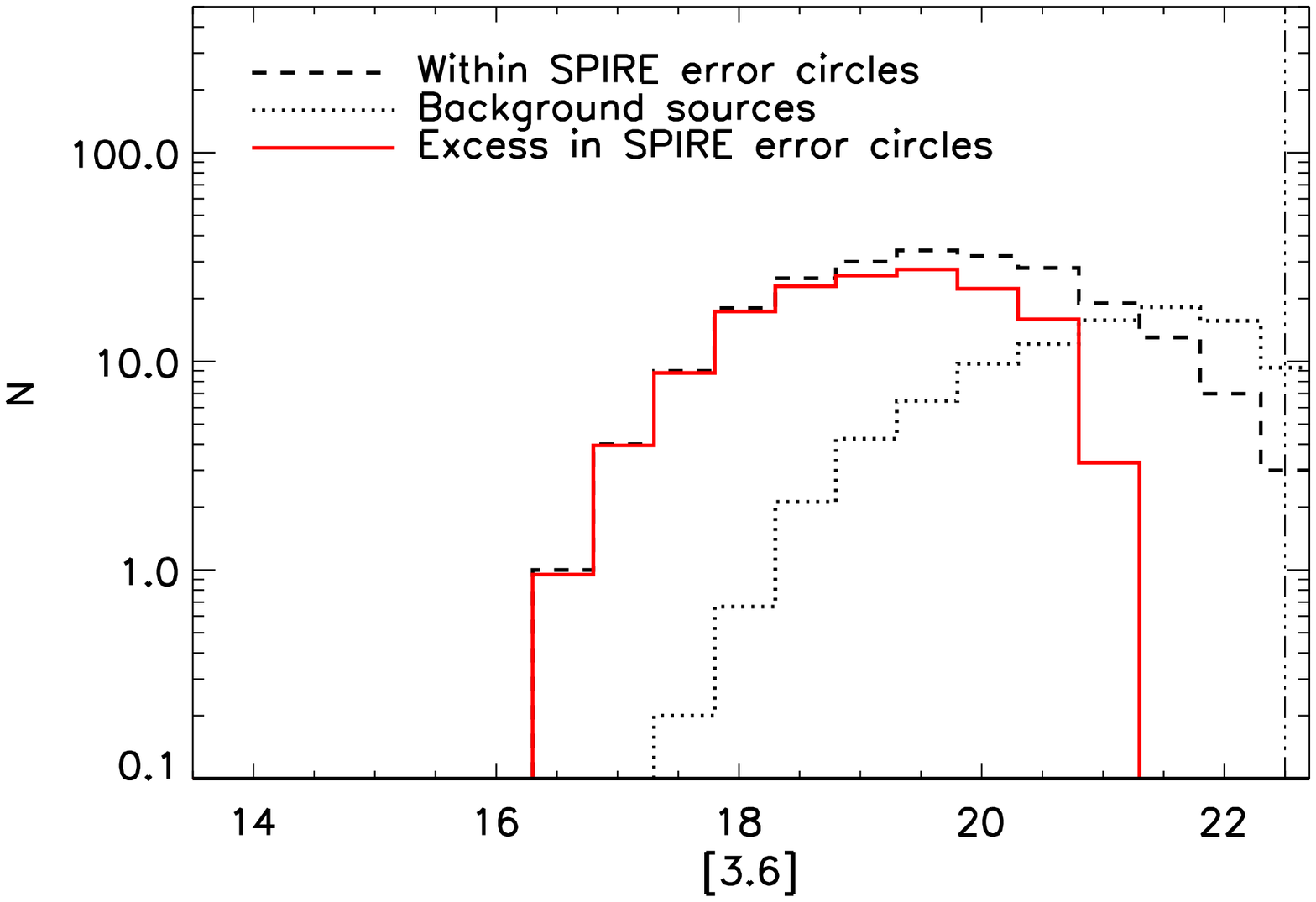}
  \includegraphics[width=9cm]{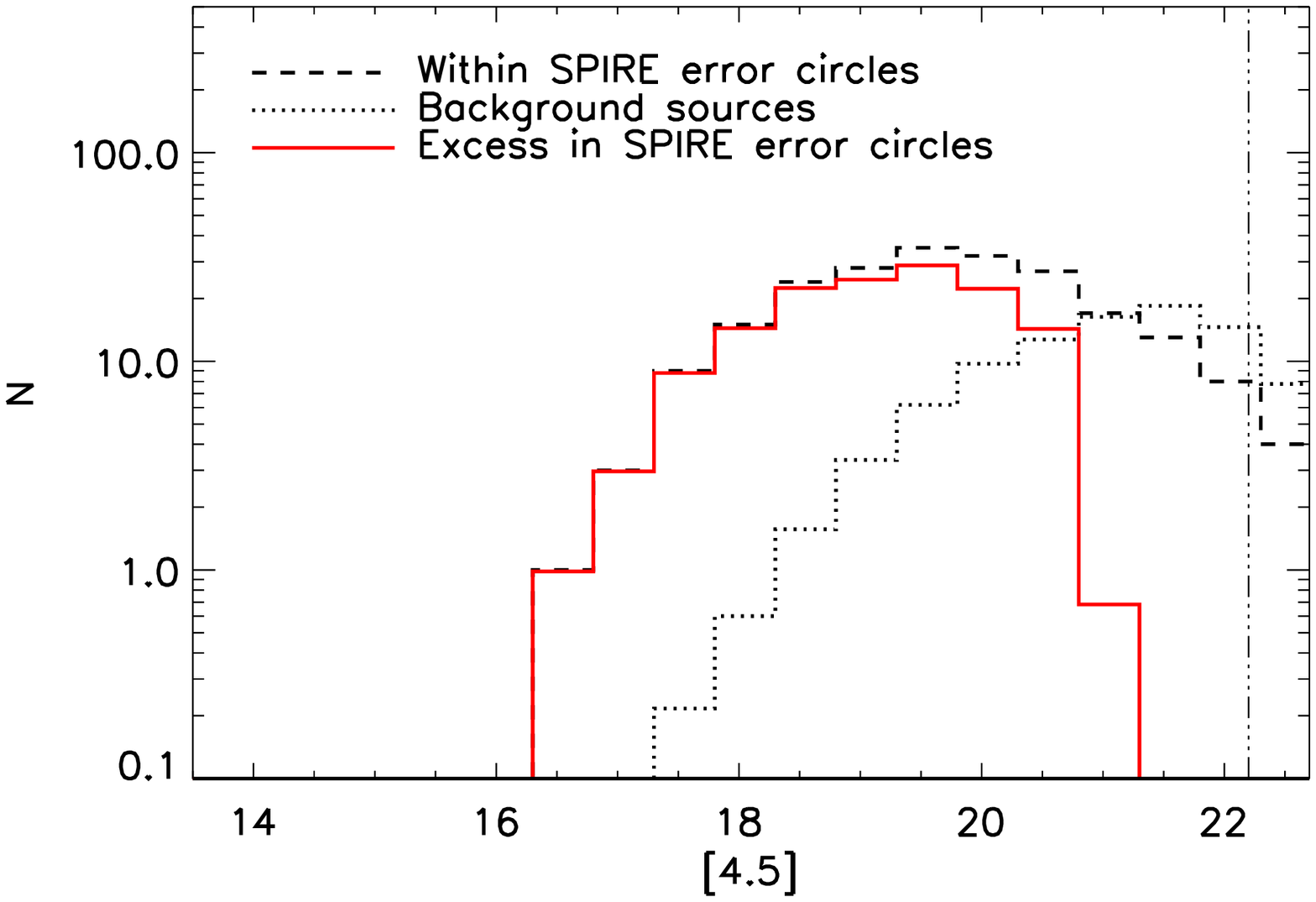}
  \caption{The magnitude distribution of IRAC galaxies at 3.6 ({\it
      top}) and 4.5\um\ ({\it bottom}) within the 7.2\arcsec\
    counterpart search radius around SPIRE sources. We calculate the
    distribution of background galaxies and subtract these to reveal
    an excess around SPIRE sources which peaks at 19.0 mag at both 3.6
    and 4.5\um\ . These are the magnitude distributions of detected
    IRAC galaxies that are associated with SPIRE sources.  The
    magnitude distribution of background sources is obtained by
    averaging the counts within 7.2\arcsec\ of 100 different
    realizations of 159 random positions that are each $>$ 15\arcsec\
    from the nearest SPIRE centroid.  To reduce the Poisson
    fluctuations associated with the small sample of 159 SPIRE
    centroids, the magnitude distribution of IRAC counts within SPIRE
    error circles is averaged over two bins.  The 50\% \spitzer-IRAC
    completeness limits of 22.5 and 22.2~mag at 3.6 and 4.5\um,
    respectively, are indicated by vertical dash-dot-dot lines.  All
    of the excess galaxies that are associated with SPIRE sources are
    brighter than these detection limits indicating that the IRAC data
    are sufficiently deep to detect counterparts to all typical SPIRE
    sources.  }
  \label{fig:hist}
\end{figure}

The LR, $L$, is the ratio of the  probability that the IRAC source is
the correct SPIRE counterpart with the equivalent probability for an
unassociated background source. Following Sutherland \& Saunders (1992),
$L$ is calculated as:
\begin{equation}
 L = \frac{q(m)f(r)}{n(m)} \,  ,
  \label{eq:lr}
 \end{equation}  
 where $q(m)$ and $n(m)$ are the normalized magnitude distributions of
 counterparts and background sources, respectively.  The radial probability
 distribution of the separation between the SPIRE 250\um\ and IRAC
 positions is denoted by $f(r)$.
 
We estimate $n(m)$, the normalized magnitude distribution of
  background sources, by averaging the source counts in areas of
   circle within 7.2\arcsec\ radius of 100 different realizations of 159 random
  positions.  The radius is  three times  the positional uncertainty of
 the SPIRE catalog. We use $\sigma_{\rm pos}=$ 2.4 $\pm$ 0.9\arcsec\ (Smith
 et al.\ 2011) calculated in the \atlas\ SDP field
 from the positional offsets between SDSS DR7 galaxies
 and 6621 SPIRE sources.   
  The center of each random circle is required to be
  at least 15\arcsec\ away from the nearest SPIRE centroid to minimize possible 
  contamination of real association.  The
  quantity $q(m)$, the normalized magnitude distribution of true IRAC
  counterparts to SPIRE sources cannot be directly derived.  We
  empirically estimate $q(m)$ by first subtracting the magnitude
  distribution of background galaxies from the magnitude distribution
  of all IRAC sources within search radius.

  This results in an estimate of the
  magnitude distribution of sources that are in excess of the
  background, $N_{\rm excess}(m)$, and assumed to be the true
  counterparts to SPIRE sources, Then $q(m)$ is calculated via:
    \begin{equation}
    q(m) =  \frac {N_{\rm excess}(m)}{\sum_m N_{\rm excess}(m)} \times Q_0\, , 
  \end{equation}
where  $Q_0$ is  the fraction of true counterparts  above the IRAC detection limit.    
The $[3.6]$ and $[4.5]$ distributions of $n(m)$ and $N_{\rm excess}(m)$ are shown in
  figure~\ref{fig:hist}.

The radial probability distribution, $f(r)$, is given by
\begin{equation}
 f(r)=\frac{1}{2\pi {{\sigma}^2_{\rm pos} }}~{\rm exp}\left(\frac{-r^2}{2{\sigma}^2_{\rm pos}}\right) \, ,
  \label{eq:f}
 \end{equation} 
 where $r$ is the separation between the SPIRE 250\um\ and IRAC
 centroids, and $\sigma_{\rm pos}$ is the positional uncertainty of
 the SPIRE catalog.   The median 250\,\micron\
   signal-to-noise ratio of the SPIRE sources in our sample is 6.33 
   and only 43 (27\%) have SNR$>8$. Therefore, we ignore the
   improvement in $\sigma_{\rm pos}$ for sources with high
   signal-to-noise ratio (Smith et al. 2011).  

 Each \atlas\ source may have several potential
 counterparts. Therefore, we calculate the reliability, $R_j$,
 of each IRAC galaxy, $j$, in SPIRE counterpart search
 radius. $R_j$ is the probability that the galaxy, $j$, is the
   correct IRAC counterpart to the SPIRE source. Following Sutherland
 \& Saunders (1992):
\begin{equation}
R_j=\frac{L_j}{\sum_i L_i + (1-Q_0)} \, ,
  \label{eq:r}
\end{equation}
where $i$ represents each IRAC source in the search radius.

To accept a potential counterpart as reliable we require that $R_j
\geq $ 0.8, i.e. there is $<$ 20\% chance of a false association.  This
criterion was used by Smith et al.\ (2011) for the counterpart search
with SDSS data and our results on the identification rate with IRAC
data can be directly compared to their results based on optical
imaging.  We note that the choice of the exact limiting value of $R$
does not strongly affect our conclusions. Of the 123 SPIRE sources
with $R\geq$ 0.8 counterparts (section~\ref{sec:lrresult}) 107 (87\%)
have $R\geq$ 0.9, and 101 (82\%) have $R\ge$ 0.95.

For a catalog in which identified sources are required to have
$R\geq$ 0.8 the overall false detection rate is
\begin{equation}
N({\rm false}) = \sum_{R_j \geq 0.8} (1-R_j) \, .
  \label{eq:n}
\end{equation}

\subsection{$Q_0$ calculation}
\label{sec:q0}

\begin{figure}
  \includegraphics[width=9cm]{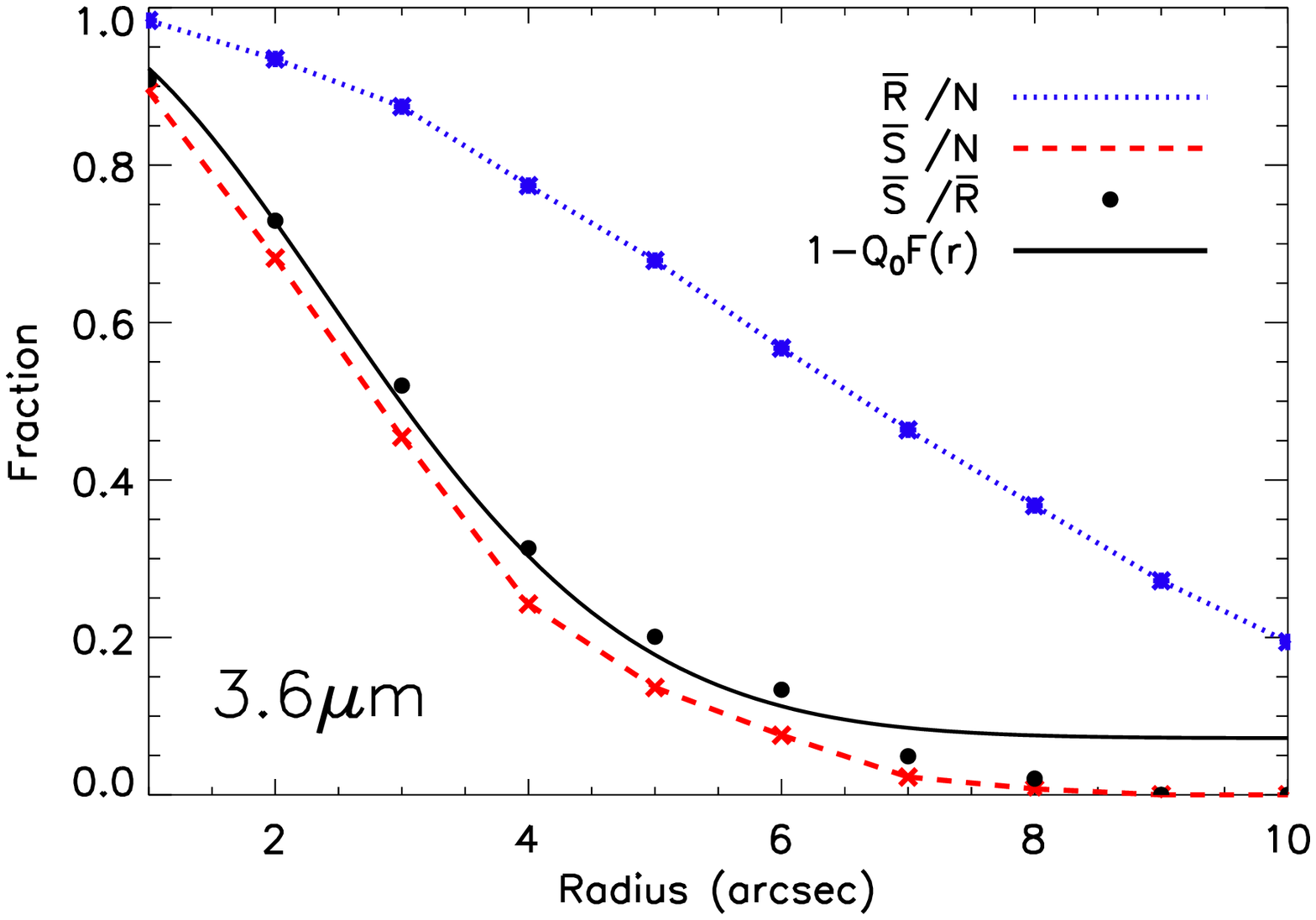}
  \includegraphics[width=9cm]{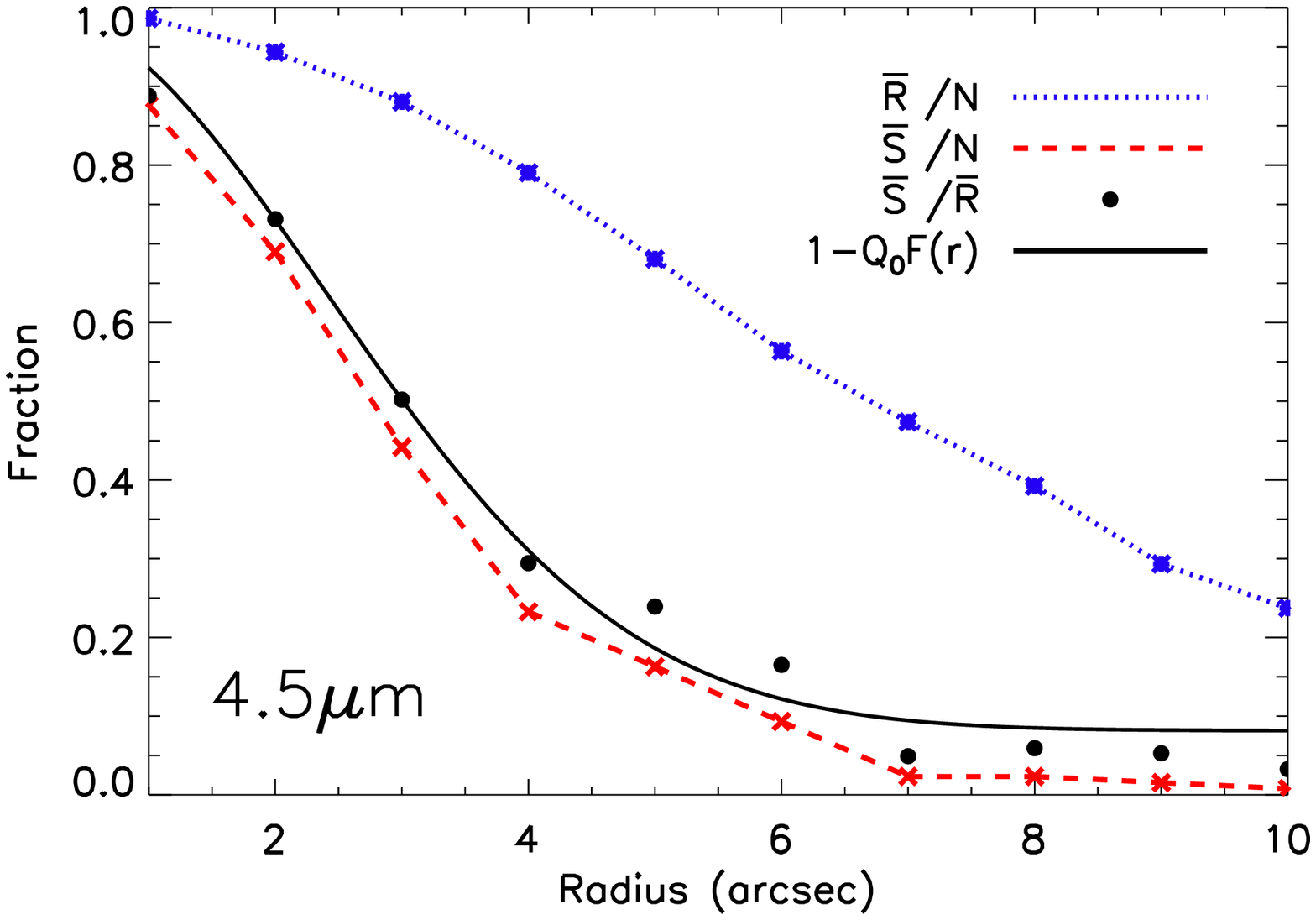}
  \caption{ Quantities employed in the calculation of
      $Q_0$. $\bar{S}(r)/\bar{R}(r)$, the ratio of SPIRE blanks to random
      blanks,  is fitted with $1-Q_0F(r)$
      in order to calculate $Q_0$. The best fit values are
    $Q_0=$ 0.93 $\pm$ 0.09 and  0.92 $\pm$ 0.08  at 3.6 and 4.5\um, respectively.
    We also show $\bar{R}(r)/N$ and $\bar{S}(r)/N$, the rate of random
      positions with at least one IRAC source and the rate of SPIRE
      blanks, respectively. }
  \label{fig:simone}
\end{figure}

In order to apply the LR method to the SPIRE and IRAC catalogs we must
first compute $Q_0$.  Smith et al.\ (2011) calculated $Q_0$ from the
total number of SDSS galaxies in the error circles of SPIRE
sources. However, this method requires some a priori chosen radius and
is only valid if each SPIRE source is associated with only one SDSS
galaxy.  Instead, we make use of an alternative method, derived by
Fleuren et al.\ (2012), which is radius independent and unaffected by
the clustering of IRAC sources. Clustering can be a significant
complication because SPIRE sources are expected to reside in overdense
regions similar to environments of sub-mm galaxies (e.g. Aravena et
al.\ 2010b; Hickox et al.\ 2012).

For the purposes of this discussion we define a SPIRE error circle
without a true IRAC counterpart as a {\it ``blank''}.  Directly
observed blanks arise due to counterparts below the IRAC detection
limit or beyond the search radius, $r$. However, the total number of
blanks must also be statistically corrected for contamination by
unassociated foreground or background galaxies. Thus, the true number of 
blanks ($\bar{S}_t$) is equal to the number of observed SPIRE blanks
($\bar{S}(r)$) and the number of SPIRE sources that are falsely
matched with  random background galaxies.  We define $R_o(r)$ as the
number of random positions that contain a IRAC galaxy while
$\bar{R}(r)= N - R_o(r)$ is the number of random blanks.  Then,
\begin{equation}
\bar{S}_t = \bar{S}(r) +[\bar{S}_t \times\frac{R_o(r)}{N}] \, ,
\label{eq:st}
\end{equation}
with $R_o(r)$ and $\bar{R}(r)$ calculated from 100 random realizations of
$N=159$ error circles, which are all located at least
15\arcsec\ away from SPIRE centroids.

$Q_0$ is defined as the fraction of true counterparts above the
  IRAC detection limit. Thus,  for an infinite search radius, the
  rate of true blanks, $\bar{S}_t/N$ is simply equal to
  $1-Q_0$. Fleuren et al.\ (2012) demonstrate that
  equation~\ref{eq:st} can be rearranged to show that
  $\bar{S}_t/N=\bar{S}(r)/\bar{R}(r)$, and thus
  $1-Q_0=\bar{S}(r)/\bar{R}(r)$. 

However, there is also the possibility that the true counterpart is
  outside of the examined area and this must be accounted for in the
  $Q_0$ estimate.  The probability that the real SPIRE source is
  outside the search radius can be derived analytically from the
  normalized SPIRE source distribution, $f(r)$ (equation~\ref{eq:f}).
  This probability, $F(r)$, is 
 \begin{equation}
F(r) =   \int_0^r P(r^\prime)\,\mathrm{d}r^\prime =1-e^{-\frac{r^2}{2\sigma^2}}  \, ,
\end{equation}
where $P(r) = 2\pi r f(r)$ (Fleuren et al.\ 2012). Assuming
that the probability of a SPIRE source being a true blank
($1-Q_0$) and the probability that the detected counterpart is
outside of the search radius ($1-F(r)$) are independent, it
  can be shown that the total probability that there is no
counterpart is (see Fleuren et al.\ 2012)
\begin{equation}
\frac{\bar{S}_t}{N} = 1- Q_0 F(r) \, .
  \label{eq:blank}
\end{equation}
However, we have already shown that
  $\bar{S}_t/N=\bar{S}(r)/\bar{R}(r)$. Thus, using the observables $\bar{S}(r)$
  and $\bar{R}(r)$ one can calculate $Q_0$ via $1- Q_0 F(r) =
\bar{S}(r)/\bar{R}(r)$.  Figure~\ref{fig:simone} shows $\bar{S}(r)/\bar{R}(r)$,
  $\bar{R}(r)/N$, and $\bar{S}(r)/N$, observed for search radii of one to
  10\arcsec. We calculate the constant, $Q_0$, by fitting $1-Q_0F(R)$
  to $\bar{S}(r)/\bar{R}(r)$. $\chi^2$ minimization  
  yields $Q_0 =$  0.93 $\pm$ 0.09,  0.92 $\pm$ 0.08 at 3.6 and 4.5\um,
respectively.

\begin{figure}
  \includegraphics[width=9cm]{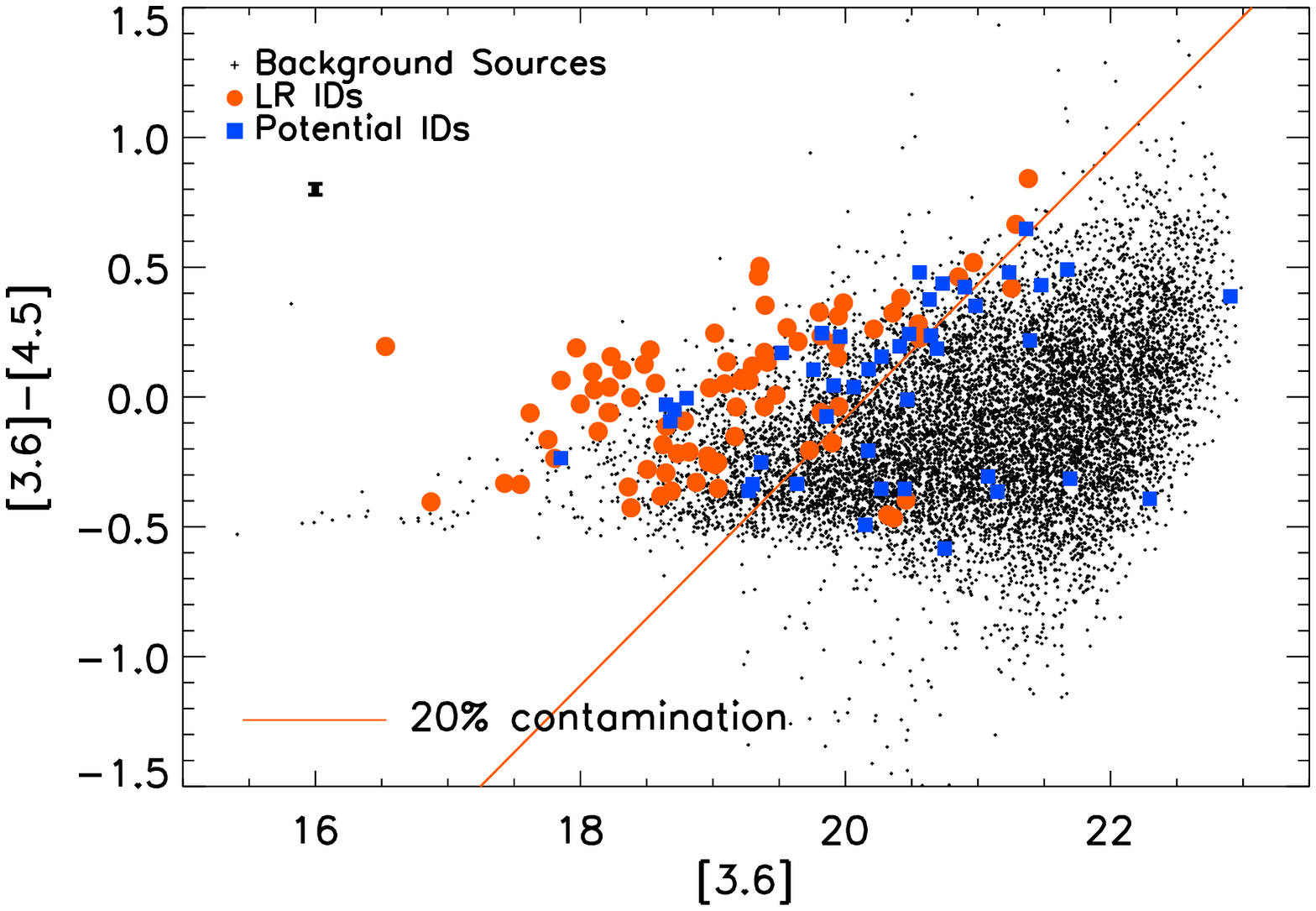}
    \includegraphics[width=9cm]{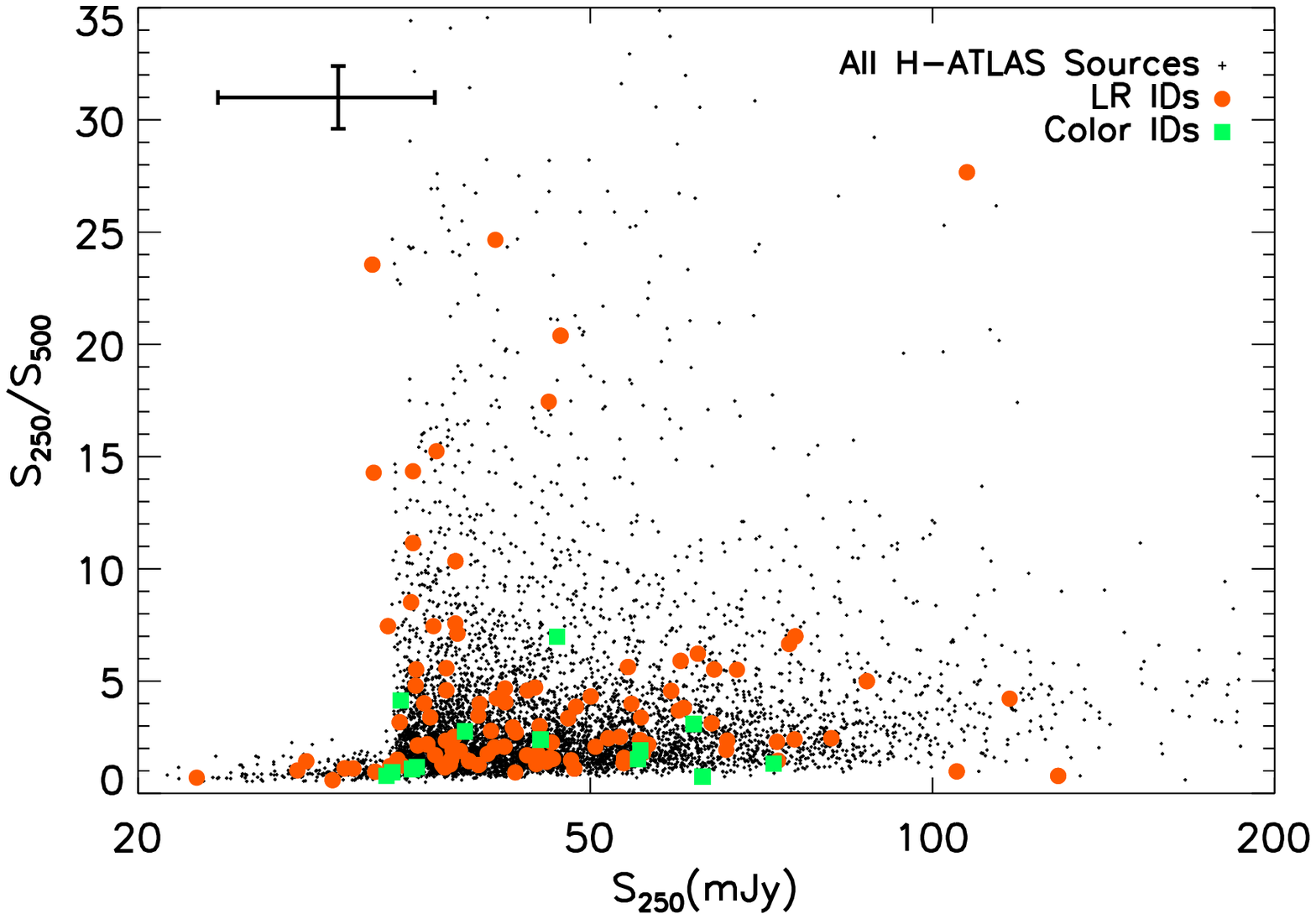}

  \caption{ {\bf Top:} IRAC color-magnitude diagram.  All LR
    counterparts with $R \geq$ 0.8 and potential counterparts are
      highlighted. Potential counterparts are IRAC galaxies within
      7.2\arcsec\ of SPIRE sources that are not otherwise identified
      with the LR method.  The LR counterparts typically have
    $[3.6]-[4.5]> -$ 0.4 and are brighter the background IRAC population.
    IRAC galaxies above the solid line and within 7.2\arcsec\ of SPIRE
    positions have a probability of being random associations of
    $\le$ 20\%. This discriminator is used to identify an additional 23
    IRAC counterparts to 13 SPIRE sources (``color IDs'';
      section~\ref{sec:color}). {\bf Bottom:} SPIRE color-flux diagram for all sources in the
    \atlas\ SDP field. We highlight sources with identified IRAC
    counterparts using the LR and color-magnitude methods.
    Sources that have formal 500\,\micron\ fluxes below the
      1$\sigma$ detection limit are shown as lower limits and are
      excluded from the statistical analysis. A two dimensional KS
      test between the background population and the identified SPIRE
      sources yields the probability that the two datasets are drawn
      from the same parent population of $p=$ 0.172. Thus, we conclude
      that the identified galaxies are not a strongly biased subsample
      of SPIRE sources, although there may be a small selection
      effect. Average error bars are shown at the top-left
    hand corner of both plots. }
   \label{fig:color}
\end{figure}

\begin{table*}
  \begin{center}
    \small
    \caption{Summary of IRAC counterparts to SPIRE sources}
 \label{tab:summary} 
    \begin{tabular}{lccccccc}
      \hline
      \hline
      Method & $N_{\rm SPIRE}$$^a$ & $N_{\rm ID}$$^b$ & ID rate$^c$ & $N_{\rm IRAC}$$^d$& $N_{\rm multiple}$$^e$  & Contamination rate  \\
      \hline   
      LR     &   159             &   123           &   77\%      &   123       &   0   &   1.9\%                             \\
      Color-magnitude&   19      &   13            &   68\%      &   23        &   7   &   12.6\%                            \\
      \hline                                                                           
      Total  &   159             &   136           &    86\%     &   146       &   7   &   3.6\%                             \\
      \hline\hline
    \end{tabular}
  \end{center}
  \tablecomments{  
    $^a$ Number of SPIRE sources considered. 
    $^b$ Number of SPIRE sources with at least one counterpart. 
    $^c$ Fraction of SPIRE sources with at least one identified counterpart. 
    $^d$ Number of IRAC counterparts identified. 
    $^e$ Number of SPIRE sources with multiple IRAC counterparts.}
\end{table*}

\subsection{LR counterparts}
\label{sec:lrresult}

We apply the LR technique outlined in section~\ref{sec:lr} and ~\ref{sec:q0} at 3.6 and 4.5\um,
  to identify IRAC counterparts to
123 of 159 \atlas\ SPIRE sources (Tables~\ref{tab:summary} and
\ref{tab:photom}). $q(m)$ and $n(m)$ are calculated in a radius of
3$\sigma_{\rm pos}=$ 7.2\arcsec\ (although $Q_0$ is radius independent;
section~\ref{sec:q0}) and only IRAC galaxies within 7.2\arcsec\ of
SPIRE sources are considered to be potential counterparts. This limit
includes 99.7\% of true counterparts, whilst minimizing the potential
contamination from unassociated sources.  The analysis is performed
separately at 3.6 and 4.5\um\ and any IRAC galaxy with $R\geq$ 0.8 at
either wavelength is considered to be a SPIRE counterpart.  In
addition, where both 3.6 and 4.5\um\ data is available we combine the
probabilities ($R$) from each wavelength, and include five
counterparts that have a combined probability of being considered to
be reliable identifications $\geq$ 0.8, but have $R<$ 0.8 at 3.6 and
4.5\,\micron\ individually.  All 123 counterparts are presented in
table~\ref{tab:photom}, including 81 galaxies with both 3.6 and
4.5\um\ photometry. The false identification rate is 1.9\% or
approximately two sources (equation~\ref{eq:n}); in the case of
counterparts that are identified at both 3.6 and 4.5\um\ we use the
highest of the two $R_j$ in this calculation.

The identification rate of 77\% is significantly higher than that from optical
(37\% in SDSS; Smith et al.\,2011) or near-IR analyses (51\% in VIKING;
Fleuren et al.\,2012) of \atlas\ sources.  This difference is
  likely to be primarily driven by the wavelengths of the study, which
  are less sensitive to $K$-correction and dust absorption than optical
  data. 

The counterpart identification rate is
  also significantly higher than the 50\% obtained with WISE
  3.4-\micron\ data (Bond et al. 2012); this is true even if we
  only consider the 3.6\,\micron\ data, where 97 of the 129 SPIRE
  sources with 3.6\,\micron\ coverage (75\%) are identified.  WISE is
  shallower than our IRAC observations -- 19.7~mag at 3.4\,\micron\
  compared to 22.5~mag at 3.6\,\micron\ in IRAC. However, 79\% of the
  3.6-\micron\ identified counterparts have $[3.6]\le$ 19.7 mag,
  suggesting that IRAC data to this depth would identify counterparts
  to $\sim$ 60\% of SPIRE sources. The remaining difference in the IRAC
  and WISE identification rates is likely to be due to the resolutions
  of the instruments -- the WISE 3.4\,\micron\ beam is $\sim3$ times
  larger than IRAC -- although Cosmic Variance may also contribute.
We note that if instead the data were limited to $[3.6] <$ 20.5,
corresponding to the 5$\sigma$ point-source detection limit reached in
a 120 sec integration with IRAC at 3.6\um, the expected identification rate
drops from 77\% to $\sim$ 70\%.

The results presented here are insensitive to the exact value of
  $Q_0$. When using a 1$\sigma$ lower value for $Q_0$ (i.e. $Q_0=$ 0.84
  at both 3.6 and 4.5\,\micron), and comparing to the results presented
  in section~\ref{sec:lrresult} and table~\ref{tab:photom}, the
  counterparts to 116 sources (95\%) are unchanged. In this case there are four
  (3\%) previously identified sources that no longer have reliable
  counterparts, and three (2\%) previously unidentified SPIRE sources
  that now have $R\geq$ 0.8 counterparts.

\subsection{Counterparts identified in IRAC color-magnitude space}
\label{sec:color}

While we reliably identify 77\% of the SPIRE sources with the LR
method, the values of $Q_0$ indicate that $\sim$ 90\% of SPIRE counterparts are
  detected in IRAC.  Figure~\ref{fig:color} shows the IRAC
color-magnitude diagram for galaxies in the \atlas\ SDP field, with
SPIRE counterparts identified with the LR method
highlighted. These galaxies occupy a distinct region of IRAC
color-magnitude space -- they typically have $[3.6]-[4.5]\ga -$ 0.4 and
are brighter than the background population.  We use this property to
identify counterparts to the SPIRE sources that have both 3.6 and
4.5-\um\ IRAC coverage but no $R\geq$  0.8 counterparts from the LR
method. A similar analysis was performed by Biggs et al.\,(2011) at 3.6
and 5.8-\micron\ to identify to LABOCA 870-\um\ sources.

We begin by determining the region of color-magnitude space in which
there is a minimal chance of contamination by background sources (the
ID region). A 20\% contamination limit is used because this equivalent
to $R\geq$ 0.8 for the LR method. We demarcate the ID region with the
simplest reasonable function -- a single diagonal line. The placement
of the line is determined by calculating the gradients and intercepts
that would yield 20\% contamination from background sources within
7.2\arcsec\ radius of the LR counterparts. The number of LR
counterparts returned is maximized to yield a best-fit gradient of
$0.515$ for a line that intercepts $[3.6]=$ 16.0 mag at
  $[3.6]-[4.5]= -$2.142. This division is shown as a solid line in
  figure~\ref{fig:color}.  IRAC galaxies that lie above this line and
are within 7.2\arcsec\ of SPIRE sources that are otherwise
unidentified, have $>$ 80\% probability of being physically associated
with the SPIRE source, i.e., $<$ 20\% chance of being an unrelated
foreground or background galaxies, and are considered counterparts.

There are 36 SPIRE sources that do not have LR counterparts. Of
  these 19 have both 3.6 and 4.5\,\micron\ data, and 23 IRAC
  counterparts to 13 of these sources are identified with the
  color-magnitude method (seven have multiple counterparts). These
  sources are presented in table~\ref{tab:photom}. The
  contamination rate for this method is 12.6\% (approximately two IRAC
  galaxies), calculated by summing the probabilities of finding a
  background galaxy within 7.2\arcsec\ of a SPIRE source 
  {occupying a distinct region of   color-magnitude space. }

In figure~\ref{fig:color} we show the $S_{250}/S_{500}$ color-flux
plot for \atlas\ SDP sources and consider whether sources with
identified IRAC counterparts are representative of the whole SPIRE
population.  A 2D Kolmogorov-Smirnov (KS) test between the background
\atlas\ SPIRE population and the sources with identified counterparts
has $p=$ 0.172, suggesting that the two samples are drawn from similar,
but not necessarily identical, parent populations. Considering the
$S_{250}/S_{500}$ color and 250\,\micron\ fluxes separately yields
$p=0.710$ and $p=0.289$, respectively. Thus we conclude that the
sources with identified IRAC counterparts are very similar to the
whole SPIRE population, but may have a slightly different distribution
of 250\,\micron\ flux.

\subsection{Astrometric offset}
\label{sec:offset}
\begin{figure}
  \includegraphics[width=8cm]{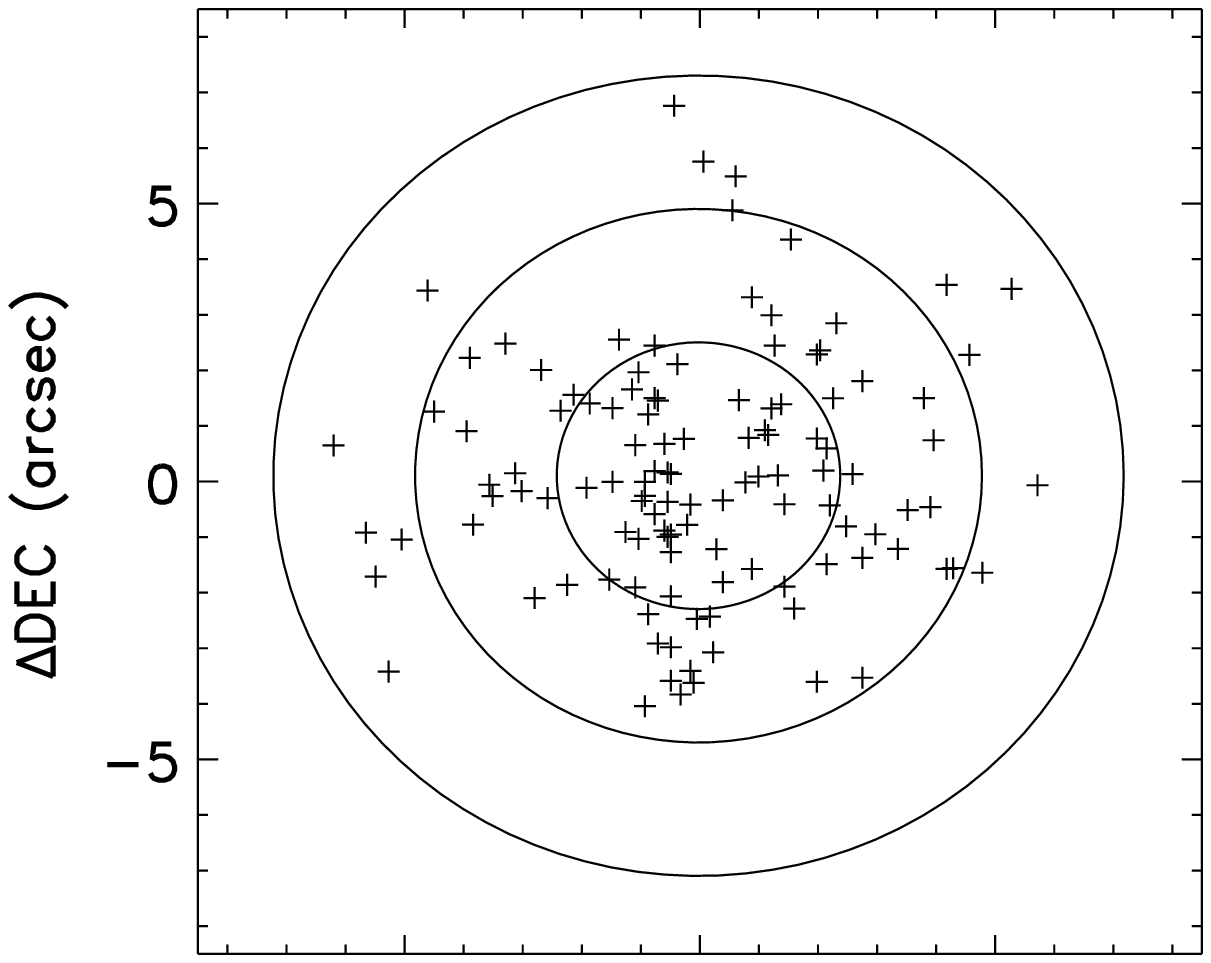}
  \caption{
 The distribution of $\Delta$RA  and  $\Delta$Dec offsets between SPIRE sources and the corresponding IRAC counterparts identified in this work.  
The  astrometric median  separation between  SPIRE sources  and  IRAC  is  0.10\arcsec\  which is consistent  with the overall IRAC astrometric 
uncertainty tied to the SDSS astrometry.   
The three concentric circles  have radii of  1, 2 and 3$\sigma_{\rm pos}$ (7.2\arcsec) from the SPIRE centroid. The overall offsets in RA and Dec show
 that there is no systematic offset between input  SPIRE and IRAC sources.    
 }
  \label{fig:offset}
\end{figure}

 To check whether there is any noticeable astrometric shift among the input catalogs for LR analysis,  in figure~\ref{fig:offset} we plot the positional differences between the SPIRE sources and the corresponding IRAC counterparts. 
 The  median  separation between SPIRE  and  IRAC  sources is  0.10\arcsec.  
We also investigate the overall astrometric reference frame difference between IRAC and SDSS catalogs, using $\sim$ 40\% of the IRAC sample that
is also identified with SDSS galaxies within in 2$''$. The positional difference  between IRAC and SDSS is  0.19 $\pm$ 0.17\arcsec
and we found no systematic offset either in RA or Dec.

Our IRAC data are complemented with VIKING $Z,  Y, J, H$
and $K_{\rm s}$-band photometry, which is measured in 2\arcsec\
  diameter apertures (Sutherland et al.\,in prep). We
cross-match the IRAC (FWHM $\sim$ 1.6\arcsec) to the nearest neighbor  VIKING (FWHM $\sim$ 0.9\arcsec) source
within 2\arcsec\ search radius.  Indeed, the mean offset between the
IRAC and VIKING positions is 0.09 $\pm$ 0.50\arcsec\ and 0.28 $\pm$ 0.28\arcsec\ in
RA and Dec, respectively. 
We find no statistically significant astrometric error between  two catalogs.

\begin{figure*}
\begin{minipage}{17.5cm}
\begin{center}
\subfigure{
\includegraphics[width=4.5cm]{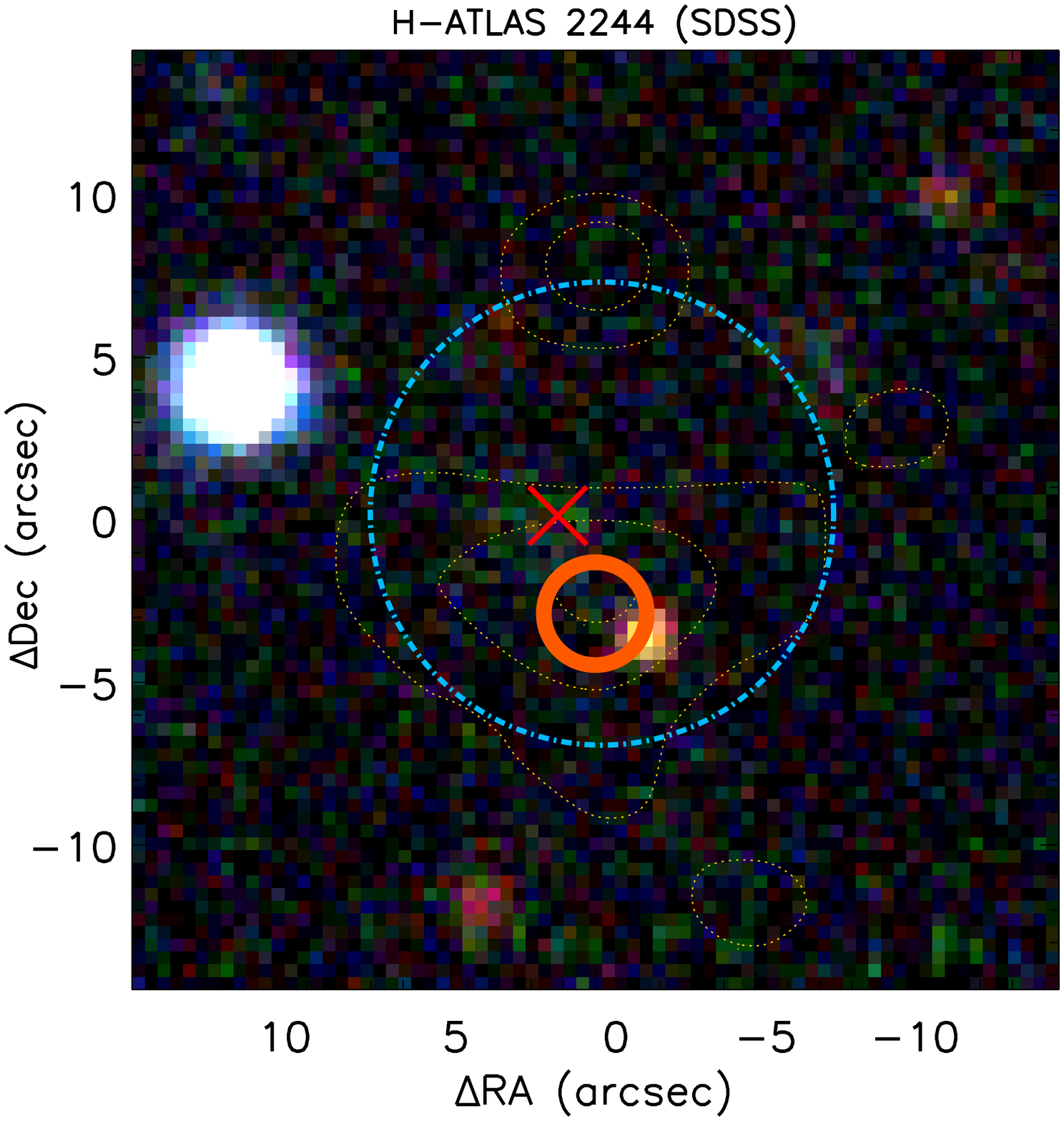} }                \quad
\subfigure{
\includegraphics[width=4.5cm]{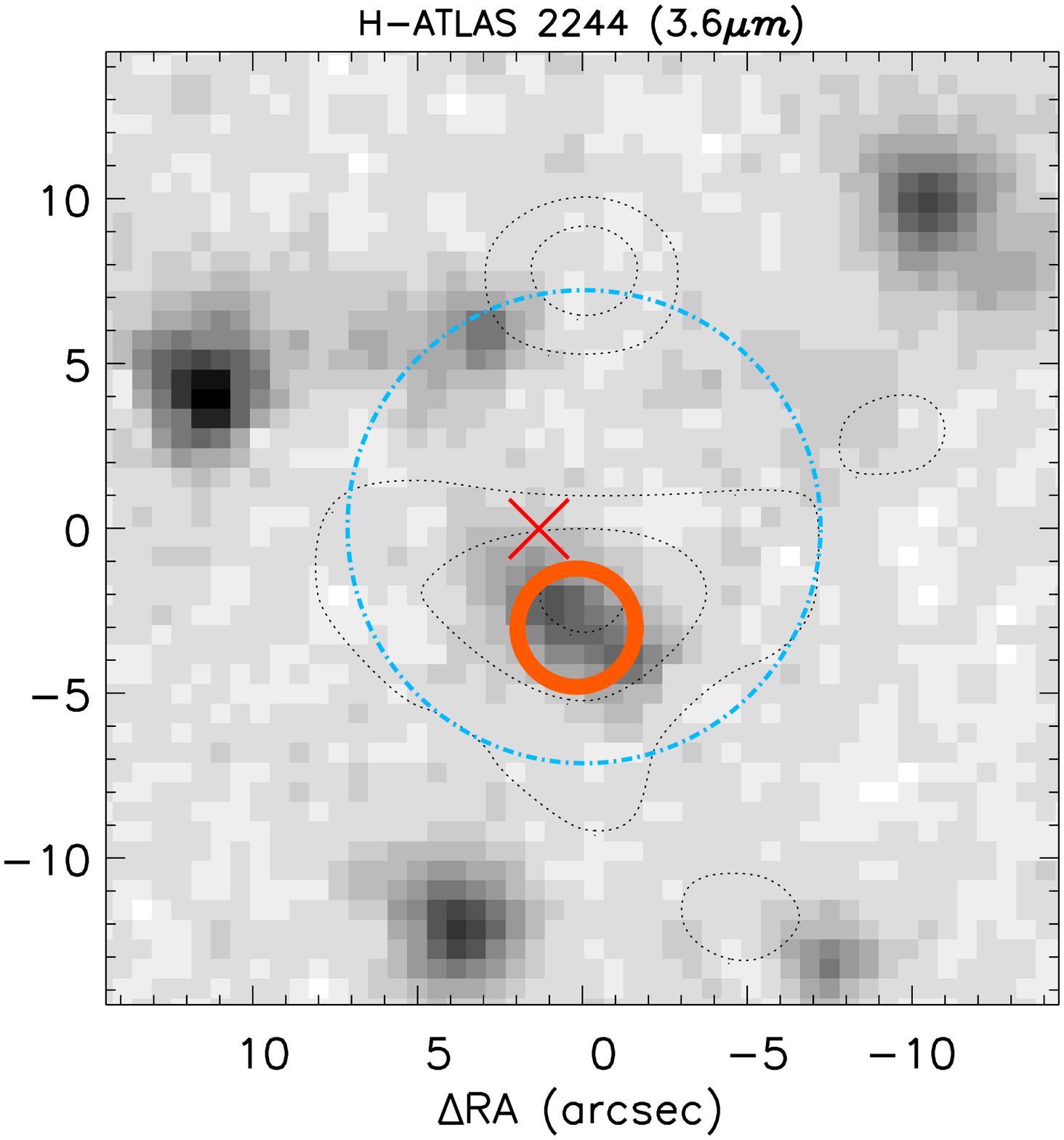} } \quad
\subfigure{
\includegraphics[width=4.5cm]{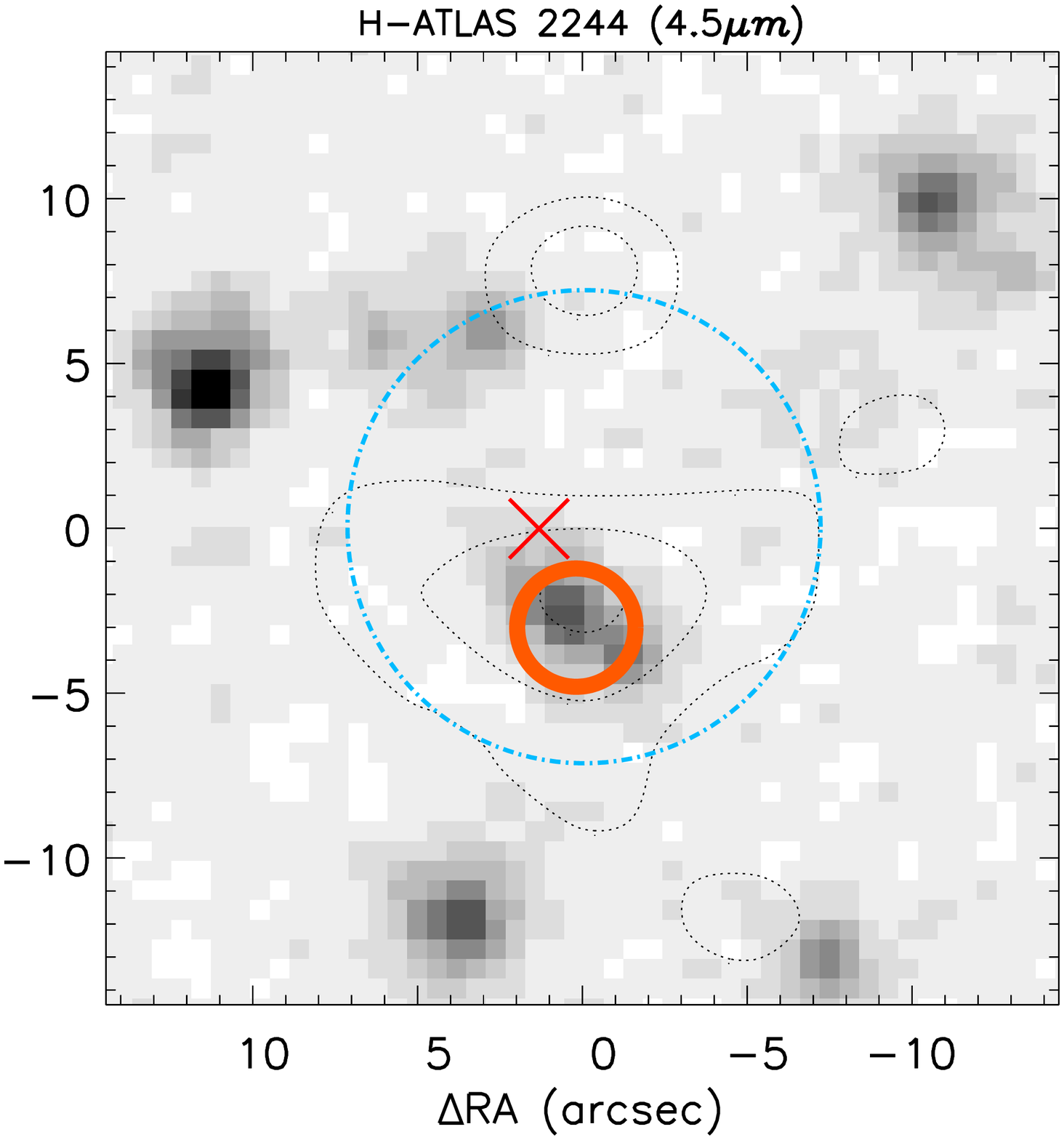} } 
\end{center}
\end{minipage}
\caption{ H-ATLAS ID-2244: an example of source that is identified
  using the LR method, but where the IRAC counterpart ($R=$ 0.99 at
  both 3.6 and 4.5\,\micron) is different from the SDSS
  counterpart (SDP.2244, $R=$ 0.97; Smith et al.\ 2011).  From left-to-right we
  show a three-color SDSS image ($g$, $r$, and $i$ bands), the
3.6\,\micron, and the 4.5\,\micron\ data.  The galaxies identified as
the SDSS (SDP.2244) and IRAC (H-ATLAS ID-2244) counterparts to the SPIRE source are marked with an
`X' and a small circle, respectively.  The large circle has
7.2\arcsec\ radius and encompasses the SPIRE 3$\sigma_{\rm pos}$
area in which counterparts are identified. Contours show the SPIRE
250\,\micron\ emission at 5, 7, 9, 11$\sigma$ levels.  In the
absence of high-resolution sub-mm imaging we cannot determine whether
this SPIRE source is the results of blended emission from the two
identified galaxies, or whether one of those counterparts is a chance
association.}  
\label{fig:postage1}
\end{figure*}

\begin{figure*}
\begin{minipage}{17.5cm}
\begin{center}
\subfigure{
\includegraphics[width=4.5cm]{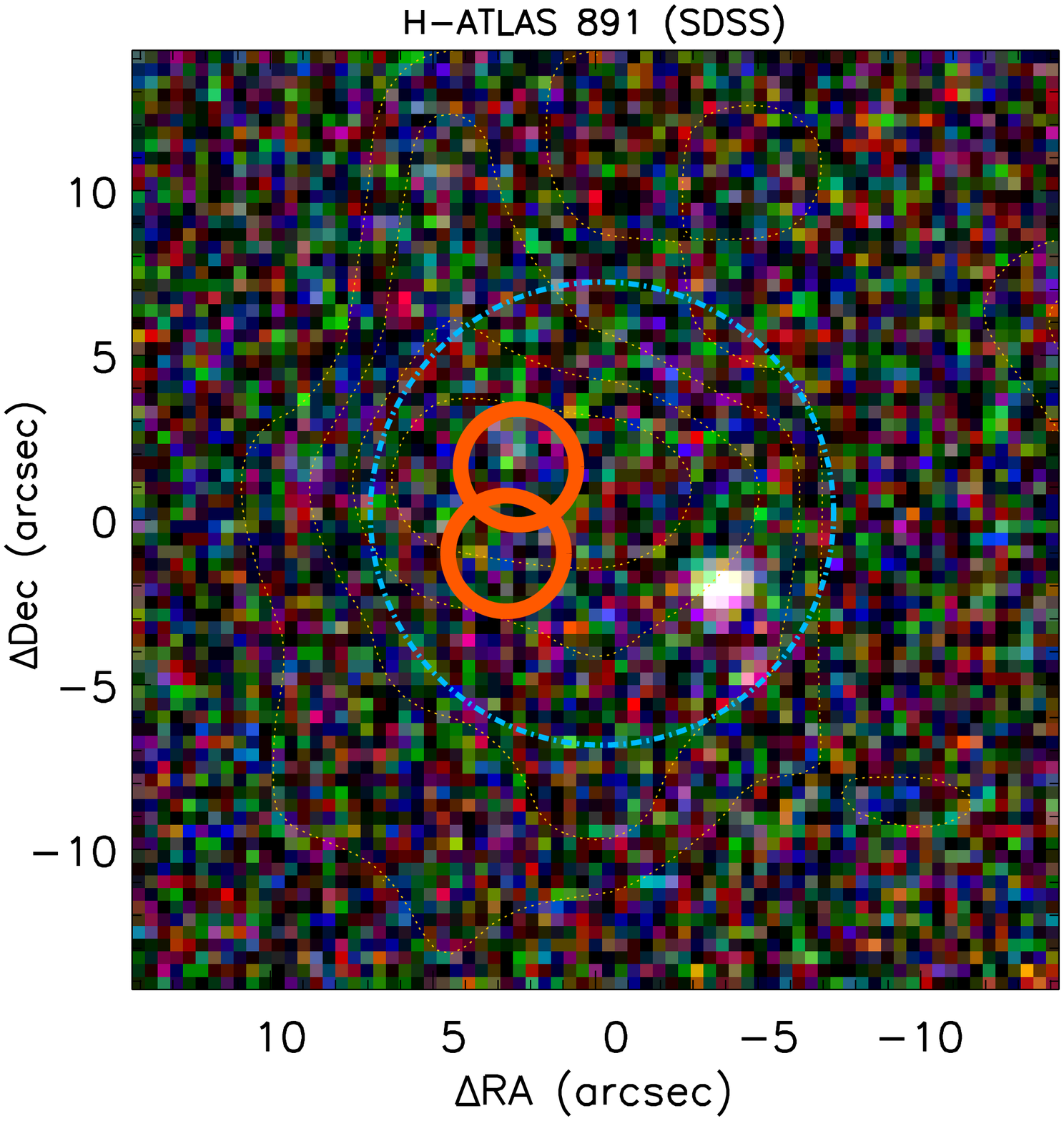} }                \quad
\subfigure{
\includegraphics[width=4.5cm]{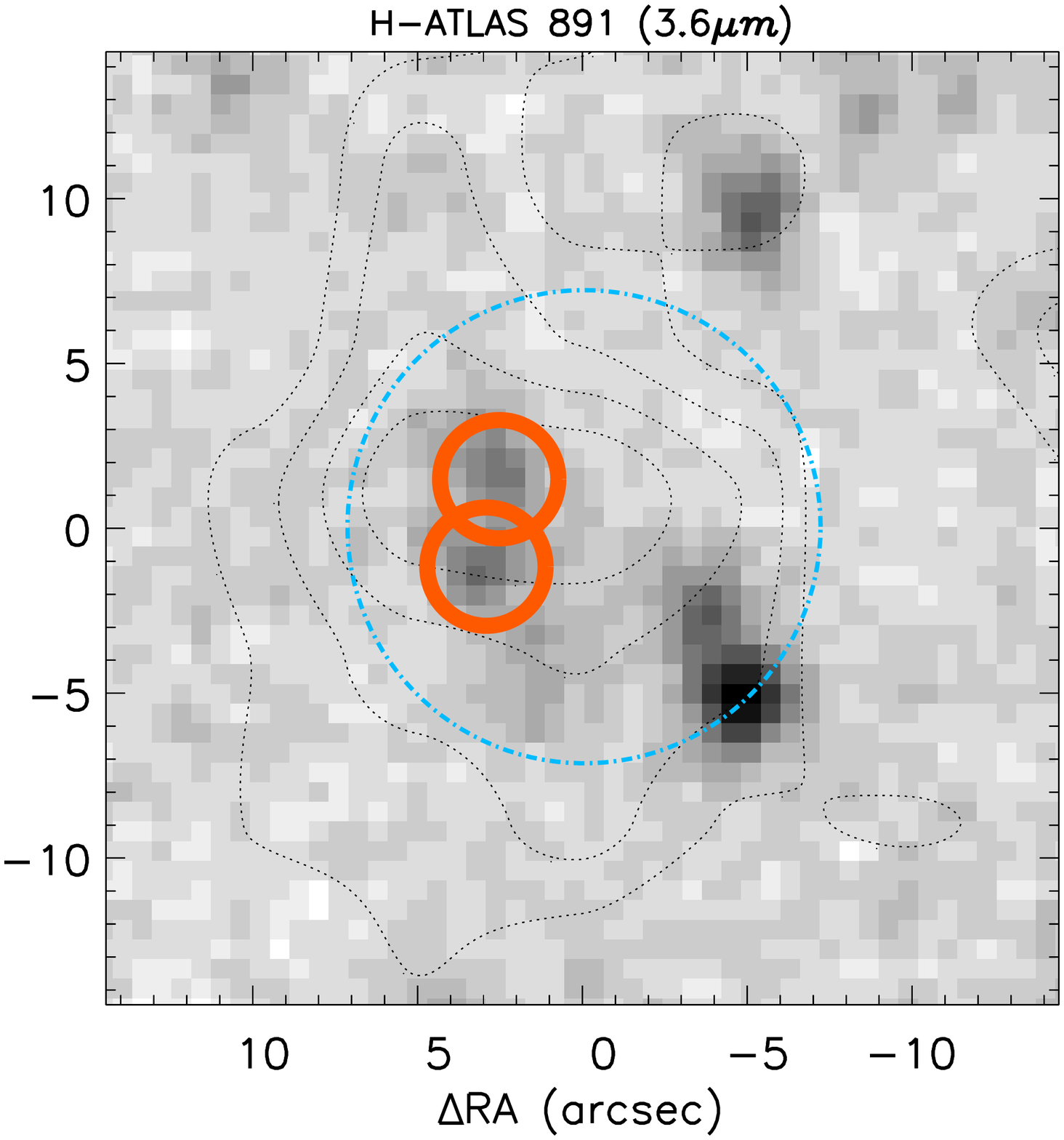} } \quad
\subfigure{
\includegraphics[width=4.5cm]{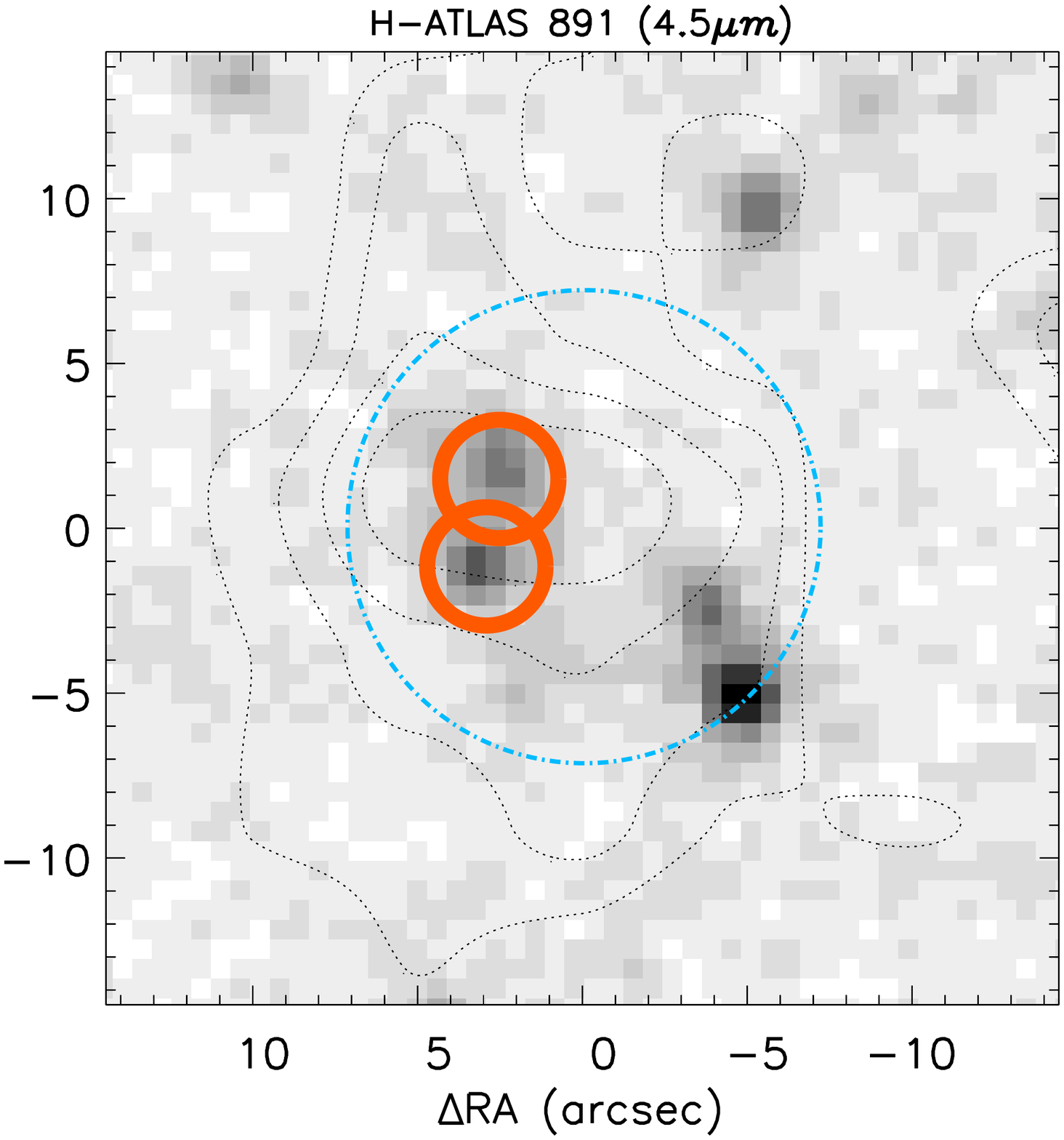} } 
\end{center}
\end{minipage}
\caption{H-ATLAS ID-891: an example of counterpart identification using
  the color-magnitude method.  Symbols are as
  figure~\ref{fig:postage1}. There is no LR counterpart in the SDSS
  $r$-band (Smith et al.\ 2011) or with IRAC at 3.6 or 4.5\um.
  However, there are two galaxies that are both identified as
  SPIRE counterparts on the basis of their IRAC colors and
    magnitudes. These two galaxies have $R=$ 0.41 and  0.47  at
  3.6\,\micron, and $R=$ 0.44 and  0.50 at 4.5\,\micron\ and neither
  is detected in SDSS.  It is probable that the SPIRE source is a blend
  of the emission from these two IRAC galaxies although, in the
    absence of spectroscopic redshifts it is unclear whether the
  configuration is caused by a line-of-sight alignment or an
  interaction between the two IRAC galaxies.  }
\label{fig:postage2}
\end{figure*}

\begin{figure*}
\begin{minipage}{17.5cm}
\begin{center}
\subfigure{
\includegraphics[width=4.5cm]{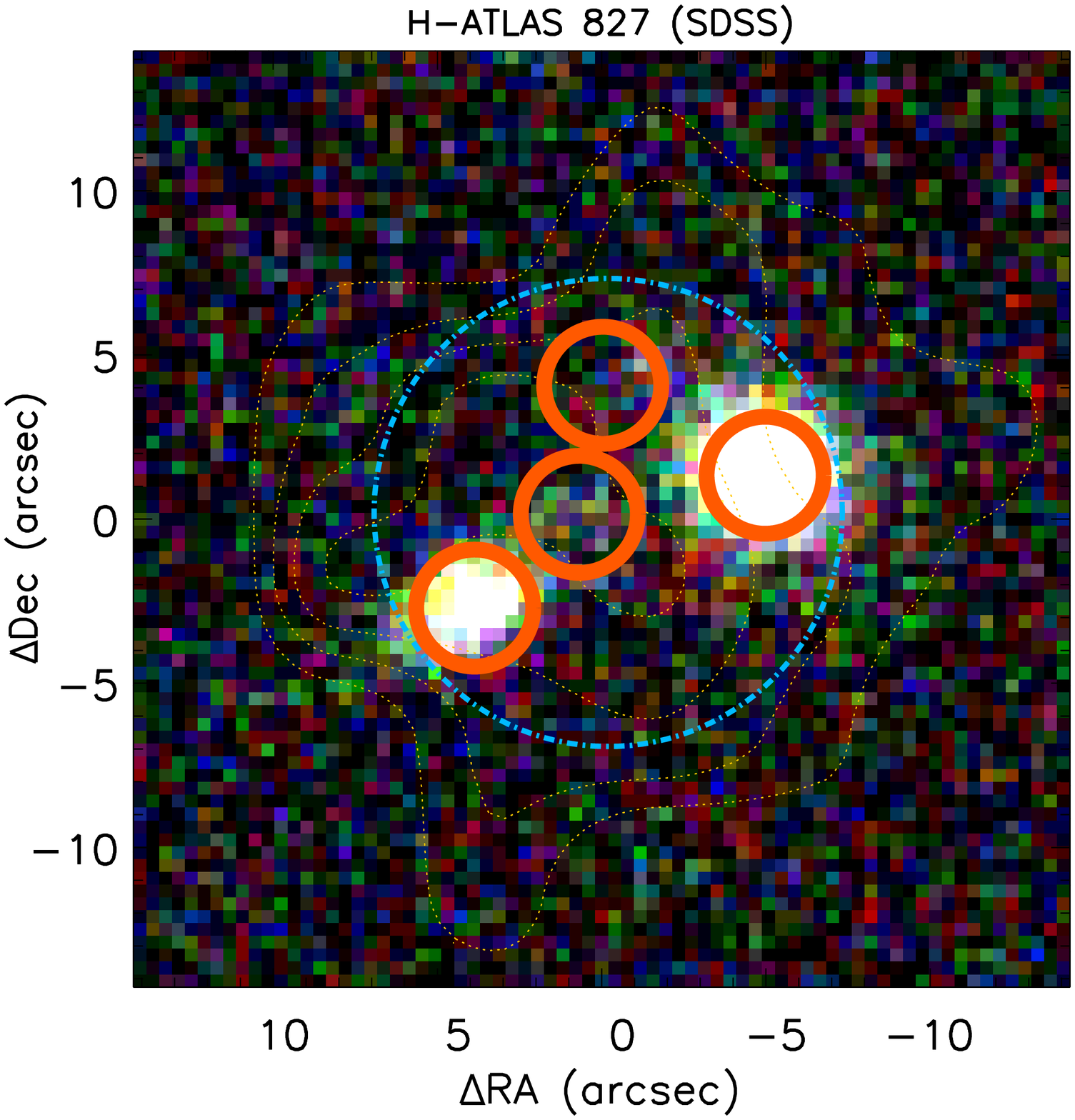} }                \quad
\subfigure{
\includegraphics[width=4.5cm]{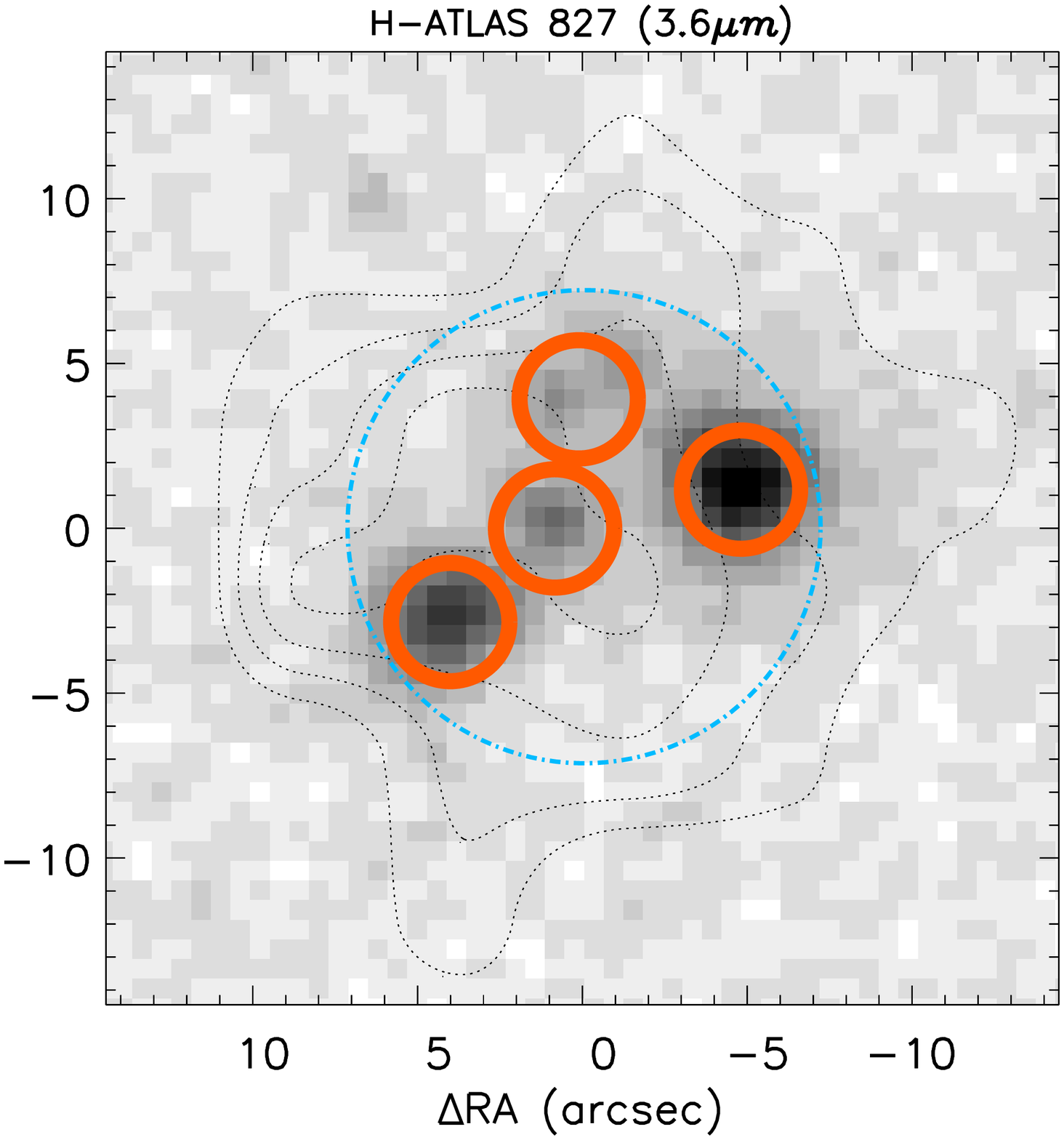} } \quad
\subfigure{
\includegraphics[width=4.5cm]{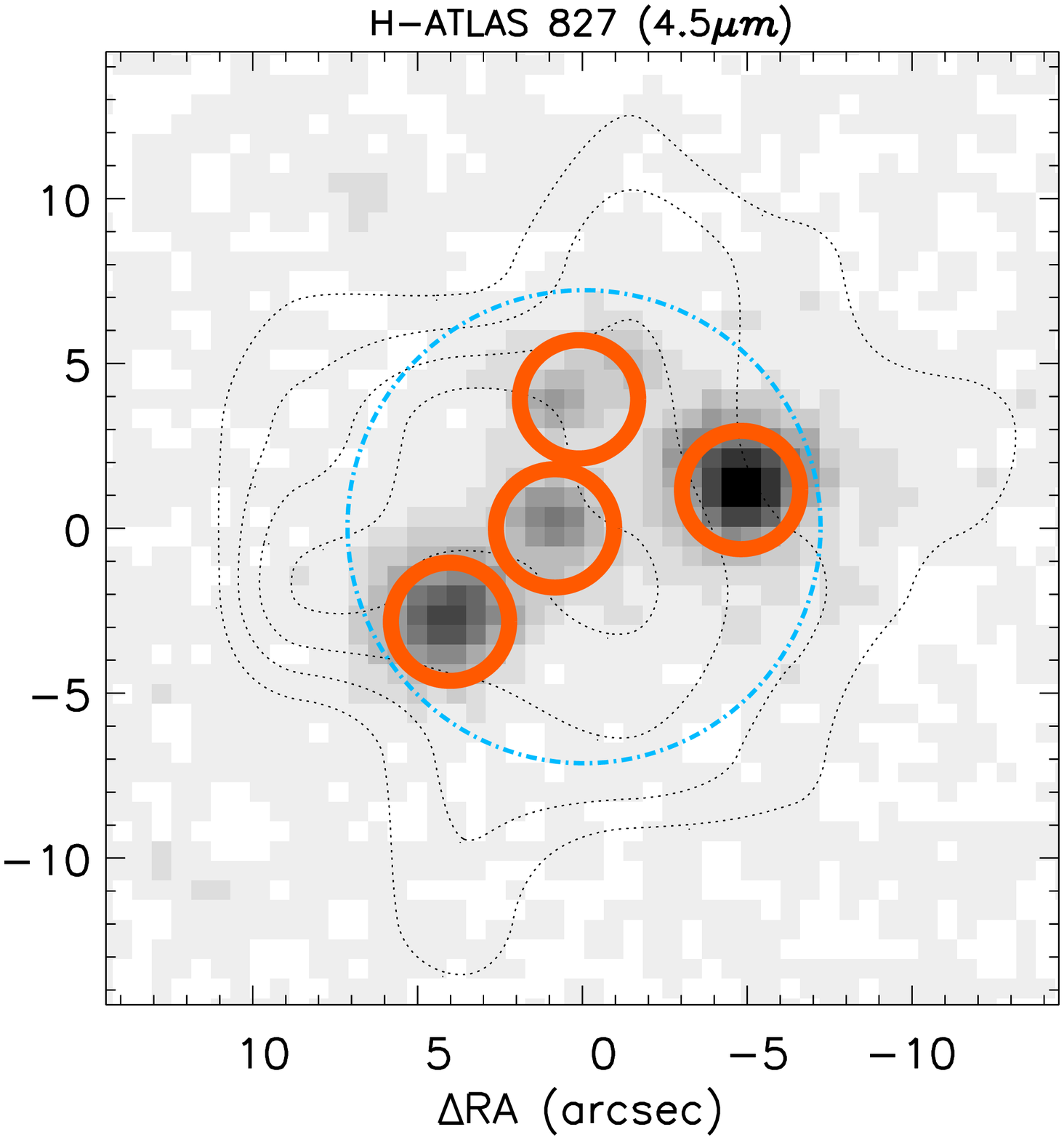} } 
\end{center}
\end{minipage}
\caption{H-ATLAS ID-827 has four counterparts identified with the
    IRAC color-magnitude method and is the most complex system in our
    sample. The westernmost and easternmost galaxies are detected by
    SDSS (but are not identified as counterparts in these data; Smith
    et al.\ 2011), and have photometric redshifts of $z=$ 0.07 $\pm$ 0.01
    and $z=$ 0.20 $\pm$ 0.03, respectively. The two fainter counterparts
    are not detected in SDSS but have $[3.6]-[4.5]>$ 0, indicating that
    they may be at $z\gtrsim$ 1.4 (Papovich 2008). We conclude that the
    SPIRE source is most likely to be comprised of blended emission
    from the four galaxies. Symbols are as
    figure~\ref{fig:postage1}.}
\label{fig:postage3}
\end{figure*}

\section{Discussion}
\label{sec:discussion}

As described in sections~\ref{sec:lrresult} and \ref{sec:color} we
identified IRAC counterparts to 136 (86\%) of the SPIRE sources. 123
of these are identified with the LR method and 23 counterparts to 13
SPIRE sources are identified with the color-magnitude method
(Table~\ref{tab:summary}).  The contamination rate of 1.9\% for the LR
method is smaller than that from the color-magnitude method (12.6\%)
but corresponds to a similar number of IRAC galaxies -- two to three.
Combining the results from the two identification methods, we expect
that the total false identification rate of the catalog presented in
Table~\ref{tab:photom} is 3.6\%, such that five to six of the IRAC
counterparts presented in table~\ref{tab:photom} are false.

\begin{figure}
  \includegraphics[width=9cm]{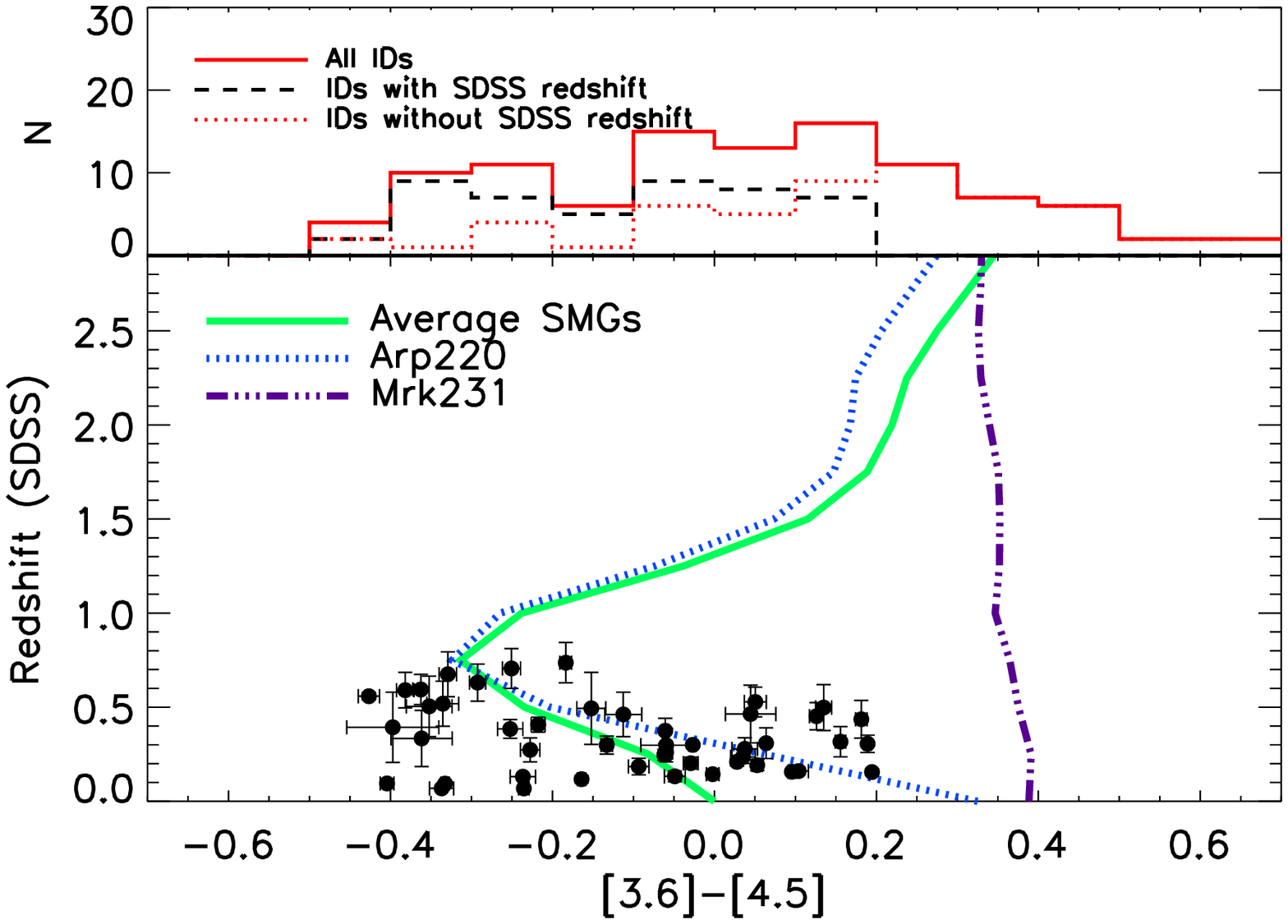}
  \caption{The distribution of IRAC $[3.6]-[4.5]$ color for SPIRE
    counterparts with photometry at both wavelengths.  For the 47
    sources with SDSS spectroscopic or photometric redshifts we also
    plot the $[3.6]-[4.5]$ color against redshift. The 57 SPIRE-IRAC
    identifications without reliable redshifts are either faint or
    undetected in SDSS and 43 of them have $[3.6] - [4.5] >$ 0,
    an indicator of high redshift source ($z\gtrsim 1.4$; Papovich
    2008).  We expect that more than half of the sub-mm bright sources
    reside at $z \gtrsim$ 1.4.  For reference the color-redshift
    tracks of Arp~220 (Silva et al.\,1998), Mrk231 (Berta\, 2006) and the average sub-mm galaxy  SED (Wardlow et
    al.\, 2011) are shown.  }
  \label{fig:stat}
\end{figure}

\subsection{Statistics of Identified Sources}

We next compare the counterparts identified in the IRAC data with
results based on SDSS (Smith et al.\ 2011).  Of the 146 IRAC
counterparts, 102 (70\%) are undetected in SDSS, 52 (36\%) agree with
the SDSS identification, and two (1\%) are different galaxies to the
SDSS counterparts. On the basis of the false-identification rates of
the two studies (4.2\% for SDSS and 3.6\% for IRAC) approximately four
counterparts are expected to disagree between the two surveys, which
is consistent with that observed here.
  
The two sources with different counterparts identified in IRAC and
SDSS are H-ATLAS ID-2244 and 6962. We show one of these sources,
H-ATLAS ID-2244, as an example in figure~\ref{fig:postage1}. In this
case the SDSS (SDP.2244) and IRAC (H-ATLAS ID-2244) counterparts are separated by 3.2\arcsec\ and
both have a $\le$ 3\% probability of being false identifications:
$R=$ 0.97 in SDSS and $R=$  0.99 at both 3.6 and 4.5\,\micron. In the
absence of additional data, such as sub-mm or radio interferometry,
the true nature of this source is unclear. It is possible that one of
the counterparts is a chance association, although we cannot say which
one. It is also possible that the SPIRE source is comprised of a blend
of emission from both counterparts in a single SPIRE beam.

Blending is also an important consideration for counterparts that
  are identified with the color-magnitude method.  Seven of the 13
  SPIRE sources identified with this method have multiple IRAC
counterparts; conversely the LR method has none. This apparent
disparity is not surprising because the LR method implicitly assumes
that each SPIRE source has a single IRAC counterpart. Where there are
$N$ multiple counterparts contributing equally to the sub-mm flux
the average $R$ cannot exceed $1/N$, and will typically be $\sim
  1/N$.  Due to the large SPIRE beam (FWHM = 18.1\arcsec\ at 250\um)
it is likely than in at least a few cases a single SPIRE source
may be composed of blended emission from multiple galaxies or from
galaxy interactions. The color-magnitude method does
not consider the presence of other IRAC sources and is therefore not
biased against multiple counterparts. In addition, beyond requiring
that counterparts to lie within the SPIRE counterpart search radius,
the color-magnitude method does not consider the separation between
the SPIRE and IRAC centroids and therefore it is not biased against
wide-separation counterparts. However, it does require both 3.6 and
4.5\um\ fluxes and it assumes that all SPIRE sources have a similar
color-magnitude distribution at 3.6 and 4.5\um.

The seven SPIRE sources in the sample that have multiple IRAC
counterparts are likely to be a combination of blended SPIRE sources
and individual sub-mm galaxies  associated with multiple interacting IRAC
sources. \atlas\ ID-891 is shown as an example in
figure~\ref{fig:postage2}. In this case all the detected SDSS galaxies
are $\ga 3.5\arcsec$ away from the SPIRE centroid and all have $R<$ 0.8
(Smith et al.\,2011).  There are two potential counterparts in the
IRAC data that are not detected in SDSS.  Using the LR method alone
these two nearby sources have reliability of $R=$ 0.41 and $0.44$ at
3.6\um\ and $R=$ 0.47  and  0.50  at 4.5\um.  However, both galaxies
have $[3.6]$ and $[4.5]$ that place them in the region of
color-magnitude space with $<$ 20\% probability of being random
associations with the SPIRE emission and as such they are both
identified as counterparts.  The two galaxies have similar colors,
with $([3.6]-[4.5])=$ 0.38 and 0.43 mag, consistent with a redshift
of $z\gtrsim$ 1.4 (Papovich 2008).  The two IRAC sources are separated by
2.9\arcsec\ (24.9 kpc at $z=$ 2, 23.0 kpc at $z=$ 3.  Furthermore, the
SPIRE source is bright ($S_{500}=$ 84 $\pm$ 9mJy) and red in the sub-mm
bands, with a rising spectrum from 250 to 500\um, indicating a sub-mm
photometric redshift of $z\ga$ 3 (e.g.  Lapi et al.\,2011), although
we caution that the dust temperature and redshift are degenerate in
the sub-mm bands.  It is likely that \atlas\ ID-891 is a blend of
emission from the two identified counterparts and it is possible that
the sub-mm emission is the result of a merger or interaction of two
gas-rich galaxies (e.g.\ Aravena et al.\ 2010b). Spectroscopic data are
required to verify this scenario. and we identify ID-891 as a target
for additional follow-up, especially to identify the exact nature of
the sub-mm emission.

H-ATLAS ID-827, shown in figure~\ref{fig:postage3}, is the most
complex system in our sample. In this case there are four IRAC
counterparts identified on the basis of their colors and
magnitudes. The four galaxies have a wide range of $[3.6]$ with values
from 17.8~mag to 20.2~mag. The counterparts do not appeared to be
clustered, as may be expected for gravitational lensing or a
multi-component interaction, and each is separated from its nearest
neighbor by $\sim$ 4\arcsec. The two brightest galaxies are detected by
SDSS, but were not identified as SPIRE counterparts by Smith et al.\
(2011); they have photometric redshifts of $z=$ 0.07 $\pm$ 0.01 and
$z=$ 0.20 $\pm$ 0.03 for the westernmost and easternmost, respectively. The
two fainter counterparts are not detected in SDSS and both have
$[3.6]-[4.5]>$ 0, indicating that they are likely to be at $z\gtrsim$ 1.4
(Papovich 2008). The morphology, astrometry and available redshift
information suggests that gravitational lensing in unlikely and that
\atlas\ ID-827 is most likely to be a blended source, which is
comprised of two low-redshift and two high-redshift components.

\begin{figure}
  \includegraphics[width=9cm]{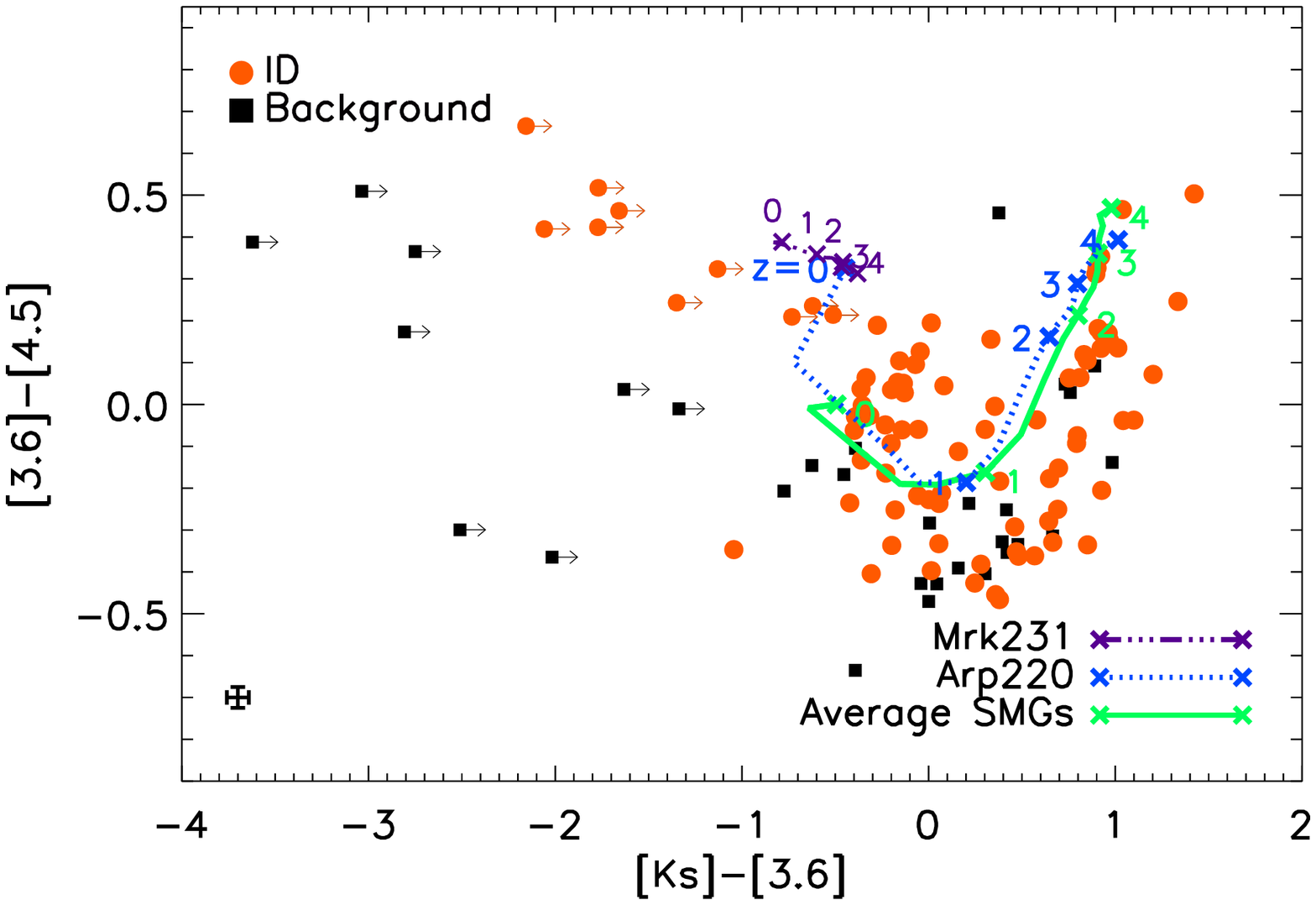}
  \includegraphics[width=9cm]{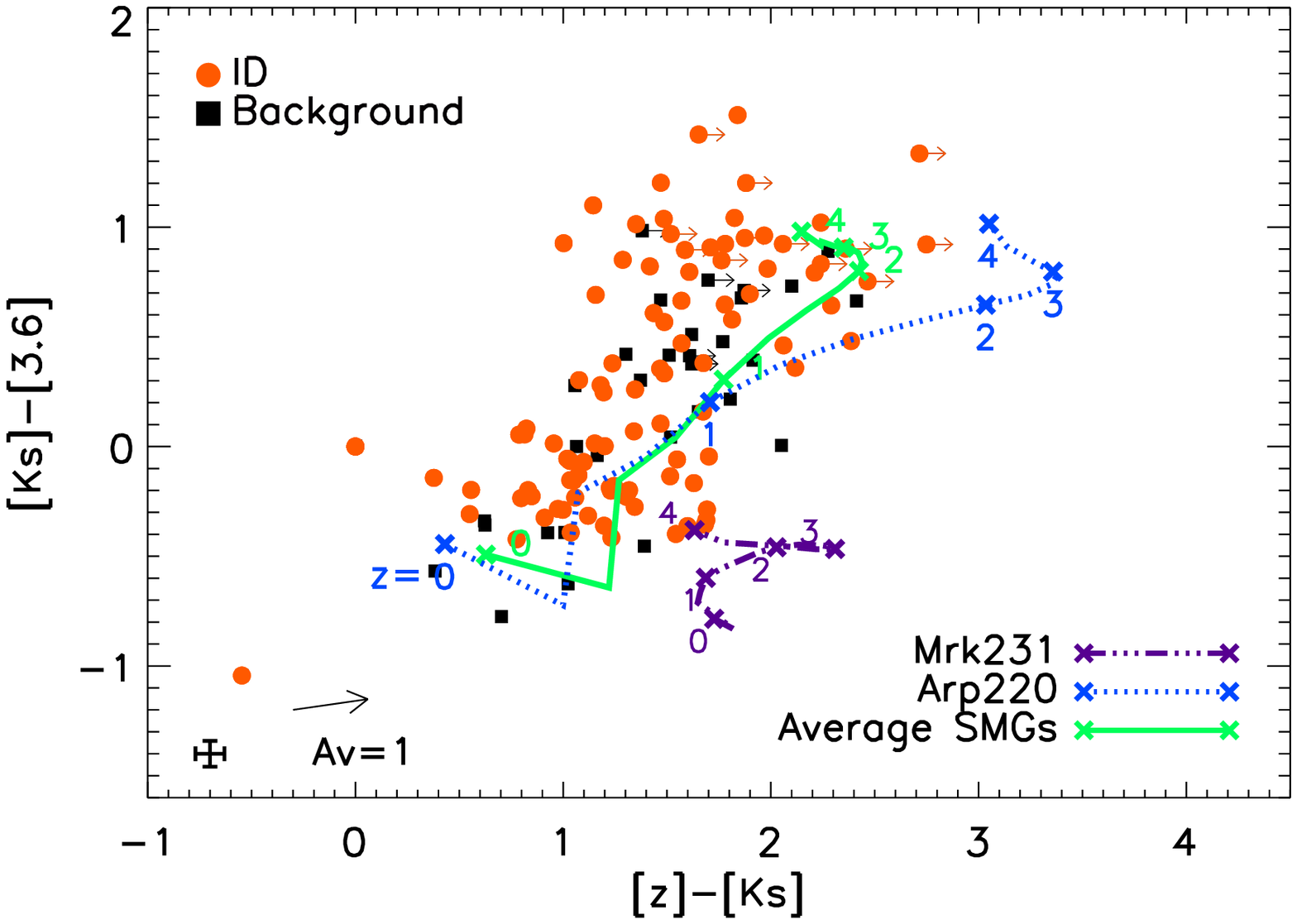}
  \caption{IRAC and VIKING color-color diagrams. {\bf Top:} [$K_{\rm
      s}$], $[3.6]$ and $[4.5]$. {\bf Bottom:} [$z^{\prime}$], [$K_{\rm s}$] and
    $[3.6]$.  We highlight SPIRE counterparts and background sources
    that are within 7.2\arcsec\ of SPIRE centroids.  Tracks for Arp~220 
    (Silva et al.\,1998), Mrk~231 (Berta\, 2006) and 
      the average 870-\um\ selected sub-mm galaxies (Wardlow et al.\,2011) 
    are shown, and an $A_{V} = 1$~mag reddening vector and the
    median error bar are plotted in the lower left-hand corner.  The
    SPIRE counterparts are slightly bluer than both these tracks
    suggesting that they may have less dust reddening and $A_V$ values
    that are 0.5 to 1 magnitudes smaller than the SED templates.  }
  \label{fig:viking}
\end{figure}

\begin{figure}
  \includegraphics[width=9cm]{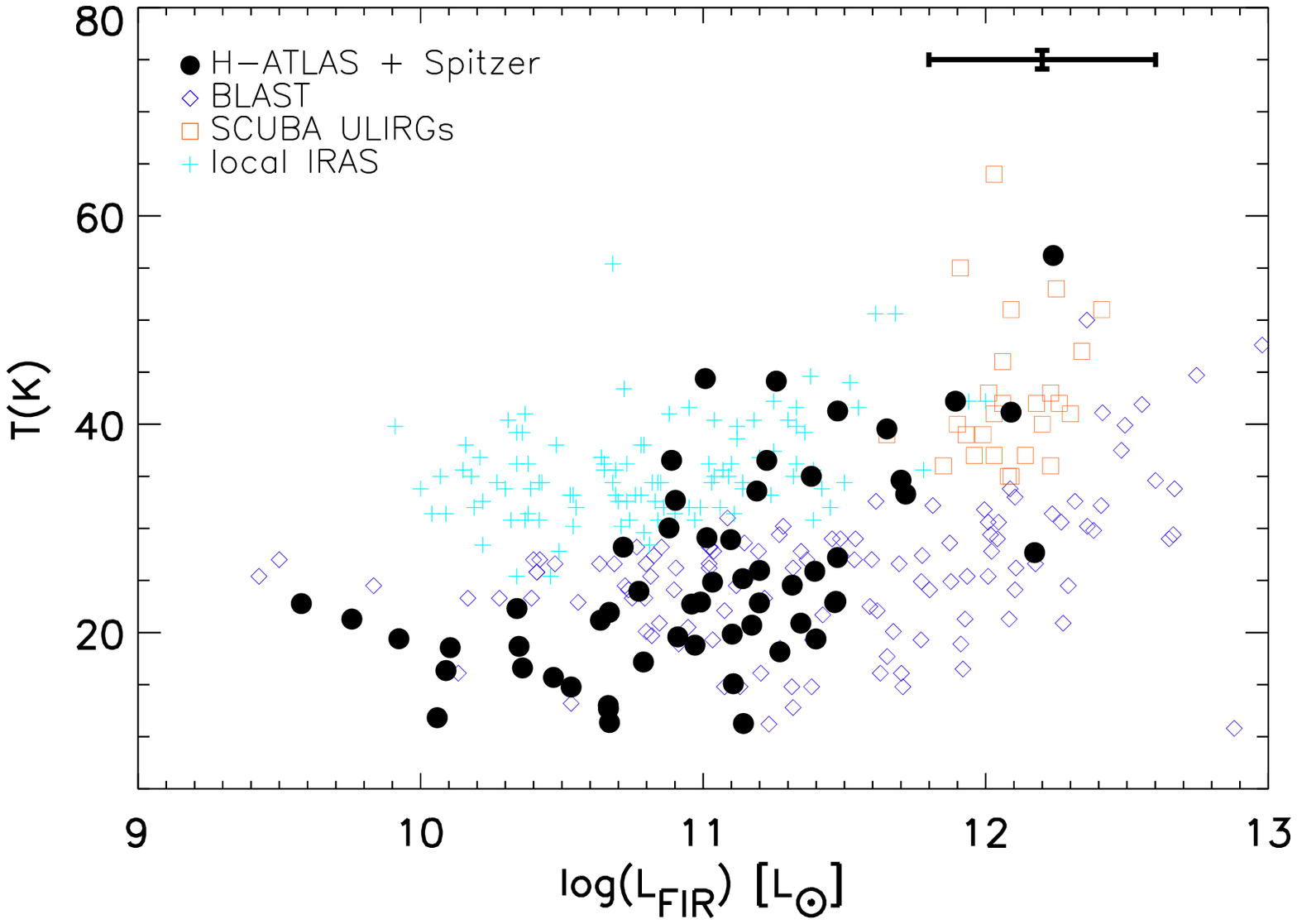}
       \includegraphics[width=9cm]{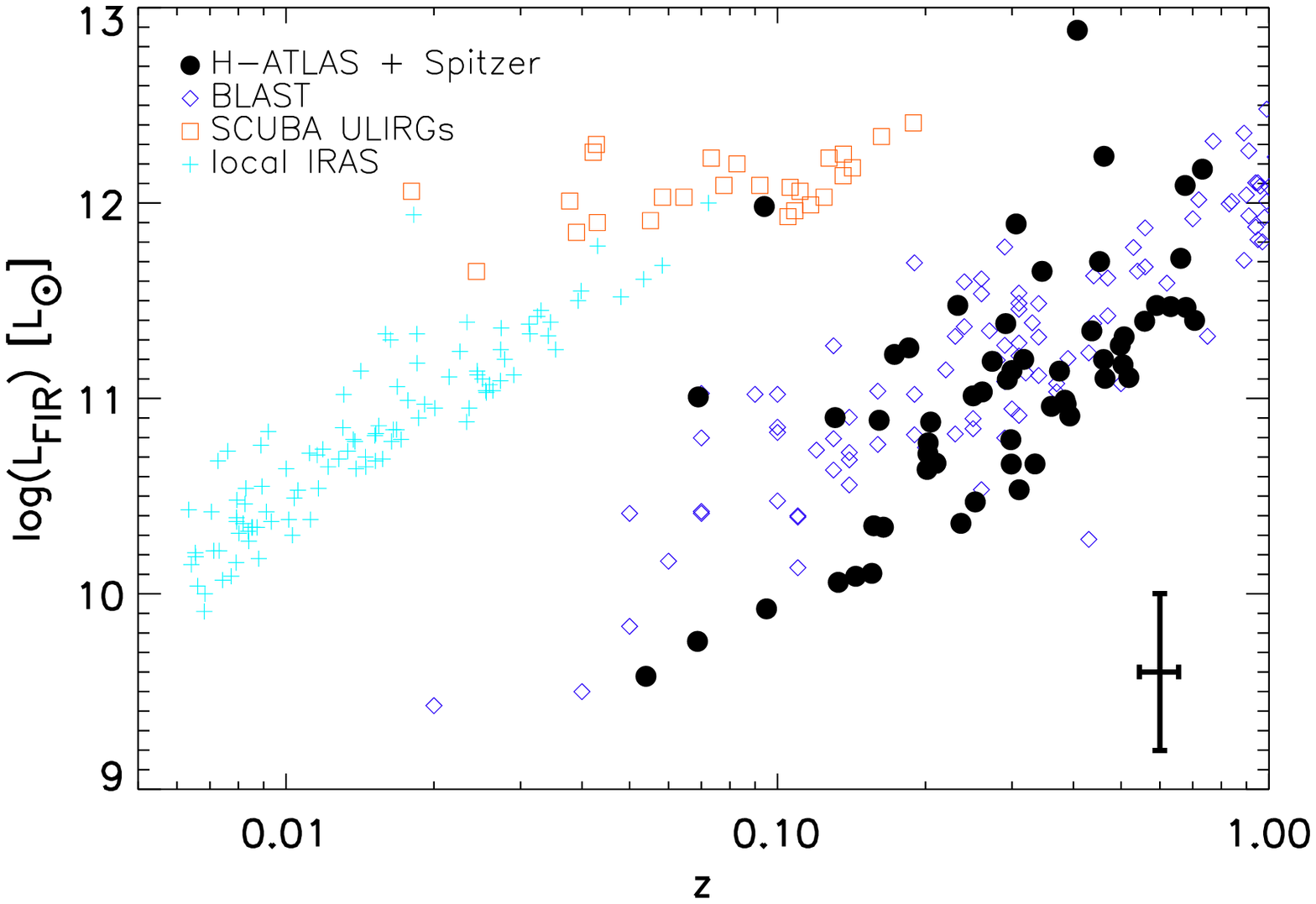}
  \caption{
{\bf Top:} 
    Dust temperature as a function of the infrared luminosity   $L_{\rm IR}$ (8 - 1100\um)  
    for H-ATLAS/IRAC galaxies with SDSS redshifts.   $L_{\rm IR}$ 
    is obtained by fitting the SPIRE flux with $\beta$ = 1.5 in isothermal modified black-body model (equation~\ref{eq:modified}). Error bars are median 1$\sigma$ standard deviation   from   the best-fit models with dust temperature ($T_{\rm d}$) and total IR luminosities  $L_{\rm IR}$ as free parameters.   
{\bf Bottom:}   Dust temperature as a function of the redshift for H-ATLAS/IRAC galaxies. See text for the comparison galaxy samples plotted here.
  }
  \label{fig:lfir}
\end{figure}

\subsection{Redshifts of Identified Sources}

Knowledge of the redshift distribution of SPIRE sources is central to
understanding their role in the Universe. However, only 50\% of the IRAC
counterparts are detected in SDSS and 74\% have VIKING
photometry. Therefore, any redshift distribution derived from the
optical data will be biased to low redshifts where the detection rates
in these surveys are high.  Instead, in figure~\ref{fig:stat} we plot
the $[3.6]-[4.5]$ color distribution, and $[3.6]-[4.5]$ against
redshift for the counterparts with spectroscopic or photometric
redshift from SDSS.  This plot includes the 104 IRAC counterparts that
have both 3.6 and 4.5\um\ coverage, 47 of which have an optical
redshift. All of the optical redshift are $z<$ 0.8; 32 (68\%) of these
galaxies have $([3.6]-[4.5]) <$ 0, while 15 (32\%) have $([3.6]-[4.5])
>$ 0.  The 57 identifications without reliable redshifts from the
optical are undetected in SDSS and 43 of them have $([3.6]-[4.5]) >$ 0,
a crude indicator that $z\gtrsim1.4$ (e.g. Papovich 2008). Thus, up to
$\sim$ 40\% of the identified SPIRE population may lie at high
redshifts. Deep spectroscopy and photometry, particularly at near-IR
wavelengths, is required to determine more precise redshift
information.

We also investigate the redshift distribution of the SPIRE sources by
calculating sub-mm photometric redshifts from the fluxes of the three
SPIRE bands. We employ a template fitting method which uses $\chi^2$
minimization, comparing the observed fluxes to the SEDs of Arp~220 (Silva et al.\, 1998),
 Mrk~231 (Berta\, 2006).  
We caution that the sub-mm photometric redshifts are limited
by the absence of longer wavelength data, and the degeneracy between
redshift and dust temperature. Furthermore, flux boosting is an
important consideration and will add significant errors to the fluxes
of the faintest sources.  Due to these uncertainties, we do not
consider the sub-mm photometric redshifts on a source-by-source basis,
but instead examine the overall distribution.

\begin{figure}
  \includegraphics[width=9cm]{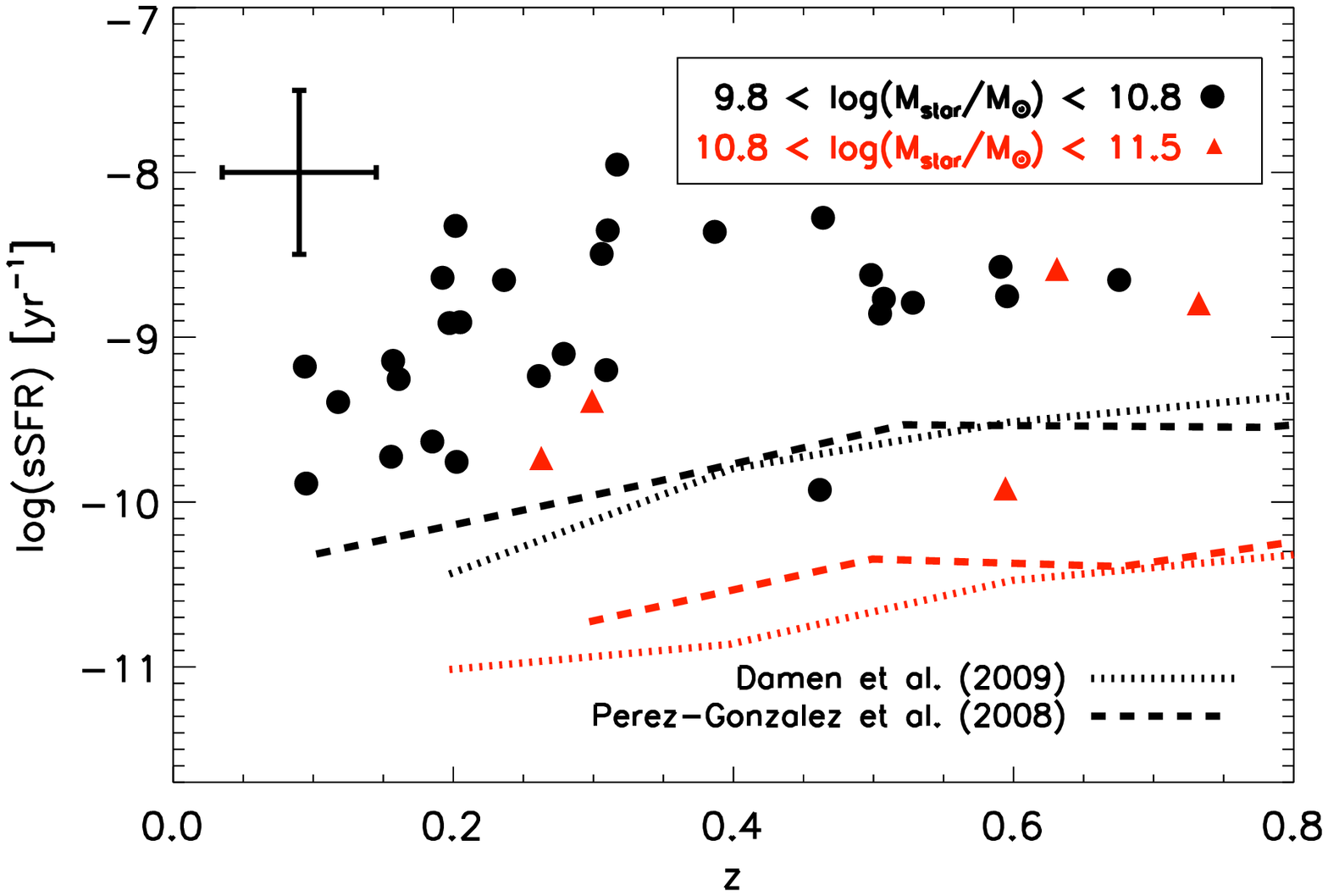}
    \includegraphics[width=9cm]{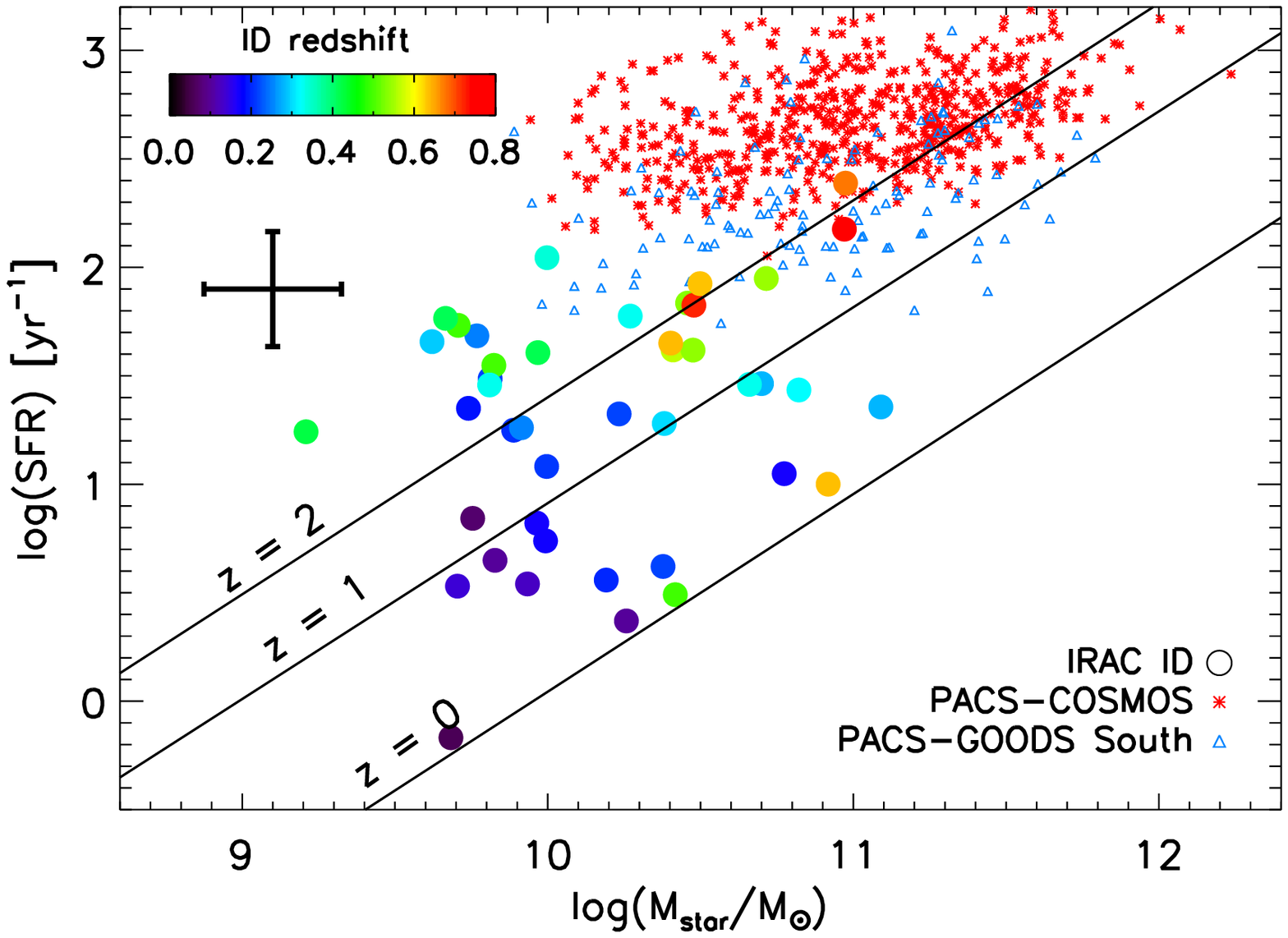}
  \caption{
 {\bf  Top:}
 Specific star formation rate (sSFR) for our IRAC-identified SPIRE galaxies as a function of the redshift.  
 Filled circles  and triangles  represent  H-ATLAS galaxies in
 stellar mass bins  10$^{9.8}$-  10$^{10.8}$ $M_\odot$ and 
 10$^{10.8}$-  10$^{11.5}$ $M_\odot$, respectively.   Our data are compared to  P\'erez-Goz\'alez et al.\,(2008)  and  Damen et al.\,(2009). 
  Upper (lower)  two lines  are for the  smaller (larger)  mass bin.  The error bar on the top left shows the 
  average   68\% confidence range on  $log$(sSFR).   
{\bf Bottom:}  Star formation rate as a function of the stellar mass. We plot
results from a previous study for comparison:  PACS-COSMOS   and 
  PACS-GOODS South (Rodighiero et al. 2011).    
  We also show the mean  SFR and stellar mass correlation  at $z=2$ (Daddi et al. 2007),  $1$  and $0$ (Elbaz et al. 2007). 
The majority of the H-ATLAS/IRAC galaxies  fall  between $z = 0$ $\sim 2$  in SFR-$M_\star$ relation 
showing a slight excess of  star formation rate  
compared with populations at similar redshifts, but selected at optical and near-IR wavelengths.
 }
  \label{fig:ssfr_z}
\end{figure}

This sub-mm photometric redshift distribution peaks at $z\sim 2$ with a
tail out to $z\sim$ 4, and suggests that the majority of the IRAC
counterparts without SDSS redshifts lie at $z\sim 2$. This result is
consistent with the photometric redshifts of the whole SPIRE
population (e.g. Amblard et al.\,2010; Lapi et al.\,2011) and the
redshift distribution of 850\um\ sources (e.g.  Chapman et al.\,2005;
Wardlow et al.\,2011). Now that the counterparts to a significant
majority of SPIRE sources are known,  follow-up spectroscopic campaigns
to establish the redshifts of all the sources are feasible.

\subsection{Properties of Identified Sources}

{\it Near-IR Colors}: In figure~\ref{fig:viking} we show IRAC and
VIKING color-color diagrams, highlighting identified SPIRE
counterparts in comparison with background sources that are within
7.2\arcsec\ of the SPIRE centroids.  
As we show in figure~\ref{fig:color} the SPIRE counterparts have distinct
$[3.6]-[4.5]$ color compared to  the background sources, but the two populations are
indistinguishable in $[K_{\rm s}]-[3.6]$ and $[{\rm z}]-[K_{\rm s}]$.
  We also compare to the redshift tracks of Arp~220 (Silva et al.\,1998),  
Mrk~231 (Berta\, 2006) and the average
870-\um\ selected sub-mm galaxy (Wardlow et al.\,2011).  The SPIRE counterparts
are slightly bluer than both these tracks suggesting that they may
have less dust reddening and $A_V$ values that are 0.5 to 1 magnitudes
smaller than the SED templates.

{\it IR Luminosities and Dust Temperatures}: 
To understand  properties of the identified H-ATLAS/IRAC galaxies associated with SDSS redshifts, we first consider a SED analysis of the SPIRE data making use of
the redshifts based on the IRAC identification. 
  First we consider the far-IR/sub-mm portion of the SED with 3-band SPIRE fluxes
at 250, 350, and 500\um\ and reproduce the analysis of Amblard et al. (2010) where the dust temperature ($T_{\rm d}$) and IR luminosities 
$L_{\rm IR}$ were studied. 
We make use of a modified black-body with isothermal dust temperature to describe the dust emission, in which flux can be written as
\begin{equation} 
f_\mu \propto \mu^{3+\beta}e^{(h\mu/kT_{\rm d}-1)} \,.
\label{eq:modified}
\end{equation}
To be consistent with previous other estimates from the literature we fix $\beta=1.5$ and fit to the SPIRE fluxes 
to establish the dust temperature
and the overall normalization to the SED.  The IR luminosity is estimated from the best-fit SED over the wavelength range of
8 to 1100\um. Since we only fit to the dust emission at wavelengths greater than 250\um\ our luminosities are likely low by a factor of 2 to 3
if there are any warm dust, heated by AGNs, present in these galaxies. However, our overall uncertainty 
on the IR luminosity estimates
is at least a factor of 5. 
For the sample as a whole we find an average dust temperature of  26 $\pm$ 9 K, 
which compares well with the average dust temperature of
27 $\pm$ 8 K for $z<  $ 0.1 SPIRE galaxies based on SDSS identifications only (Amblard et al. 2010).

We summarize our results related to the FIR SED analysis in figure~\ref{fig:lfir} 
where we show the  $L_{\rm IR}$ - $T_{\rm d}$ relation
and the $L_{\rm IR}$ as a function of the redshift, determined from the reliable H-ATLAS IDs with either SDSS spectroscopic or photometric redshift.
We also compare the H-ATLAS dust temperatures and IR luminosities with sub-mm bright galaxy samples in the literature.
Samples include  the sources in BLAST with COMBO-17 
(Wolf et al. 2004) or a SWIRE photometric redshift (Rowan-Robinson et al. 2008) 
and {\it Spitzer}-MIPS 70 and 160\um\ fluxes (Dye et al.
2009),   local ULIRGS observed with SCUBA at 450 and 850\um\  (Clements et al. 2010), 
 and local IRAS-selected galaxies with 60 and 100\um\ fluxes along with SCUBA 850\um\ (Dunne et al. 2000). 

In figure~\ref{fig:lfir}  we find that the SPIRE-selected galaxies 
in H-ATLAS with IRAC-based identifications for redshifts span the
luminosity range of 10$^{10}$ to 10$^{12}$  $L_{\sun}$ from sub-LIRG luminosities to ULIRG conditions. 
These luminosities and the ranges
are comparable to BLAST-detected sources, while they span lower than the local IRAS-selected galaxies, 
which tend to be at lower redshifts
$z<0.05$ and have higher luminosities; such a difference is understandable since IRAS is 
an all-sky shallow survey and detects primarily the rare,
bright galaxies in the near-by universe.

{\it Stellar Masses and Star-Formation Rates}: 
With the cross-identification  we can expand the SED analysis over
3 orders of magnitude in wavelength from optical to sub-mm. For optical and near-IR data we make use of SDSS, UKIDSS (Lawrence et al.\,2007),
VIKING (Sutherland et al.\ in prep), and IRAC 3.6 and 4.5\um\ fluxes. The analysis of full optical to sub-mm SEDs is similar in spirit to
Dunne et al. (2011) where the SEDs of SPIRE sources with reliable identifications using SDSS data were analyzed. 
We make use of the {\sc MAGPHYS}
SED modelling code (da Cunha, Charlot \& Elbaz 2008) for this work. The SED modelling in the code  
is done in a energy-balance manner such that the
absorbed light in shorter UV and optical wavelengths is accounted for by the thermal 
re-radiation at
  far-infrared and sub-mm wavelengths.  For the H-ATLAS/IRAC galaxy sample we have 5 optical fluxes from SDSS ($u, g, r, i, z$), 5 near-IR fluxes from UKIDSS+VIKING ($Z, Y, J, H, K_{\rm s}$), and 2 IRAC channels (3.6, 4.5\um), in addition to 3 SPIRE bands.
Redshift information of the sources are   from SDSS spectroscopic observations ($z= $0 - 0.8).

For comparative work with other sub-mm galaxy samples we focus on results related to stellar mass $M_\star$ and star-formation rate (SFR), as given by {\sc MAGPHYS} best-fit models. In figure~\ref{fig:ssfr_z} (top figure) 
we plot the specific star-formation rate (sSFR) given as SFR/$M_{\star}$ vs. redshift with our sample sub-divided to two stellar mass bins.
We find that the sSFR of  IRAC IDs  marginally show an anticorrelation with galaxy  stellar mass. Such a behavior may be
 related to the {\it downsizing}  scenario (Cowie et al. 1996).

However, when compared  to Damen et al.\,(2009) and P\'erez-Goz\'alez et al.\,(2008),  
IRAC IDs shows  0.4 $\sim $ 0.5 dex higher sSFRs. The Damen et al. (2009) galaxy sample is 
an IRAC-selected sample of galaxies out to $z \sim $ 1.8 
in ECDFS with photometry supplemented with optical, near-IR, and MIPS 24\um\ fluxes. 
The P\'erez-Goz\'alez et al.\,(2008) sample 
is selected with IRAC at 3.6 and 4.5\um\ in a variety of fields. 
While our SPIRE-selected sample has IRAC identifications,  they are not likely to
be the majority of the galaxies in both these studies. 
This is also clear from  figure~\ref{fig:color}  where 
we show that  most of the identified SPIRE sources are distributed  differently from the 
background population in the color-magnitude space. They occupy
the top-end of the star-formation in galaxies selected at near-IR wavelengths.

In Fig.~12 (bottom) we plot the SFR vs. the stellar mass of our sample.  
The SFR of   {\it normal} local  star-forming galaxies at $z \sim 1$  is known to correlate
strongly with the stellar mass and form the so-called
{\it star-formation main-sequence} (Brinchmann et al. 2004; Salim et
al. 2007; Peng et al. 2010). Such a correlation is also observed at
higher redshifts, albeit with a different normalization (e.g. Daddi et
al. 2007; Elbaz et al. 2007; Noeske et al.  2007; Pannella et al. 2009;
Daddi et al. 2009; Gonzalez et al. 2010; Rodighiero et al. 2010; Karim
et al. 2011).   However,  the  typical dispersion  of correlation is known to be  $\sim$ 0.3 dex  over 
a wide range of redshifts $z=$ 0 $\sim$ 2.   

Local (U)LIRGs  and high-redshift SMGs  have SFRs in excess of the main-sequence
and are defined as starburst galaxies (e.g. Elbaz et al. 2007, 2011).  These galaxies generally occupy a region that is
$\sim$ $\times$10 above the  main-sequence. However, there is also   evidence that some high-redshift (U)LIRGs and SMGs do
not have enhanced star-formation rates relative to their stellar mass and that the starburst fraction may decrease at high redshifts
(e.g. Tacconi et al. 2008; Rodighiero et al. 2011). 
This suggests changes to the mode of star-formation in (U)LIRGs between $z=  $ 0  and $z= $ 2
away from the starburst mechanism (e.g. Daddi et al. 2010b; Genzel et
al. 2010; Elbaz et al. 2011; Krumholz et al. 2012; Melbourne et al. 2012).   
In our work, $\sim$ 50\% of the subsample of IRAC-identified SPIRE galaxies  ($z < $ 1) 
lie near the starburst region. Thus, it is possible that half of the IRAC-identified SPIRE galaxies with SDSS redshifts
 are in fact starbursts that lie off the main-sequence.  

\section{Summary}
\label{sec:conclusion}

We have identified \spitzer-IRAC counterparts to sources selected with
\herschel-SPIRE at 250\um\ in the H-ATLAS survey.  Among 159 SPIRE
centroids, we found 123 reliable IRAC counterparts using a likelihood
ratio analysis.  The identified SPIRE sources are  distributed 
differently  
in IRAC color-magnitude  space compared to the field
population. Therefore, we made an additional selection based on 
the IRAC ID  locality to yield a further 23 counterparts to
13 SPIRE sources. Seven SPIRE sources have multiple IRAC
counterparts. These are likely to be due to blended emission in the
SPIRE beam.

In total 146 reliable IRAC counterparts to 136 SPIRE sources were
identified, including seven that have more than one IRAC
counterpart. The identification rate of 86\% is higher than that of
wide-field ground-based optical and near-IR imaging of \herschel\
fields. The galaxies with unknown redshifts and that are not detected
in SDSS and VIKING imaging data have SPIRE colors indicative of high
redshift sources and $[3.6]-[4.5]>$ 0, suggesting that they are likely
to be at $z\gtrsim$ 1.4. We estimate the $\sim$  40\% of SPIRE sources lie at
high redshifts, although the exact redshift distribution of SPIRE
sources remains elusive. The counterparts presented here can now be
pursued for followup data to further investigate the nature of SPIRE
galaxies. 
The majority of our detected galaxies with sub-LIRG to LIRG luminosities
are not intense starbursting galaxies in the local universe, 
though they have above average specific star-formation rates.

\vspace*{-2pt}   
\acknowledgements

{\it Herschel}-ATLAS is a project which uses  {\it Herschel},  an ESA
space observatory with science instruments provided by European-led
Principal Investigator consortia and with important participation from
NASA. The H-ATLAS website is http://www.h-atlas.org/.  
This work is based in part on observations
made with the {\it Spitzer} Space Telescope, which is operated by the
Jet Propulsion Laboratory, California Institute of Technology under a
contract with NASA.  Support for this work was provided by NASA through
an award issued by JPL/Caltech.  US participants in H-ATLAS also
acknowledge support from the  NASA {\it Herschel} Science Center through a
contract from JPL/Caltech.  

Funding for the SDSS and SDSS-II has been provided by the Alfred
P. Sloan Foundation, the Participating Institutions, the National
Science Foundation, the U.S. Department of Energy, the National
Aeronautics and Space Administration, the Japanese Monbukagakusho, the
Max Planck Society, and the Higher Education Funding Council for
England. The SDSS Web Site is http://www.sdss.org/.

The SDSS is managed by the Astrophysical Research Consortium for the
Participating Institutions. The Participating Institutions are the
American Museum of Natural History, Astrophysical Institute Potsdam,
University of Basel, University of Cambridge, Case Western Reserve
University, University of Chicago, Drexel University, Fermilab, the
Institute for Advanced Study, the Japan Participation Group, Johns
Hopkins University, the Joint Institute for Nuclear Astrophysics, the
Kavli Institute for Particle Astrophysics and Cosmology, the Korean
Scientist Group, the Chinese Academy of Sciences (LAMOST), Los Alamos
National Laboratory, the Max-Planck-Institute for Astronomy (MPIA),
the Max-Planck-Institute for Astrophysics (MPA), New Mexico State
University, Ohio State University, University of Pittsburgh,
University of Portsmouth, Princeton University, the United States
Naval Observatory, and the University of Washington.

This work used data from VISTA at the ESO Paranal Observatory, programme 179.A-2004. VISTA is an ESO near- infrared survey telescope in Chile, conceived and developed by a consortium of universities in the United Kingdom, led by Queen Mary University, London.

We thank Giulia Rodighiero for providing us electronic tables of COSMOS and GOODS SED fitting results.
The UCI group acknowledges support from NSF CAREER AST-0645427, NASA NNX10AD42G, and an award from Caltech/JPL for US participation in
Herschel-ATLAS.

\LongTables
\begin{sidewaystable*}
\begin{landscape}
\appendix
\caption{Combined photometry for  IRAC IDs.}
\label{tab:photom}
\setlength{\tabcolsep}{0.7 mm}
\include{lr_photometry_table0}
\clearpage
\end{landscape}
\end{sidewaystable*}

\begin{sidewaystable*}
\begin{landscape}
\begin{center}
TABLE 2 -- continued. 
\end{center}
\setlength{\tabcolsep}{0.7 mm}
\include{lr_photometry_table1}
\clearpage
\end{landscape}
\end{sidewaystable*}

\begin{sidewaystable*}
\begin{landscape}
\begin{center}
TABLE 2 -- continued. 
\end{center}
\setlength{\tabcolsep}{0.7 mm}
\include{lr_photometry_table2}
\clearpage
\end{landscape}
\end{sidewaystable*}

\end{document}

%% file: lr_photometry_table0.tex
\begin{tiny}
\begin{tabular}{llccccccccccccccccc}
\hline
IAU ID & H-ATLAS  & RA$^{a}$  & Dec$^{a}$ &  $Z$$^{b}$ & {\it Y$^{b}$} & {\it J$^{b}$} & {\it H$^{b}$} & $K_{\rm s}$$^{b}$ & 3.6$\,\mu$m & 4.5$\,\mu$m
&$S_{250}$$^c$ & $S_{350}$$^c$ & $S_{500}$$^c$ & $R^d$ & $R^d$ & z$^e$ & Separation$^f$  \\
       & ID & (J2000) & (J2000) &   (mag)  & (mag)  & (mag) & (mag) &  (mag) & (mag) & (mag) & (mJy) & (mJy) &(mJy) &(3.6\micron) &(4.5\micron) & & (arcsec)\\
\hline
J090311.6+003906 &           81 & $ 0           9^{\rm m} 0           3^{\rm h}           11\fs           57$ & $+0           0^\circ          39^\prime 0           6\farcs            4$ & $18.80\pm18.80$ & $ 18.51\pm 0.02$ & $ 18.25\pm 0.01$ & $ 17.98\pm 0.01$ & $ 17.68\pm 0.01$ & $ 18.00\pm 0.01$ & $ 18.03\pm 0.01$ & $129.00\pm 6.56$ & $181.86\pm 7.34$ & $165.92\pm 9.27$ & $1.000$ &$0.999$ & $0.30\pm0.00$ & $ 0.66$\\
J090818.2+004720 &          106 & $ 0           9^{\rm m} 0           8^{\rm h}           18\fs           31$ & $+0           0^\circ          47^\prime           18\farcs            9$ & $20.59\pm20.59$ & $ 20.12\pm 0.04$ & $ 20.08\pm 0.04$ & $ 19.62\pm 0.06$ & $ 19.00\pm 0.03$ & $...$ & $ 17.01\pm 0.01$ & $116.93\pm 6.54$ & $ 81.56\pm 7.40$ & $ 27.71\pm 8.96$ & $...$ &$1.000$ & $0.73\pm0.20$ & $ 1.80$\\
J091304.9-005343 &          130 & $ 0           9^{\rm m} 0           8^{\rm h} 0           5\fs 0           9$ & $-0           0^\circ          53^\prime           43\farcs            2$ & $19.05\pm19.05$ & $ 18.78\pm 0.01$ & $ 18.53\pm 0.01$ & $ 18.25\pm 0.02$ & $ 17.97\pm 0.01$ & $ 18.10\pm 0.01$ & $ 18.08\pm 0.01$ & $105.04\pm 6.47$ & $127.57\pm 7.21$ & $107.52\pm 9.01$ & $0.999$ &$0.999$ & $0.21\pm0.02$ & $ 2.23$\\
J090918.3+002420 &          177 & $ 0           9^{\rm m} 0           9^{\rm h}           18\fs           46$ & $+0           0^\circ          24^\prime           21\farcs            0$ & $17.91\pm17.91$ & $ 17.72\pm 0.01$ & $ 17.55\pm 0.01$ & $ 17.29\pm 0.01$ & $ 17.35\pm 0.01$ & $ 17.55\pm 0.01$ & $ 17.88\pm 0.01$ & $107.24\pm 7.83$ & $ 45.96\pm 7.20$ & $  3.88\pm 9.00$ & $1.000$ &$0.999$ & $0.07\pm0.00$ & $ 1.43$\\
J090916.1+003208 &          282 & $ 0           9^{\rm m} 0           9^{\rm h}           16\fs           10$ & $+0           0^\circ          32^\prime 0           7\farcs            1$ & $>23.06$ & $> 22.35$ & $ 21.90\pm 0.19$ & $ 20.81\pm 0.15$ & $ 20.35\pm 0.12$ & $ 19.01\pm 0.01$ & $ 18.77\pm 0.01$ & $ 81.47\pm 6.43$ & $ 60.26\pm 7.13$ & $ 33.18\pm 9.02$ & $0.970$ &$0.989$ & $...$ & $ 1.07$\\
J090916.7+002808 &          317 & $ 0           9^{\rm m} 0           9^{\rm h}           16\fs           82$ & $+0           0^\circ          28^\prime 0           7\farcs            8$ & $21.44\pm21.44$ & $ 21.04\pm 0.06$ & $ 20.47\pm 0.05$ & $ 19.85\pm 0.06$ & $ 19.15\pm 0.04$ & $ 18.50\pm 0.01$ & $ 18.78\pm 0.01$ & $ 87.56\pm 6.45$ & $ 64.38\pm 7.23$ & $ 17.53\pm 9.02$ & $0.998$ &$0.999$ & $...$ & $ 0.51$\\
J090835.8+004139 &          458 & $ 0           9^{\rm m} 0           8^{\rm h}           35\fs           82$ & $+0           0^\circ          41^\prime           40\farcs            6$ & $19.04\pm19.04$ & $ 18.80\pm 0.01$ & $ 18.41\pm 0.01$ & $ 18.02\pm 0.01$ & $ 17.70\pm 0.01$ & $ 17.97\pm 0.01$ & $ 17.78\pm 0.01$ & $ 75.75\pm 6.49$ & $ 32.32\pm 7.13$ & $ 10.83\pm 8.79$ & $1.000$ &$0.999$ & $0.31\pm0.05$ & $ 0.81$\\
J090930.4+002224 &          462$^{\star}$ & $ 0           9^{\rm m} 0           9^{\rm h}           30\fs           51$ & $+0           0^\circ          22^\prime           30\farcs            8$ & $19.54\pm19.54$ & $ 19.35\pm 0.02$ & $ 19.05\pm 0.02$ & $ 18.74\pm 0.02$ & $ 18.48\pm 0.02$ & $ 18.71\pm 0.01$ & $ 18.76\pm 0.01$ & $ 72.48\pm 6.36$ & $ 76.71\pm 7.21$ & $ 54.92\pm 8.92$ & $0.121$ &$0.159$ & $0.13\pm0.03$ & $ 6.73$\\
J090327.7+004119 &          466 & $ 0           9^{\rm m} 0           3^{\rm h}           27\fs           65$ & $+0           0^\circ          41^\prime           18\farcs            1$ & $18.68\pm18.68$ & $ 18.32\pm 0.01$ & $ 18.17\pm 0.01$ & $ 17.95\pm 0.01$ & $ 17.86\pm 0.01$ & $ 17.81\pm 0.01$ & $ 18.04\pm 0.01$ & $ 74.80\pm 6.50$ & $ 33.52\pm 7.21$ & $ 11.24\pm 9.00$ & $0.999$ &$0.871$ & $0.13\pm0.00$ & $ 2.92$\\
J090816.8+003211 &          493 & $ 0           9^{\rm m} 0           8^{\rm h}           16\fs           90$ & $+0           0^\circ          32^\prime           11\farcs            6$ & $>22.43$ & $> 22.02$ & $ 21.77\pm 0.18$ & $ 21.23\pm 0.25$ & $ 20.78\pm 0.16$ & $ 19.35\pm 0.01$ & $ 18.85\pm 0.01$ & $ 75.63\pm 6.54$ & $ 61.45\pm 7.20$ & $ 31.51\pm 9.16$ & $0.997$ &$0.999$ & $...$ & $ 0.77$\\
J090829.1+002714 &          511 & $ 0           9^{\rm m} 0           8^{\rm h}           28\fs           96$ & $+0           0^\circ          27^\prime           12\farcs            6$ & $21.14\pm21.14$ & $ 20.76\pm 0.07$ & $ 20.37\pm 0.05$ & $ 19.92\pm 0.08$ & $ 19.43\pm 0.05$ & $ 18.53\pm 0.01$ & $ 18.34\pm 0.01$ & $ 72.97\pm 6.50$ & $ 48.30\pm 7.14$ & $ 31.77\pm 8.95$ & $0.995$ &$0.998$ & $0.44\pm0.10$ & $ 3.50$\\
J090925.3+004438 &          558 & $ 0           9^{\rm m} 0           9^{\rm h}           25\fs           32$ & $+0           0^\circ          44^\prime           37\farcs            9$ & $>22.54$ & $> 21.73$ & $> 21.05$ & $> 19.74$ & $> 19.13$ & $...$ & $ 20.37\pm 0.04$ & $ 73.14\pm 6.54$ & $ 65.94\pm 7.22$ & $ 50.22\pm 9.10$ & $...$ &$0.960$ & $...$ & $ 0.81$\\
J090828.0+005005 &          641 & $ 0           9^{\rm m} 0           8^{\rm h}           28\fs 0           0$ & $+0           0^\circ          50^\prime 0           4\farcs            8$ & $18.40\pm18.40$ & $ 18.34\pm 0.01$ & $ 18.06\pm 0.01$ & $ 17.84\pm 0.01$ & $ 17.70\pm 0.01$ & $...$ & $ 17.96\pm 0.01$ & $ 67.29\pm 6.53$ & $ 28.12\pm 7.15$ & $ 12.22\pm 8.95$ & $...$ &$0.999$ & $0.17\pm0.00$ & $ 1.05$\\
J090803.6+003737 &          711 & $ 0           9^{\rm m} 0           8^{\rm h} 0           3\fs           55$ & $+0           0^\circ          37^\prime           36\farcs            7$ & $19.59\pm19.59$ & $ 19.03\pm 0.02$ & $ 19.13\pm 0.02$ & $ 18.79\pm 0.03$ & $ 18.51\pm 0.02$ & $ 18.21\pm 0.01$ & $ 18.27\pm 0.01$ & $ 64.26\pm 6.51$ & $ 36.95\pm 7.29$ & $ 11.65\pm 9.15$ & $0.999$ &$0.999$ & $0.26\pm0.04$ & $ 1.55$\\
J090308.5+004146 &          754 & $ 0           9^{\rm m} 0           3^{\rm h} 0           8\fs           55$ & $+0           0^\circ          41^\prime           46\farcs            0$ & $...$ & $...$ & $...$ & $...$ & $...$ & $ 19.98\pm 0.02$ & $ 19.62\pm 0.02$ & $ 65.99\pm 6.55$ & $ 46.14\pm 7.25$ & $ 28.08\pm 9.21$ & $0.984$ &$0.990$ & $...$ & $ 0.45$\\
J090910.2+004345 &          776 & $ 0           9^{\rm m} 0           9^{\rm h}           10\fs           35$ & $+0           0^\circ          43^\prime           42\farcs            1$ & $...$ & $...$ & $...$ & $...$ & $...$ & $...$ & $ 20.10\pm 0.03$ & $ 65.86\pm 6.44$ & $ 51.68\pm 7.18$ & $ 34.00\pm 8.98$ & $...$ &$0.926$ & $...$ & $ 4.11$\\
J090844.0+003256 &          827$^{\star}$ & $ 0           9^{\rm m} 0           8^{\rm h}           44\fs 0           6$ & $+0           0^\circ          32^\prime           56\farcs            7$ & $...$ & $...$ & $...$ & $...$ & $...$ & $ 19.82\pm 0.04$ & $ 19.58\pm 0.03$ & $ 61.63\pm 6.35$ & $ 30.69\pm 7.27$ & $ 19.89\pm 9.18$ & $0.353$ &$0.454$ & $...$ & $ 0.88$\\
J090844.0+003256 &          827$^{\star}$ & $ 0           9^{\rm m} 0           8^{\rm h}           44\fs 0           2$ & $+0           0^\circ          33^\prime 0           0\farcs            7$ & $...$ & $...$ & $...$ & $...$ & $...$ & $ 20.18\pm 0.04$ & $ 20.07\pm 0.04$ & $ 61.63\pm 6.35$ & $ 30.69\pm 7.27$ & $ 19.89\pm 9.18$ & $0.055$ &$0.065$ & $...$ & $ 3.88$\\
J090844.0+003256 &          827$^{\star}$ & $ 0           9^{\rm m} 0           8^{\rm h}           44\fs           28$ & $+0           0^\circ          32^\prime           53\farcs            9$ & $19.29\pm19.29$ & $ 19.20\pm 0.02$ & $ 18.89\pm 0.01$ & $ 18.57\pm 0.02$ & $ 18.25\pm 0.02$ & $ 18.65\pm 0.01$ & $ 18.68\pm 0.01$ & $ 61.63\pm 6.35$ & $ 30.69\pm 7.27$ & $ 19.89\pm 9.18$ & $0.109$ &$0.170$ & $0.20\pm0.03$ & $ 5.00$\\
J090844.0+003256 &          827$^{\star}$ & $ 0           9^{\rm m} 0           8^{\rm h}           43\fs           69$ & $+0           0^\circ          32^\prime           57\farcs            9$ & $18.21\pm18.21$ & $ 18.04\pm 0.01$ & $ 17.77\pm 0.01$ & $ 17.50\pm 0.01$ & $ 17.43\pm 0.01$ & $ 17.85\pm 0.01$ & $ 18.09\pm 0.01$ & $ 61.63\pm 6.35$ & $ 30.69\pm 7.27$ & $ 19.89\pm 9.18$ & $0.482$ &$0.309$ & $0.07\pm0.01$ & $ 4.91$\\
J090929.1+002440 &          861 & $ 0           9^{\rm m} 0           9^{\rm h}           29\fs           33$ & $+0           0^\circ          24^\prime           42\farcs            3$ & $19.12\pm19.12$ & $ 18.96\pm 0.01$ & $ 18.61\pm 0.01$ & $ 18.31\pm 0.02$ & $ 18.02\pm 0.02$ & $ 18.09\pm 0.01$ & $ 18.00\pm 0.01$ & $ 63.93\pm 6.52$ & $ 40.59\pm 7.21$ & $ 20.48\pm 9.11$ & $0.998$ &$0.999$ & $0.16\pm0.02$ & $ 3.02$\\
J090814.6+002707 &          865 & $ 0           9^{\rm m} 0           8^{\rm h}           14\fs           59$ & $+0           0^\circ          27^\prime 0           4\farcs            6$ & $>22.43$ & $ 21.91\pm 0.18$ & $ 21.42\pm 0.13$ & $ 20.75\pm 0.16$ & $ 19.96\pm 0.08$ & $ 19.21\pm 0.01$ & $ 19.15\pm 0.01$ & $ 60.43\pm 6.52$ & $ 43.91\pm 7.40$ & $ 15.89\pm 9.19$ & $0.994$ &$0.994$ & $...$ & $ 3.00$\\
J090849.8+003203 &          891$^{\star}$ & $ 0           9^{\rm m} 0           8^{\rm h}           50\fs 0           9$ & $+0           0^\circ          32^\prime 0           2\farcs            6$ & $...$ & $...$ & $...$ & $...$ & $...$ & $ 20.64\pm 0.03$ & $ 20.26\pm 0.02$ & $ 62.76\pm 6.48$ & $ 75.75\pm 7.23$ & $ 84.45\pm 9.07$ & $0.407$ &$0.437$ & $...$ & $ 3.21$\\
J090849.8+003203 &          891$^{\star}$ & $ 0           9^{\rm m} 0           8^{\rm h}           50\fs 0           6$ & $+0           0^\circ          32^\prime 0           5\farcs            2$ & $...$ & $...$ & $...$ & $...$ & $...$ & $ 20.73\pm 0.03$ & $ 20.30\pm 0.03$ & $ 62.76\pm 6.48$ & $ 75.75\pm 7.23$ & $ 84.45\pm 9.07$ & $0.465$ &$0.500$ & $...$ & $ 2.96$\\
J090925.5+001941 &          946 & $ 0           9^{\rm m} 0           9^{\rm h}           25\fs           44$ & $+0           0^\circ          19^\prime           39\farcs            6$ & $21.63\pm21.63$ & $ 21.44\pm 0.09$ & $ 21.47\pm 0.13$ & $ 20.73\pm 0.14$ & $ 20.49\pm 0.13$ & $ 19.39\pm 0.01$ & $ 19.43\pm 0.01$ & $ 62.16\pm 6.45$ & $ 29.23\pm 7.19$ & $  9.99\pm 8.83$ & $0.901$ &$0.916$ & $...$ & $ 2.34$\\
J090800.6+002651 &          977 & $ 0           9^{\rm m} 0           8^{\rm h} 0           0\fs           46$ & $+0           0^\circ          26^\prime           51\farcs            1$ & $17.12\pm17.12$ & $ 16.81\pm 0.00$ & $ 16.73\pm 0.00$ & $ 16.52\pm 0.00$ & $ 16.57\pm 0.00$ & $ 16.87\pm 0.01$ & $ 17.28\pm 0.01$ & $ 58.89\pm 6.48$ & $ 36.59\pm 7.19$ & $ 12.91\pm 8.91$ & $0.933$ &$0.984$ & $0.10\pm0.00$ & $ 3.52$\\
J090856.6+003813 &         1147 & $ 0           9^{\rm m} 0           8^{\rm h}           56\fs           63$ & $+0           0^\circ          38^\prime           15\farcs            3$ & $20.05\pm20.05$ & $ 19.64\pm 0.03$ & $ 19.38\pm 0.02$ & $ 19.00\pm 0.03$ & $ 18.56\pm 0.02$ & $ 18.23\pm 0.01$ & $ 18.08\pm 0.01$ & $ 60.04\pm 6.34$ & $ 34.84\pm 7.16$ & $ 10.17\pm 9.03$ & $0.940$ &$0.955$ & $0.32\pm0.08$ & $ 1.68$\\
J090841.3+001559 &         1178 & $ 0           9^{\rm m} 0           8^{\rm h}           41\fs           27$ & $+0           0^\circ          15^\prime           59\farcs            2$ & $18.84\pm18.84$ & $ 18.89\pm 0.02$ & $ 18.57\pm 0.01$ & $ 18.31\pm 0.02$ & $ 18.04\pm 0.01$ & $ 18.28\pm 0.01$ & $...$ & $ 59.80\pm 6.40$ & $ 35.59\pm 7.26$ & $ 16.22\pm 8.96$ & $0.999$ &$...$ & $0.20\pm0.03$ & $ 1.48$\\
J090847.4+003459 &         1221 & $ 0           9^{\rm m} 0           8^{\rm h}           47\fs           42$ & $+0           0^\circ          35^\prime 0           5\farcs            8$ & $20.78\pm20.78$ & $ 20.48\pm 0.05$ & $ 20.43\pm 0.05$ & $ 20.08\pm 0.09$ & $ 19.76\pm 0.06$ & $ 19.81\pm 0.02$ & $ 19.87\pm 0.03$ & $ 55.73\pm 6.48$ & $ 42.34\pm 7.16$ & $ 26.26\pm 8.97$ & $0.816$ &$0.815$ & $0.30\pm0.09$ & $ 6.77$\\
J090920.5+004420 &         1248 & $ 0           9^{\rm m} 0           9^{\rm h}           20\fs           45$ & $+0           0^\circ          44^\prime           19\farcs            5$ & $16.90\pm16.90$ & $ 16.72\pm 0.00$ & $ 16.50\pm 0.00$ & $ 16.26\pm 0.00$ & $ 16.35\pm 0.00$ & $...$ & $ 17.16\pm 0.01$ & $ 56.25\pm 7.83$ & $ 21.45\pm 7.24$ & $ 25.88\pm 8.81$ & $...$ &$1.000$ & $0.05\pm0.00$ & $ 1.47$\\
J090932.7+003705 &         1264 & $ 0           9^{\rm m} 0           9^{\rm h}           32\fs           77$ & $+0           0^\circ          37^\prime 0           6\farcs            8$ & $22.32\pm22.32$ & $ 22.22\pm 0.18$ & $ 21.34\pm 0.11$ & $ 21.19\pm 0.21$ & $ 20.55\pm 0.14$ & $ 19.90\pm 0.02$ & $ 20.08\pm 0.03$ & $ 56.01\pm 6.47$ & $ 40.49\pm 7.16$ & $ 23.38\pm 8.97$ & $0.787$ &$0.717$ & $...$ & $ 1.58$\\
J090824.5+002927 &         1288 & $ 0           9^{\rm m} 0           8^{\rm h}           24\fs           84$ & $+0           0^\circ          29^\prime           29\farcs            4$ & $>22.37$ & $> 22.02$ & $ 21.56\pm 0.15$ & $> 20.99$ & $ 20.31\pm 0.10$ & $ 19.39\pm 0.01$ & $ 19.22\pm 0.01$ & $ 55.29\pm 6.54$ & $ 46.57\pm 7.23$ & $ 23.28\pm 9.01$ & $0.988$ &$0.989$ & $...$ & $ 4.08$\\
J090913.6+003256 &         1293$^{\star}$ & $ 0           9^{\rm m} 0           9^{\rm h}           14\fs 0           4$ & $+0           0^\circ          32^\prime           59\farcs            0$ & $>22.54$ & $> 21.73$ & $> 21.05$ & $> 19.74$ & $> 19.13$ & $ 20.48\pm 0.03$ & $ 20.24\pm 0.02$ & $ 55.08\pm 6.48$ & $ 52.82\pm 7.22$ & $ 36.26\pm 9.16$ & $0.272$ &$0.530$ & $...$ & $ 6.70$\\
J090851.4+003955 &         1340 & $ 0           9^{\rm m} 0           8^{\rm h}           51\fs           38$ & $+0           0^\circ          39^\prime           54\farcs 0$ & $21.79\pm21.79$ & $ 21.35\pm 0.11$ & $ 20.68\pm 0.07$ & $ 20.27\pm 0.10$ & $ 19.58\pm 0.05$ & $ 18.78\pm 0.01$ & $ 18.88\pm 0.01$ & $ 53.96\pm 6.51$ & $ 38.57\pm 7.21$ & $  9.59\pm 8.96$ & $0.998$ &$0.999$ & $...$ & $ 0.96$\\
J090818.5+002043 &         1366 & $ 0           9^{\rm m} 0           8^{\rm h}           18\fs           74$ & $+0           0^\circ          20^\prime           40\farcs            1$ & $20.02\pm20.02$ & $ 19.68\pm 0.03$ & $ 19.56\pm 0.03$ & $ 19.21\pm 0.04$ & $ 18.99\pm 0.03$ & $ 19.06\pm 0.01$ & $...$ & $ 55.40\pm 6.46$ & $ 31.22\pm 7.15$ & $ 16.40\pm 9.04$ & $0.888$ &$...$ & $0.21\pm0.03$ & $ 4.48$\\
J090755.2+002254 &         1381 & $ 0           9^{\rm m} 0           7^{\rm h}           54\fs           99$ & $+0           0^\circ          22^\prime           55\farcs            3$ & $>22.37$ & $ 21.87\pm 0.18$ & $ 21.56\pm 0.15$ & $ 21.04\pm 0.21$ & $ 20.50\pm 0.12$ & $ 19.54\pm 0.01$ & $...$ & $ 55.67\pm 6.39$ & $ 39.10\pm 7.28$ & $ 30.98\pm 9.13$ & $0.969$ &$...$ & $...$ & $ 4.05$\\
J090845.7+002805 &         1397 & $ 0           9^{\rm m} 0           8^{\rm h}           45\fs           88$ & $+0           0^\circ          28^\prime 0           6\farcs 0$ & $21.47\pm21.47$ & $ 20.99\pm 0.08$ & $ 21.31\pm 0.12$ & $ 20.57\pm 0.14$ & $ 20.12\pm 0.09$ & $ 19.11\pm 0.01$ & $ 18.97\pm 0.01$ & $ 55.06\pm 6.60$ & $ 44.21\pm 7.30$ & $ 33.07\pm 9.01$ & $0.996$ &$0.998$ & $0.50\pm0.12$ & $ 1.96$\\
J090934.9+002840 &         1398$^{\star}$ & $ 0           9^{\rm m} 0           9^{\rm h}           34\fs           57$ & $+0           0^\circ          28^\prime           40\farcs            3$ & $22.26\pm22.26$ & $ 22.62\pm 0.26$ & $ 21.63\pm 0.15$ & $ 20.87\pm 0.16$ & $ 20.65\pm 0.15$ & $ 19.86\pm 0.02$ & $ 19.93\pm 0.02$ & $ 55.33\pm 6.56$ & $ 44.18\pm 7.16$ & $ 28.99\pm 9.06$ & $0.346$ &$0.366$ & $...$ & $ 5.33$\\
J090934.9+002840 &         1398$^{\star}$ & $ 0           9^{\rm m} 0           9^{\rm h}           35\fs 0           7$ & $+0           0^\circ          28^\prime           43\farcs            5$ & $...$ & $...$ & $...$ & $...$ & $...$ & $ 20.06\pm 0.02$ & $ 20.02\pm 0.03$ & $ 55.33\pm 6.56$ & $ 44.18\pm 7.16$ & $ 28.99\pm 9.06$ & $0.637$ &$0.615$ & $...$ & $ 3.77$\\
J090817.7+002106 &         1426 & $ 0           9^{\rm m} 0           8^{\rm h}           17\fs           68$ & $+0           0^\circ          21^\prime 0           8\farcs            3$ & $22.36\pm22.36$ & $ 21.96\pm 0.19$ & $ 21.77\pm 0.18$ & $> 20.99$ & $ 20.52\pm 0.12$ & $ 19.00\pm 0.01$ & $...$ & $ 54.34\pm 6.53$ & $ 32.19\pm 7.14$ & $ 13.61\pm 8.92$ & $0.996$ &$...$ & $...$ & $ 1.62$\\
J090854.3+003248 &         1434 & $ 0           9^{\rm m} 0           8^{\rm h}           54\fs           48$ & $+0           0^\circ          32^\prime           48\farcs            4$ & $19.21\pm19.21$ & $ 19.02\pm 0.02$ & $ 18.48\pm 0.01$ & $ 18.00\pm 0.01$ & $ 17.52\pm 0.01$ & $ 17.85\pm 0.01$ & $ 17.79\pm 0.01$ & $ 47.07\pm 6.42$ & $ 11.82\pm 7.16$ & $  2.31\pm 9.10$ & $0.994$ &$0.999$ & $0.31\pm0.08$ & $ 1.33$\\
J090924.3+004156 &         1483 & $ 0           9^{\rm m} 0           9^{\rm h}           24\fs           32$ & $+0           0^\circ          41^\prime           54\farcs            3$ & $...$ & $...$ & $...$ & $...$ & $...$ & $...$ & $ 20.06\pm 0.02$ & $ 50.52\pm 6.42$ & $ 36.97\pm 7.21$ & $ 24.34\pm 8.95$ & $...$ &$0.987$ & $...$ & $ 2.20$\\
J090945.6+003714 &         1559 & $ 0           9^{\rm m} 0           9^{\rm h}           45\fs           59$ & $+0           0^\circ          37^\prime           14\farcs            1$ & $...$ & $...$ & $...$ & $...$ & $...$ & $ 19.56\pm 0.02$ & $ 19.29\pm 0.01$ & $ 53.50\pm 6.44$ & $ 45.17\pm 7.24$ & $ 39.72\pm 9.05$ & $0.995$ &$0.997$ & $...$ & $ 1.07$\\
J090823.8+004611 &         1685 & $ 0           9^{\rm m} 0           8^{\rm h}           23\fs           78$ & $+0           0^\circ          46^\prime           10\farcs            3$ & $16.77\pm16.77$ & $ 16.78\pm 0.00$ & $ 16.83\pm 0.00$ & $ 16.95\pm 0.01$ & $ 17.32\pm 0.01$ & $ 18.36\pm 0.01$ & $ 18.71\pm 0.01$ & $ 53.11\pm 6.49$ & $ 31.00\pm 7.12$ & $ 21.12\pm 9.27$ & $0.999$ &$0.998$ & $...$ & $ 1.14$\\
J090822.6+004200 &         1750 & $ 0           9^{\rm m} 0           8^{\rm h}           22\fs           88$ & $+0           0^\circ          42^\prime 0           4\farcs            2$ & $19.71\pm19.71$ & $ 19.33\pm 0.02$ & $ 18.89\pm 0.01$ & $ 18.45\pm 0.02$ & $ 18.02\pm 0.01$ & $ 18.38\pm 0.01$ & $ 18.38\pm 0.01$ & $ 51.85\pm 6.51$ & $ 36.36\pm 7.26$ & $ 21.11\pm 8.90$ & $0.913$ &$0.985$ & $0.14\pm0.02$ & $ 5.47$\\
J090307.6+004012 &         1762 & $ 0           9^{\rm m} 0           3^{\rm h} 0           7\fs           79$ & $+0           0^\circ          40^\prime           16\farcs            7$ & $20.23\pm20.23$ & $ 19.91\pm 0.05$ & $ 19.74\pm 0.03$ & $ 19.28\pm 0.03$ & $ 18.89\pm 0.03$ & $ 18.82\pm 0.01$ & $ 19.03\pm 0.01$ & $ 53.54\pm 6.56$ & $ 49.59\pm 7.17$ & $ 33.94\pm 8.91$ & $0.990$ &$0.983$ & $...$ & $ 4.62$\\
J090920.4+002316 &         1976 & $ 0           9^{\rm m} 0           9^{\rm h}           20\fs           46$ & $+0           0^\circ          23^\prime           21\farcs            1$ & $20.14\pm20.14$ & $ 19.95\pm 0.03$ & $ 19.51\pm 0.02$ & $ 18.91\pm 0.03$ & $ 18.44\pm 0.02$ & $ 18.48\pm 0.01$ & $ 18.36\pm 0.01$ & $ 50.05\pm 6.57$ & $ 23.40\pm 7.29$ & $ 11.60\pm 9.11$ & $0.940$ &$0.948$ & $0.45\pm0.07$ & $ 4.91$\\
J090314.5+003359 &         2121 & $ 0           9^{\rm m} 0           3^{\rm h}           14\fs           73$ & $+0           0^\circ          33^\prime           58\farcs            6$ & $18.95\pm18.95$ & $ 18.88\pm 0.02$ & $ 18.67\pm 0.01$ & $ 18.30\pm 0.02$ & $ 18.04\pm 0.01$ & $ 18.37\pm 0.01$ & $...$ & $ 48.57\pm 6.51$ & $ 17.97\pm 7.26$ & $ 12.58\pm 8.86$ & $0.999$ &$...$ & $0.23\pm0.04$ & $ 2.60$\\
J090920.5+003742 &         2180 & $ 0           9^{\rm m} 0           9^{\rm h}           20\fs           31$ & $+0           0^\circ          37^\prime           42\farcs            5$ & $18.97\pm18.97$ & $ 18.74\pm 0.01$ & $ 18.39\pm 0.01$ & $ 18.05\pm 0.01$ & $ 17.77\pm 0.01$ & $ 18.13\pm 0.01$ & $ 18.27\pm 0.01$ & $ 48.41\pm 6.43$ & $ 49.96\pm 7.18$ & $ 44.04\pm 8.93$ & $0.998$ &$0.998$ & $0.30\pm0.05$ & $ 3.13$\\
J090913.5+002736 &         2244 & $ 0           9^{\rm m} 0           9^{\rm h}           13\fs           54$ & $+0           0^\circ          27^\prime           33\farcs            9$ & $...$ & $...$ & $...$ & $...$ & $...$ & $ 19.02\pm 0.01$ & $ 19.28\pm 0.01$ & $ 48.09\pm 6.51$ & $ 39.94\pm 7.22$ & $ 32.63\pm 9.21$ & $0.993$ &$0.994$ & $...$ & $ 3.08$\\
J090855.6+001526 &         2277 & $ 0           9^{\rm m} 0           8^{\rm h}           55\fs           92$ & $+0           0^\circ          15^\prime           25\farcs            2$ & $18.98\pm18.98$ & $ 18.77\pm 0.01$ & $ 18.35\pm 0.01$ & $ 18.02\pm 0.01$ & $ 17.75\pm 0.01$ & $ 17.94\pm 0.01$ & $...$ & $ 46.36\pm 6.47$ & $ 16.00\pm 7.20$ & $ -2.30\pm 9.00$ & $0.998$ &$...$ & $0.20\pm0.04$ & $ 4.56$\\
\hline
\end{tabular}
\end{tiny}
\\Note. -- 
$^{\star}$~ID with color-magnitude method.
$^a$~Coordinates are the position of IRAC counterparts.
$^b$~Photometry from the VIKING survey ($Z, Y ,J , H,  K_{\rm s}$) is extracted in $2\arcsec$ diameter apertures (Sutherland et al.\ in prep).
$^c$~The reported SPIRE 350 and 500~\micron\ fluxes are measured at the postions of 250~\micron\ sources; no SNR cut is applied.
$^d$~R is reliability of the counterparts calculated separately at 3.6 and 4.5~\micron\ from the LR method (section~\ref{sec:lr}).
$^e$~Redshifts are from the SDSS catalogue for sources detected in that survey.
$^f$~The separation between the centroid of the SPIRE emission and the IRAC counterpart. 
\vspace{3mm}

%% file: lr_photometry_table1.tex
\begin{tiny}
\begin{tabular}{llccccccccccccccccc}
\hline
IAU ID & H-ATLAS  & RA$^{a}$  & Dec$^{a}$ &  $Z$$^{b}$ & {\it Y} & {\it J$^{b}$} & {\it H$^{b}$} & $K_{\rm s}$$^{b}$ & 3.6$\,\mu$m & 4.5$\,\mu$m
&$S_{250}$$^c$ & $S_{350}$$^c$ & $S_{500}$$^c$ & $R^d$ & $R^d$ & z$^e$ & Separation$^f$  \\
       & ID & (J2000) & (J2000) &   (mag)  & (mag)  & (mag) & (mag) &  (mag) & (mag) & (mag) & (mJy) & (mJy) &(mJy) &(3.6\micron) &(4.5\micron) & & (arcsec)\\
\hline
J090925.2+003224 &         2281 & $ 0           9^{\rm m} 0           9^{\rm h}           25\fs           11$ & $+0           0^\circ          32^\prime           26\farcs            8$ & $>23.06$ & $> 22.35$ & $ 21.43\pm 0.12$ & $ 20.87\pm 0.16$ & $ 20.32\pm 0.11$ & $ 19.39\pm 0.01$ & $ 19.04\pm 0.01$ & $ 47.80\pm 6.50$ & $ 41.33\pm 7.27$ & $ 14.29\pm 9.27$ & $0.992$ &$0.993$ & $...$ & $ 3.35$\\
J091257.8-005508 &         2343 & $ 0           9^{\rm m} 0           9^{\rm h}           57\fs           82$ & $-0           0^\circ          55^\prime 0           2\farcs            7$ & $18.46\pm18.46$ & $ 18.13\pm 0.01$ & $ 17.98\pm 0.01$ & $ 17.91\pm 0.01$ & $ 18.08\pm 0.02$ & $ 18.22\pm 0.01$ & $ 18.28\pm 0.01$ & $ 42.07\pm 6.55$ & $ 24.56\pm 7.29$ & $ 10.37\pm 9.30$ & $0.927$ &$0.949$ & $0.38\pm0.07$ & $ 5.75$\\
J090841.3+002005 &         2432 & $ 0           9^{\rm m} 0           8^{\rm h}           41\fs           38$ & $+0           0^\circ          20^\prime 0           6\farcs            6$ & $20.27\pm20.27$ & $ 19.97\pm 0.03$ & $ 19.79\pm 0.03$ & $ 19.25\pm 0.04$ & $ 18.80\pm 0.03$ & $ 18.70\pm 0.01$ & $...$ & $ 46.36\pm 6.45$ & $ 27.64\pm 7.15$ & $ 30.30\pm 9.12$ & $0.998$ &$...$ & $0.39\pm0.07$ & $ 1.44$\\
J090752.3+002100 &         2437 & $ 0           9^{\rm m} 0           7^{\rm h}           52\fs           22$ & $+0           0^\circ          21^\prime 0           0\farcs            3$ & $...$ & $...$ & $...$ & $...$ & $...$ & $ 20.05\pm 0.02$ & $...$ & $ 46.26\pm 6.46$ & $ 34.17\pm 7.25$ & $ 20.42\pm 9.32$ & $0.989$ &$...$ & $...$ & $ 1.92$\\
J090850.0+004309 &         2459$^{\star}$ & $ 0           9^{\rm m} 0           8^{\rm h}           49\fs           79$ & $+0           0^\circ          43^\prime 0           7\farcs            7$ & $20.62\pm20.62$ & $ 20.11\pm 0.04$ & $ 20.02\pm 0.04$ & $ 19.56\pm 0.05$ & $ 19.16\pm 0.04$ & $ 18.80\pm 0.01$ & $ 18.81\pm 0.01$ & $ 46.74\pm 6.55$ & $ 30.97\pm 7.16$ & $  6.70\pm 8.99$ & $0.354$ &$0.399$ & $...$ & $ 4.45$\\
J090850.0+004309 &         2459$^{\star}$ & $ 0           9^{\rm m} 0           8^{\rm h}           50\fs           16$ & $+0           0^\circ          43^\prime 0           9\farcs            7$ & $>22.37$ & $ 21.98\pm 0.20$ & $ 22.17\pm 0.26$ & $ 21.09\pm 0.22$ & $ 20.61\pm 0.14$ & $ 19.76\pm 0.03$ & $ 19.65\pm 0.01$ & $ 46.74\pm 6.55$ & $ 30.97\pm 7.16$ & $  6.70\pm 8.99$ & $0.643$ &$0.598$ & $...$ & $ 1.50$\\
J090803.8+002250 &         2549 & $ 0           9^{\rm m} 0           8^{\rm h} 0           3\fs           57$ & $+0           0^\circ          22^\prime           51\farcs            6$ & $19.52\pm19.52$ & $ 18.90\pm 0.02$ & $ 18.62\pm 0.01$ & $ 18.18\pm 0.02$ & $ 17.83\pm 0.01$ & $ 18.12\pm 0.01$ & $...$ & $ 45.96\pm 6.55$ & $ 23.21\pm 7.15$ & $  2.63\pm 9.01$ & $0.995$ &$...$ & $0.29\pm0.08$ & $ 4.67$\\
J090855.5+002808 &         2565 & $ 0           9^{\rm m} 0           8^{\rm h}           55\fs           59$ & $+0           0^\circ          28^\prime 0           6\farcs            8$ & $>21.79$ & $> 21.40$ & $> 20.94$ & $>-99.00$ & $> 19.20$ & $ 21.25\pm 0.06$ & $ 20.83\pm 0.04$ & $ 45.75\pm 6.51$ & $ 38.65\pm 7.14$ & $ 32.42\pm 8.92$ & $0.896$ &$0.977$ & $...$ & $ 1.85$\\
J090930.2+002755 &         2680 & $ 0           9^{\rm m} 0           9^{\rm h}           30\fs           22$ & $+0           0^\circ          27^\prime           55\farcs 0$ & $22.07\pm22.07$ & $ 21.55\pm 0.10$ & $ 20.99\pm 0.08$ & $ 20.49\pm 0.11$ & $ 20.08\pm 0.09$ & $ 19.27\pm 0.01$ & $ 19.21\pm 0.01$ & $ 45.12\pm 6.36$ & $ 37.32\pm 7.15$ & $ 15.07\pm 8.93$ & $0.953$ &$0.966$ & $...$ & $ 0.51$\\
J090803.7+002921 &         2715 & $ 0           9^{\rm m} 0           8^{\rm h} 0           3\fs           82$ & $+0           0^\circ          29^\prime           22\farcs            6$ & $>21.91$ & $> 21.40$ & $> 20.94$ & $> 19.61$ & $> 19.20$ & $ 20.85\pm 0.04$ & $ 20.39\pm 0.03$ & $ 46.08\pm 6.52$ & $ 28.66\pm 7.30$ & $ 29.26\pm 9.02$ & $0.985$ &$0.990$ & $...$ & $ 1.14$\\
J090905.3+001525 &         2773 & $ 0           9^{\rm m} 0           9^{\rm h} 0           5\fs           49$ & $+0           0^\circ          15^\prime           23\farcs            6$ & $>22.54$ & $> 21.73$ & $> 21.05$ & $> 19.74$ & $> 19.20$ & $ 20.06\pm 0.02$ & $...$ & $ 44.96\pm 6.53$ & $ 38.06\pm 7.18$ & $ 25.49\pm 8.90$ & $0.836$ &$...$ & $...$ & $ 2.61$\\
J090846.0+004339 &         2793$^{\star}$ & $ 0           9^{\rm m} 0           8^{\rm h}           46\fs 0           1$ & $+0           0^\circ          43^\prime           36\farcs            8$ & $...$ & $...$ & $...$ & $...$ & $...$ & $ 20.41\pm 0.06$ & $ 20.21\pm 0.03$ & $ 45.21\pm 6.42$ & $ 39.54\pm 7.23$ & $ 18.88\pm 8.97$ & $0.613$ &$0.478$ & $...$ & $ 2.57$\\
J090846.0+004339 &         2793$^{\star}$ & $ 0           9^{\rm m} 0           8^{\rm h}           45\fs           88$ & $+0           0^\circ          43^\prime           39\farcs 0$ & $...$ & $...$ & $...$ & $...$ & $...$ & $ 20.56\pm 0.07$ & $ 20.08\pm 0.02$ & $ 45.21\pm 6.42$ & $ 39.54\pm 7.23$ & $ 18.88\pm 8.97$ & $0.378$ &$0.515$ & $...$ & $ 2.40$\\
J090943.0+004322 &         2796 & $ 0           9^{\rm m} 0           9^{\rm h}           43\fs           23$ & $+0           0^\circ          43^\prime           22\farcs            2$ & $>23.06$ & $> 22.35$ & $ 19.70\pm 0.03$ & $ 19.38\pm 0.04$ & $ 19.09\pm 0.04$ & $...$ & $ 18.66\pm 0.01$ & $ 39.85\pm 6.38$ & $ 21.33\pm 7.15$ & $ 11.52\pm 9.10$ & $...$ &$0.998$ & $0.16\pm0.17$ & $ 2.24$\\
J090819.6+003259 &         2866 & $ 0           9^{\rm m} 0           8^{\rm h}           20\fs          00$ & $+0           0^\circ          33^\prime 0           2\farcs            6$ & $>22.43$ & $> 22.02$ & $ 21.94\pm 0.21$ & $> 20.99$ & $ 20.84\pm 0.17$ & $ 19.95\pm 0.02$ & $ 19.63\pm 0.01$ & $ 44.81\pm 6.48$ & $ 44.94\pm 7.15$ & $ 34.21\pm 9.12$ & $0.882$ &$0.881$ & $...$ & $ 6.31$\\
J091302.7-004618 &         2986 & $ 0           9^{\rm m} 0           8^{\rm h} 0           2\fs           92$ & $-0           0^\circ          46^\prime           19\farcs            7$ & $19.06\pm19.06$ & $ 18.85\pm 0.02$ & $ 18.62\pm 0.01$ & $ 18.39\pm 0.02$ & $ 18.11\pm 0.02$ & $...$ & $ 18.22\pm 0.01$ & $ 44.02\pm 6.53$ & $ 21.20\pm 7.40$ & $  9.62\pm 8.99$ & $...$ &$0.928$ & $0.25\pm0.03$ & $ 3.57$\\
J090922.4+002715 &         3043 & $ 0           9^{\rm m} 0           9^{\rm h}           22\fs           19$ & $+0           0^\circ          27^\prime           15\farcs            3$ & $21.11\pm21.11$ & $ 20.80\pm 0.05$ & $ 20.49\pm 0.05$ & $ 20.15\pm 0.08$ & $ 19.54\pm 0.06$ & $ 18.87\pm 0.01$ & $ 19.20\pm 0.01$ & $ 42.74\pm 6.48$ & $ 17.48\pm 7.14$ & $ 14.55\pm 8.99$ & $0.922$ &$0.919$ & $0.68\pm0.12$ & $ 3.57$\\
J090333.3+004746 &         3056 & $ 0           9^{\rm m} 0           3^{\rm h}           33\fs           53$ & $+0           0^\circ          47^\prime           48\farcs            8$ & $...$ & $...$ & $...$ & $...$ & $...$ & $...$ & $ 19.44\pm 0.03$ & $ 43.97\pm 6.43$ & $ 48.32\pm 7.15$ & $ 25.88\pm 8.83$ & $...$ &$0.891$ & $...$ & $ 3.12$\\
J090800.5+002457 &         3084 & $ 0           9^{\rm m} 0           8^{\rm h} 0           0\fs           51$ & $+0           0^\circ          24^\prime           57\farcs            3$ & $>21.85$ & $> 21.40$ & $> 20.94$ & $> 19.61$ & $> 19.20$ & $ 21.21\pm 0.05$ & $...$ & $ 42.94\pm 6.51$ & $ 61.39\pm 7.18$ & $ 46.14\pm 8.76$ & $0.918$ &$...$ & $...$ & $ 0.57$\\
J090839.4+004107 &         3113 & $ 0           9^{\rm m} 0           8^{\rm h}           39\fs           11$ & $+0           0^\circ          41^\prime 0           6\farcs            9$ & $18.76\pm18.76$ & $ 18.39\pm 0.01$ & $ 17.98\pm 0.01$ & $ 17.57\pm 0.01$ & $ 17.22\pm 0.01$ & $ 17.62\pm 0.01$ & $ 17.68\pm 0.01$ & $ 44.18\pm 6.41$ & $ 19.63\pm 7.23$ & $ -0.42\pm 9.33$ & $0.959$ &$0.948$ & $0.25\pm0.03$ & $ 5.73$\\
J090746.7+002148 &         3161 & $ 0           9^{\rm m} 0           7^{\rm h}           46\fs           38$ & $+0           0^\circ          21^\prime           47\farcs            2$ & $19.63\pm19.63$ & $ 19.24\pm 0.02$ & $ 19.05\pm 0.02$ & $ 18.64\pm 0.02$ & $ 18.32\pm 0.02$ & $ 18.52\pm 0.01$ & $...$ & $ 44.71\pm 6.44$ & $ 18.68\pm 7.27$ & $  9.49\pm 9.12$ & $0.984$ &$...$ & $0.35\pm0.03$ & $ 5.16$\\
J090832.7+002406 &         3205 & $ 0           9^{\rm m} 0           8^{\rm h}           32\fs           85$ & $+0           0^\circ          24^\prime 0           7\farcs            1$ & $21.17\pm21.17$ & $ 20.64\pm 0.06$ & $ 20.19\pm 0.04$ & $ 19.64\pm 0.06$ & $ 19.11\pm 0.04$ & $ 18.65\pm 0.01$ & $ 18.94\pm 0.01$ & $ 40.89\pm 6.65$ & $ 30.78\pm 7.31$ & $ 14.69\pm 8.98$ & $0.998$ &$0.998$ & $0.63\pm0.10$ & $ 2.13$\\
J090829.1+001556 &         3242 & $ 0           9^{\rm m} 0           8^{\rm h}           28\fs           96$ & $+0           0^\circ          15^\prime           58\farcs            8$ & $>22.37$ & $> 22.02$ & $ 21.90\pm 0.20$ & $> 20.99$ & $ 20.49\pm 0.12$ & $ 19.29\pm 0.02$ & $...$ & $ 42.99\pm 6.59$ & $ 43.26\pm 7.16$ & $ 15.88\pm 8.91$ & $0.987$ &$...$ & $...$ & $ 4.12$\\
J090843.7+001437 &         3251 & $ 0           9^{\rm m} 0           8^{\rm h}           43\fs           61$ & $+0           0^\circ          14^\prime           38\farcs            4$ & $19.81\pm19.81$ & $ 19.59\pm 0.03$ & $ 19.28\pm 0.02$ & $ 18.82\pm 0.03$ & $ 18.46\pm 0.02$ & $ 18.20\pm 0.01$ & $...$ & $ 38.04\pm 6.61$ & $ 28.16\pm 7.14$ & $  5.02\pm 8.95$ & $0.961$ &$...$ & $0.51\pm0.12$ & $ 2.68$\\
J091303.6-004855 &         3285 & $ 0           9^{\rm m} 0           8^{\rm h} 0           3\fs           34$ & $-0           0^\circ          48^\prime           59\farcs            1$ & $18.82\pm18.82$ & $ 18.58\pm 0.01$ & $ 18.24\pm 0.01$ & $ 17.92\pm 0.01$ & $ 17.67\pm 0.01$ & $...$ & $ 18.20\pm 0.01$ & $ 42.06\pm 6.40$ & $ 11.54\pm 7.16$ & $  9.00\pm 9.00$ & $...$ &$0.975$ & $0.11\pm0.02$ & $ 6.28$\\
J090836.4+002948 &         3304 & $ 0           9^{\rm m} 0           8^{\rm h}           36\fs           62$ & $+0           0^\circ          29^\prime           48\farcs            3$ & $21.90\pm21.90$ & $ 21.38\pm 0.12$ & $ 21.04\pm 0.09$ & $ 20.94\pm 0.19$ & $ 20.43\pm 0.11$ & $ 19.22\pm 0.01$ & $ 19.15\pm 0.01$ & $ 39.87\pm 6.41$ & $ 45.74\pm 7.28$ & $ 31.52\pm 8.86$ & $0.996$ &$0.996$ & $...$ & $ 2.10$\\
J090940.3+002939 &         3318 & $ 0           9^{\rm m} 0           9^{\rm h}           40\fs           28$ & $+0           0^\circ          29^\prime           42\farcs            4$ & $...$ & $...$ & $...$ & $...$ & $...$ & $ 20.56\pm 0.04$ & $ 20.33\pm 0.03$ & $ 42.01\pm 6.55$ & $ 32.10\pm 7.21$ & $ 20.21\pm 8.92$ & $0.932$ &$0.985$ & $...$ & $ 2.56$\\
J090921.9+002556 &         3457 & $ 0           9^{\rm m} 0           9^{\rm h}           21\fs           78$ & $+0           0^\circ          25^\prime           57\farcs            6$ & $18.83\pm18.83$ & $ 18.59\pm 0.01$ & $ 18.20\pm 0.01$ & $ 17.80\pm 0.01$ & $ 17.53\pm 0.01$ & $ 17.76\pm 0.01$ & $ 17.92\pm 0.01$ & $ 41.50\pm 6.41$ & $ 19.09\pm 7.30$ & $-12.39\pm 9.13$ & $0.997$ &$0.996$ & $0.12\pm0.02$ & $ 2.33$\\
J090832.2+001938 &         3573 & $ 0           9^{\rm m} 0           8^{\rm h}           32\fs           24$ & $+0           0^\circ          19^\prime           40\farcs            5$ & $18.93\pm18.93$ & $ 18.84\pm 0.01$ & $ 18.48\pm 0.01$ & $ 18.20\pm 0.02$ & $ 17.94\pm 0.01$ & $ 18.22\pm 0.01$ & $...$ & $ 41.25\pm 6.53$ & $ 22.19\pm 7.17$ & $  1.67\pm 9.10$ & $0.967$ &$...$ & $0.21\pm0.03$ & $ 2.15$\\
J090323.9+004620 &         3574 & $ 0           9^{\rm m} 0           3^{\rm h}           23\fs           60$ & $+0           0^\circ          46^\prime           23\farcs            8$ & $19.83\pm19.83$ & $ 19.59\pm 0.03$ & $ 19.33\pm 0.02$ & $ 18.95\pm 0.03$ & $ 18.66\pm 0.02$ & $...$ & $ 18.99\pm 0.01$ & $ 41.37\pm 6.53$ & $ 20.20\pm 7.25$ & $  9.76\pm 8.79$ & $...$ &$0.976$ & $0.29\pm0.07$ & $ 5.75$\\
J091317.7-004621 &         3583 & $ 0           9^{\rm m} 0           3^{\rm h}           17\fs           75$ & $-0           0^\circ          46^\prime           16\farcs            1$ & $19.94\pm19.94$ & $ 19.69\pm 0.03$ & $ 19.61\pm 0.03$ & $ 19.30\pm 0.04$ & $ 18.98\pm 0.03$ & $...$ & $ 19.24\pm 0.01$ & $ 41.18\pm 6.55$ & $ 34.52\pm 7.14$ & $ 20.07\pm 9.18$ & $...$ &$0.963$ & $0.25\pm0.05$ & $ 5.53$\\
J090901.5+003107 &         3616 & $ 0           9^{\rm m} 0           9^{\rm h} 0           1\fs           77$ & $+0           0^\circ          31^\prime 0           8\farcs            1$ & $21.55\pm21.55$ & $ 21.01\pm 0.08$ & $ 20.54\pm 0.06$ & $ 19.82\pm 0.07$ & $ 19.17\pm 0.04$ & $ 18.69\pm 0.01$ & $ 19.05\pm 0.01$ & $ 41.09\pm 6.50$ & $ 20.53\pm 7.20$ & $ -1.89\pm 8.79$ & $0.983$ &$0.975$ & $0.59\pm0.08$ & $ 2.59$\\
J090858.8+003158 &         3714 & $ 0           9^{\rm m} 0           8^{\rm h}           58\fs           82$ & $+0           0^\circ          31^\prime           54\farcs            8$ & $20.47\pm20.47$ & $ 20.17\pm 0.04$ & $ 19.86\pm 0.03$ & $ 19.27\pm 0.04$ & $ 18.95\pm 0.03$ & $ 19.09\pm 0.01$ & $ 19.04\pm 0.01$ & $ 39.63\pm 6.46$ & $ 15.19\pm 7.24$ & $ -3.99\pm 8.95$ & $0.882$ &$0.944$ & $0.53\pm0.08$ & $ 3.62$\\
J091001.4+004024 &         3792 & $ 0           9^{\rm m} 0           8^{\rm h} 0           1\fs           44$ & $+0           0^\circ          40^\prime           21\farcs            8$ & $21.29\pm21.29$ & $ 21.16\pm 0.07$ & $ 20.61\pm 0.06$ & $ 20.22\pm 0.09$ & $ 19.73\pm 0.07$ & $...$ & $ 19.56\pm 0.04$ & $ 39.95\pm 6.38$ & $ 28.10\pm 7.18$ & $ 10.01\pm 9.26$ & $...$ &$0.984$ & $0.46\pm0.18$ & $ 3.02$\\
J090903.5+002031 &         3825 & $ 0           9^{\rm m} 0           9^{\rm h} 0           3\fs           34$ & $+0           0^\circ          20^\prime           30\farcs            8$ & $18.27\pm18.27$ & $ 18.09\pm 0.01$ & $ 17.80\pm 0.01$ & $ 17.52\pm 0.01$ & $ 17.48\pm 0.01$ & $ 17.43\pm 0.01$ & $ 17.76\pm 0.01$ & $ 38.19\pm 6.48$ & $ 12.71\pm 7.23$ & $  5.37\pm 9.19$ & $0.999$ &$0.999$ & $0.09\pm0.00$ & $ 3.02$\\
J090921.9+004307 &         3974 & $ 0           9^{\rm m} 0           9^{\rm h}           22\fs           24$ & $+0           0^\circ          43^\prime 0           8\farcs            3$ & $>23.06$ & $> 22.35$ & $ 20.60\pm 0.06$ & $ 20.12\pm 0.08$ & $ 19.73\pm 0.07$ & $...$ & $ 18.09\pm 0.01$ & $ 40.61\pm 6.49$ & $ 36.92\pm 7.22$ & $ 23.08\pm 8.92$ & $...$ &$0.963$ & $...$ & $ 4.03$\\
J090901.9+004217 &         4113 & $ 0           9^{\rm m} 0           9^{\rm h} 0           1\fs           87$ & $+0           0^\circ          42^\prime           19\farcs            9$ & $>23.06$ & $> 22.35$ & $ 21.64\pm 0.16$ & $> 21.13$ & $ 20.70\pm 0.15$ & $ 19.80\pm 0.03$ & $ 19.48\pm 0.01$ & $ 37.23\pm 6.49$ & $ 44.93\pm 7.14$ & $ 32.16\pm 8.77$ & $0.974$ &$0.986$ & $...$ & $ 2.22$\\
J090811.9+003410 &         4185 & $ 0           9^{\rm m} 0           8^{\rm h}           12\fs 0           6$ & $+0           0^\circ          34^\prime           13\farcs            2$ & $>22.43$ & $> 22.02$ & $ 22.11\pm 0.25$ & $> 20.99$ & $ 20.91\pm 0.18$ & $ 19.94\pm 0.02$ & $ 19.79\pm 0.02$ & $ 39.10\pm 6.41$ & $ 27.54\pm 7.22$ & $ 27.09\pm 9.20$ & $0.987$ &$0.987$ & $...$ & $ 3.67$\\
J090821.6+002700 &         4344 & $ 0           9^{\rm m} 0           8^{\rm h}           21\fs           24$ & $+0           0^\circ          27^\prime 0           0\farcs 0$ & $21.66\pm21.66$ & $ 21.42\pm 0.12$ & $ 21.41\pm 0.13$ & $> 20.99$ & $ 20.65\pm 0.14$ & $ 19.73\pm 0.01$ & $ 19.93\pm 0.02$ & $ 38.39\pm 6.40$ & $ 30.19\pm 7.18$ & $ 20.33\pm 8.92$ & $0.890$ &$0.889$ & $...$ & $ 6.24$\\
J090935.1+002224 &         4352 & $ 0           9^{\rm m} 0           9^{\rm h}           35\fs           26$ & $+0           0^\circ          22^\prime           22\farcs            3$ & $...$ & $...$ & $...$ & $...$ & $...$ & $ 19.48\pm 0.03$ & $ 19.47\pm 0.04$ & $ 37.71\pm 6.61$ & $ 26.04\pm 7.17$ & $ 24.95\pm 9.08$ & $0.983$ &$0.986$ & $...$ & $ 2.79$\\
J090918.3+003409 &         4366$^{\star}$ & $ 0           9^{\rm m} 0           9^{\rm h}           18\fs           30$ & $+0           0^\circ          34^\prime 0           9\farcs            2$ & $...$ & $...$ & $...$ & $...$ & $...$ & $ 19.96\pm 0.02$ & $ 19.73\pm 0.02$ & $ 38.79\pm 6.55$ & $ 31.27\pm 7.24$ & $ 14.04\pm 8.90$ & $0.545$ &$0.545$ & $...$ & $ 0.64$\\
J090918.3+003409 &         4366$^{\star}$ & $ 0           9^{\rm m} 0           9^{\rm h}           18\fs           44$ & $+0           0^\circ          34^\prime 0           9\farcs            7$ & $20.82\pm20.82$ & $ 20.58\pm 0.04$ & $ 20.50\pm 0.05$ & $ 20.25\pm 0.09$ & $ 19.99\pm 0.08$ & $ 19.91\pm 0.02$ & $ 19.87\pm 0.02$ & $ 38.79\pm 6.55$ & $ 31.27\pm 7.24$ & $ 14.04\pm 8.90$ & $0.452$ &$0.452$ & $0.46\pm0.15$ & $ 1.60$\\
J091306.9-004719 &         4404 & $ 0           9^{\rm m} 0           9^{\rm h} 0           7\fs 0           8$ & $-0           0^\circ          47^\prime           20\farcs            9$ & $22.65\pm22.65$ & $ 22.23\pm 0.26$ & $ 21.50\pm 0.16$ & $ 20.84\pm 0.18$ & $ 20.28\pm 0.10$ & $...$ & $ 19.11\pm 0.01$ & $ 37.63\pm 6.52$ & $ 32.91\pm 7.19$ & $ 29.28\pm 9.11$ & $...$ &$0.996$ & $...$ & $ 2.37$\\
J091318.1-005409 &         4520 & $ 0           9^{\rm m} 0           9^{\rm h}           18\fs           11$ & $-0           0^\circ          54^\prime           16\farcs            1$ & $21.98\pm21.98$ & $ 21.85\pm 0.18$ & $ 21.56\pm 0.17$ & $ 20.99\pm 0.20$ & $ 20.74\pm 0.16$ & $ 20.36\pm 0.04$ & $ 20.83\pm 0.07$ & $ 39.29\pm 6.62$ & $ 20.79\pm 7.25$ & $  4.80\pm 9.07$ & $0.764$ &$0.590$ & $...$ & $ 6.51$\\
J090744.7+002005 &         4524 & $ 0           9^{\rm m} 0           7^{\rm h}           44\fs           70$ & $+0           0^\circ          20^\prime 0           6\farcs 0$ & $>21.86$ & $> 21.40$ & $> 20.94$ & $> 19.61$ & $> 19.20$ & $ 21.32\pm 0.10$ & $...$ & $ 37.82\pm 6.46$ & $ 33.80\pm 7.11$ & $ 19.53\pm 9.10$ & $0.892$ &$...$ & $...$ & $ 1.98$\\
J090902.0+001936 &         4587$^{\star}$ & $ 0           9^{\rm m} 0           9^{\rm h} 0           2\fs 0           5$ & $+0           0^\circ          19^\prime           34\farcs            2$ & $...$ & $...$ & $...$ & $...$ & $...$ & $ 21.36\pm 0.08$ & $ 20.72\pm 0.10$ & $ 34.85\pm 6.33$ & $ 45.90\pm 7.21$ & $ 32.27\pm 8.92$ & $0.075$ &$0.326$ & $...$ & $ 2.35$\\
J090902.0+001936 &         4587$^{\star}$ & $ 0           9^{\rm m} 0           9^{\rm h} 0           2\fs           39$ & $+0           0^\circ          19^\prime           35\farcs            2$ & $21.32\pm21.32$ & $ 21.16\pm 0.09$ & $ 20.56\pm 0.06$ & $ 20.36\pm 0.11$ & $ 19.84\pm 0.07$ & $ 19.27\pm 0.01$ & $ 19.63\pm 0.04$ & $ 34.85\pm 6.33$ & $ 45.90\pm 7.21$ & $ 32.27\pm 8.92$ & $0.357$ &$0.219$ & $0.33\pm0.15$ & $ 5.15$\\
J090902.0+001936 &         4587$^{\star}$ & $ 0           9^{\rm m} 0           9^{\rm h} 0           1\fs           94$ & $+0           0^\circ          19^\prime           39\farcs            3$ & $...$ & $...$ & $...$ & $...$ & $...$ & $ 20.27\pm 0.03$ & $ 20.12\pm 0.05$ & $ 34.85\pm 6.33$ & $ 45.90\pm 7.21$ & $ 32.27\pm 8.92$ & $0.557$ &$0.445$ & $...$ & $ 3.19$\\
J091304.1-005141 &         4658 & $ 0           9^{\rm m} 0           9^{\rm h} 0           4\fs           37$ & $-0           0^\circ          51^\prime           42\farcs            1$ & $20.22\pm20.22$ & $ 19.94\pm 0.04$ & $ 19.55\pm 0.03$ & $ 19.08\pm 0.04$ & $ 18.67\pm 0.02$ & $ 18.73\pm 0.01$ & $ 18.95\pm 0.01$ & $ 37.35\pm 6.50$ & $ 12.10\pm 7.18$ & $  8.14\pm 9.29$ & $0.996$ &$0.997$ & $0.41\pm0.04$ & $ 3.12$\\
J090851.6+003823 &         4693 & $ 0           9^{\rm m} 0           8^{\rm h}           51\fs           66$ & $+0           0^\circ          38^\prime           22\farcs            3$ & $...$ & $...$ & $...$ & $...$ & $...$ & $ 20.21\pm 0.03$ & $ 19.95\pm 0.03$ & $ 38.05\pm 6.62$ & $ 23.97\pm 7.16$ & $  3.68\pm 9.22$ & $0.950$ &$0.926$ & $...$ & $ 1.25$\\
\hline
\end{tabular}
\end{tiny}
\\Note. -- 
$^{\star}$~ID with color-magnitude method.
$^a$~Coordinates are the position of IRAC counterparts.
$^b$~Photometry from the VIKING survey ($Z, Y ,J , H,  K_{\rm s}$) is extracted in $2\arcsec$ diameter apertures (Sutherland et al.\ in prep).
$^c$~The reported SPIRE 350 and 500~\micron\ fluxes are measured at the postions of 250~\micron\ sources; no SNR cut is applied.
$^d$~R is reliability of the counterparts calculated separately at 3.6 and 4.5~\micron\ from the LR method (section~\ref{sec:lr}).
$^e$~Redshifts are from the SDSS catalogue for sources detected in that survey.
$^f$~The separation between the centroid of the SPIRE emission and the IRAC counterpart.
\vspace{3mm}

%% file: lr_photometry_table2.tex
\begin{tiny}
\begin{tabular}{llccccccccccccccccc}
\hline
IAU ID & H-ATLAS  & RA$^{a}$  & Dec$^{a}$ &  $Z$$^{b}$ & {\it Y} & {\it J$^{b}$} & {\it H$^{b}$} & $K_{\rm s}$$^{b}$ & 3.6$\,\mu$m & 4.5$\,\mu$m
&$S_{250}$$^c$ & $S_{350}$$^c$ & $S_{500}$$^c$ & $R^d$ & $R^d$ & z$^e$ & Separation$^f$  \\
       & ID & (J2000) & (J2000) &   (mag)  & (mag)  & (mag) & (mag) &  (mag) & (mag) & (mag) & (mJy) & (mJy) &(mJy) &(3.6\micron) &(4.5\micron) & & (arcsec)\\
\hline
J090855.2+003051 &         4739 & $ 0           9^{\rm m} 0           8^{\rm h}           55\fs 0           8$ & $+0           0^\circ          30^\prime           50\farcs            7$ & $20.16\pm20.16$ & $ 19.88\pm 0.03$ & $ 19.69\pm 0.03$ & $ 19.29\pm 0.04$ & $ 18.96\pm 0.03$ & $ 18.96\pm 0.01$ & $ 19.19\pm 0.01$ & $ 37.35\pm 6.56$ & $ 17.17\pm 7.17$ & $  6.70\pm 9.03$ & $0.997$ &$0.995$ & $0.27\pm0.06$ & $ 2.60$\\
J090842.2+001815 &         4779 & $ 0           9^{\rm m} 0           8^{\rm h}           42\fs           51$ & $+0           0^\circ          18^\prime           18\farcs            2$ & $19.72\pm19.72$ & $ 19.66\pm 0.03$ & $ 19.36\pm 0.02$ & $ 19.01\pm 0.03$ & $ 18.69\pm 0.02$ & $ 18.84\pm 0.01$ & $...$ & $ 37.31\pm 6.54$ & $ 20.36\pm 7.18$ & $ 15.89\pm 9.04$ & $0.953$ &$...$ & $0.36\pm0.07$ & $ 5.10$\\
J090831.3+002251 &         4849 & $ 0           9^{\rm m} 0           8^{\rm h}           31\fs           31$ & $+0           0^\circ          22^\prime           48\farcs            2$ & $21.08\pm21.08$ & $ 20.85\pm 0.07$ & $ 20.29\pm 0.05$ & $ 20.04\pm 0.08$ & $ 19.51\pm 0.05$ & $ 19.04\pm 0.01$ & $ 19.39\pm 0.03$ & $ 36.15\pm 6.53$ & $ 30.24\pm 7.21$ & $ 10.68\pm 8.78$ & $0.939$ &$0.983$ & $0.51\pm0.16$ & $ 3.41$\\
J090833.9+004616 &         4850 & $ 0           9^{\rm m} 0           8^{\rm h}           33\fs           93$ & $+0           0^\circ          46^\prime           16\farcs            8$ & $21.31\pm21.31$ & $ 21.01\pm 0.08$ & $ 20.93\pm 0.08$ & $ 20.16\pm 0.09$ & $ 19.66\pm 0.06$ & $...$ & $ 18.60\pm 0.01$ & $ 36.38\pm 6.55$ & $ 33.06\pm 7.31$ & $  4.89\pm 9.28$ & $...$ &$0.999$ & $...$ & $ 0.90$\\
J090847.1+003247 &         4930 & $ 0           9^{\rm m} 0           8^{\rm h}           47\fs 0           7$ & $+0           0^\circ          32^\prime           45\farcs            8$ & $17.50\pm17.50$ & $ 17.34\pm 0.01$ & $ 17.03\pm 0.00$ & $ 16.75\pm 0.01$ & $ 16.54\pm 0.00$ & $ 16.53\pm 0.01$ & $ 16.33\pm 0.01$ & $ 37.94\pm 6.53$ & $ 22.44\pm 7.24$ & $ 15.05\pm 8.92$ & $0.998$ &$...$ & $0.16\pm0.02$ & $ 1.36$\\
J090931.4+003936 &         5050 & $ 0           9^{\rm m} 0           9^{\rm h}           31\fs           53$ & $+0           0^\circ          39^\prime           38\farcs            1$ & $...$ & $...$ & $...$ & $...$ & $...$ & $> 23.00$ & $ 20.07\pm 0.05$ & $ 37.04\pm 6.55$ & $ 41.24\pm 7.23$ & $ 28.78\pm 8.89$ & $...$ &$0.989$ & $...$ & $ 1.61$\\
J090741.5+002103 &         5082 & $ 0           9^{\rm m} 0           7^{\rm h}           41\fs           34$ & $+0           0^\circ          21^\prime 0           2\farcs            8$ & $21.07\pm21.07$ & $ 20.64\pm 0.06$ & $ 20.39\pm 0.05$ & $ 19.96\pm 0.08$ & $ 19.64\pm 0.06$ & $ 19.03\pm 0.01$ & $...$ & $ 36.51\pm 6.48$ & $ 23.21\pm 7.30$ & $ 21.40\pm 9.24$ & $0.973$ &$...$ & $0.68\pm0.15$ & $ 3.92$\\
J090757.3+003006 &         5157$^{\star}$ & $ 0           9^{\rm m} 0           7^{\rm h}           57\fs           73$ & $+0           0^\circ          30^\prime 0           5\farcs 0$ & $21.44\pm21.44$ & $ 21.01\pm 0.08$ & $ 20.94\pm 0.09$ & $ 20.41\pm 0.12$ & $ 20.15\pm 0.09$ & $ 19.30\pm 0.01$ & $ 19.63\pm 0.02$ & $ 35.18\pm 6.51$ & $ 40.62\pm 7.21$ & $ 30.04\pm 9.12$ & $0.462$ &$0.277$ & $0.52\pm0.12$ & $ 5.43$\\
J090846.2+004110 &         5258 & $ 0           9^{\rm m} 0           8^{\rm h}           46\fs           38$ & $+0           0^\circ          41^\prime           11\farcs 0$ & $20.03\pm20.03$ & $ 19.67\pm 0.03$ & $ 19.28\pm 0.02$ & $ 18.81\pm 0.03$ & $ 18.40\pm 0.02$ & $ 18.57\pm 0.01$ & $ 18.51\pm 0.01$ & $ 35.60\pm 6.57$ & $ 17.51\pm 7.16$ & $ -4.48\pm 9.17$ & $0.998$ &$0.998$ & $0.19\pm0.03$ & $ 1.79$\\
J090938.9+003240 &         5267 & $ 0           9^{\rm m} 0           9^{\rm h}           39\fs          00$ & $+0           0^\circ          32^\prime           40\farcs            5$ & $20.07\pm20.07$ & $ 19.82\pm 0.02$ & $ 19.58\pm 0.02$ & $ 19.15\pm 0.03$ & $ 18.89\pm 0.03$ & $ 18.61\pm 0.01$ & $ 18.99\pm 0.01$ & $ 35.90\pm 6.55$ & $ 18.28\pm 7.24$ & $ 16.48\pm 9.06$ & $0.971$ &$0.983$ & $0.59\pm0.09$ & $ 1.49$\\
J090840.6+001725 &         5327 & $ 0           9^{\rm m} 0           8^{\rm h}           40\fs           92$ & $+0           0^\circ          17^\prime           24\farcs            8$ & $19.27\pm19.27$ & $ 19.19\pm 0.02$ & $ 18.82\pm 0.01$ & $ 18.54\pm 0.02$ & $ 18.29\pm 0.02$ & $ 18.58\pm 0.01$ & $...$ & $ 34.91\pm 6.27$ & $ 19.10\pm 7.14$ & $  2.43\pm 9.07$ & $0.935$ &$...$ & $0.20\pm0.03$ & $ 3.56$\\
J090759.7+002033 &         5366 & $ 0           9^{\rm m} 0           7^{\rm h}           59\fs           67$ & $+0           0^\circ          20^\prime           34\farcs            9$ & $21.73\pm21.73$ & $ 21.39\pm 0.12$ & $ 21.12\pm 0.10$ & $ 20.86\pm 0.18$ & $ 20.32\pm 0.10$ & $ 19.49\pm 0.01$ & $...$ & $ 36.61\pm 6.37$ & $ 19.92\pm 7.26$ & $  2.40\pm 9.03$ & $0.996$ &$...$ & $...$ & $ 1.49$\\
J090932.9+002242 &         5386 & $ 0           9^{\rm m} 0           9^{\rm h}           33\fs 0           1$ & $+0           0^\circ          22^\prime           45\farcs            6$ & $...$ & $...$ & $...$ & $...$ & $...$ & $ 21.38\pm 0.09$ & $ 20.54\pm 0.05$ & $ 34.12\pm 6.43$ & $ 39.10\pm 7.26$ & $ 27.32\pm 8.96$ & $0.824$ &$0.958$ & $...$ & $ 3.23$\\
J090314.0+004235 &         5422 & $ 0           9^{\rm m} 0           3^{\rm h}           13\fs           98$ & $+0           0^\circ          42^\prime           31\farcs            2$ & $19.61\pm19.61$ & $ 19.39\pm 0.02$ & $ 19.20\pm 0.02$ & $ 19.00\pm 0.03$ & $ 18.78\pm 0.03$ & $ 18.97\pm 0.01$ & $ 18.94\pm 0.01$ & $ 33.19\pm 6.55$ & $ 32.95\pm 7.19$ & $  4.46\pm 8.98$ & $0.981$ &$0.995$ & $0.24\pm0.04$ & $ 3.85$\\
J090914.8+002041 &         5450 & $ 0           9^{\rm m} 0           9^{\rm h}           14\fs           82$ & $+0           0^\circ          20^\prime           39\farcs            5$ & $21.62\pm21.62$ & $ 21.43\pm 0.09$ & $ 21.07\pm 0.09$ & $ 20.62\pm 0.13$ & $ 20.47\pm 0.13$ & $ 20.46\pm 0.03$ & $ 20.85\pm 0.05$ & $ 35.94\pm 6.42$ & $ 24.72\pm 7.23$ & $ 16.42\pm 9.06$ & $0.987$ &$0.922$ & $0.39\pm0.19$ & $ 2.44$\\
J090929.6+003313 &         5521 & $ 0           9^{\rm m} 0           9^{\rm h}           29\fs           72$ & $+0           0^\circ          33^\prime           16\farcs            2$ & $>22.54$ & $> 21.73$ & $> 21.05$ & $> 19.74$ & $> 19.13$ & $ 19.64\pm 0.01$ & $ 19.43\pm 0.01$ & $ 35.71\pm 6.31$ & $ 24.72\pm 7.15$ & $  8.91\pm 8.99$ & $0.992$ &$0.995$ & $...$ & $ 2.75$\\
J090837.1+005002 &         5530 & $ 0           9^{\rm m} 0           8^{\rm h}           37\fs 0           4$ & $+0           0^\circ          50^\prime 0           2\farcs            2$ & $20.91\pm20.91$ & $ 20.67\pm 0.06$ & $ 20.10\pm 0.04$ & $ 19.72\pm 0.06$ & $ 19.21\pm 0.04$ & $...$ & $ 18.17\pm 0.01$ & $ 36.26\pm 6.56$ & $ 16.39\pm 7.27$ & $ -0.12\pm 9.02$ & $...$ &$0.999$ & $0.60\pm0.06$ & $ 0.93$\\
J090825.8+004217 &         5538 & $ 0           9^{\rm m} 0           8^{\rm h}           25\fs           83$ & $+0           0^\circ          42^\prime           13\farcs            7$ & $22.80\pm22.80$ & $ 22.16\pm 0.23$ & $ 21.76\pm 0.18$ & $ 21.21\pm 0.24$ & $ 20.68\pm 0.14$ & $ 20.32\pm 0.03$ & $ 20.77\pm 0.03$ & $ 35.38\pm 6.41$ & $ 17.99\pm 7.32$ & $ -5.24\pm 9.06$ & $0.736$ &$0.661$ & $...$ & $ 3.46$\\
J090821.9+002445 &         5564 & $ 0           9^{\rm m} 0           8^{\rm h}           22\fs           14$ & $+0           0^\circ          24^\prime           47\farcs            3$ & $>22.37$ & $ 22.06\pm 0.21$ & $ 21.45\pm 0.13$ & $ 20.77\pm 0.16$ & $ 20.13\pm 0.09$ & $ 19.30\pm 0.01$ & $ 19.18\pm 0.01$ & $ 35.10\pm 6.62$ & $ 19.57\pm 7.16$ & $  7.32\pm 8.97$ & $0.992$ &$0.993$ & $...$ & $ 3.29$\\
J090910.9+003517 &         5621 & $ 0           9^{\rm m} 0           9^{\rm h}           11\fs           35$ & $+0           0^\circ          35^\prime           17\farcs            4$ & $20.81\pm20.81$ & $ 20.71\pm 0.05$ & $ 20.34\pm 0.05$ & $ 20.13\pm 0.08$ & $ 19.66\pm 0.06$ & $ 18.97\pm 0.01$ & $ 19.22\pm 0.01$ & $ 33.80\pm 6.56$ & $ 25.47\pm 7.15$ & $ 26.89\pm 8.96$ & $0.973$ &$0.956$ & $0.71\pm0.11$ & $ 5.72$\\
J090848.1+002626 &         5691 & $ 0           9^{\rm m} 0           8^{\rm h}           48\fs           13$ & $+0           0^\circ          26^\prime           22\farcs            1$ & $20.10\pm20.10$ & $ 19.89\pm 0.03$ & $ 19.57\pm 0.03$ & $ 19.19\pm 0.04$ & $ 18.86\pm 0.03$ & $ 19.04\pm 0.01$ & $ 19.29\pm 0.01$ & $ 35.24\pm 6.38$ & $ 18.50\pm 7.16$ & $ 16.50\pm 8.90$ & $0.885$ &$0.890$ & $0.39\pm0.05$ & $ 4.15$\\
J090812.1+002430 &         5735 & $ 0           9^{\rm m} 0           8^{\rm h}           12\fs           26$ & $+0           0^\circ          24^\prime           30\farcs            7$ & $19.21\pm19.21$ & $ 18.78\pm 0.01$ & $ 18.66\pm 0.01$ & $ 18.45\pm 0.02$ & $ 18.16\pm 0.02$ & $ 18.31\pm 0.01$ & $ 18.21\pm 0.01$ & $ 34.91\pm 6.51$ & $ 15.87\pm 7.25$ & $  3.13\pm 8.99$ & $0.998$ &$0.997$ & $0.16\pm0.02$ & $ 1.00$\\
J090852.9+004106 &         5774 & $ 0           9^{\rm m} 0           8^{\rm h}           52\fs           90$ & $+0           0^\circ          41^\prime 0           7\farcs            1$ & $>21.85$ & $> 21.40$ & $> 20.94$ & $> 19.61$ & $> 19.20$ & $ 19.82\pm 0.02$ & $ 19.58\pm 0.01$ & $ 33.83\pm 6.45$ & $ 38.19\pm 7.17$ & $ 22.59\pm 9.26$ & $0.996$ &$0.996$ & $...$ & $ 0.79$\\
J090912.3+002129 &         6012 & $ 0           9^{\rm m} 0           9^{\rm h}           12\fs           65$ & $+0           0^\circ          21^\prime           28\farcs 0$ & $19.82\pm19.82$ & $ 19.59\pm 0.02$ & $ 19.27\pm 0.02$ & $ 18.90\pm 0.03$ & $ 18.63\pm 0.03$ & $ 18.38\pm 0.01$ & $ 18.81\pm 0.01$ & $ 35.13\pm 6.65$ & $ 24.35\pm 7.22$ & $  6.36\pm 9.07$ & $0.997$ &$0.995$ & $0.56\pm0.02$ & $ 3.93$\\
J090908.3+002545 &         6100$^{\star}$ & $ 0           9^{\rm m} 0           9^{\rm h} 0           8\fs           64$ & $+0           0^\circ          25^\prime           48\farcs            4$ & $22.45\pm22.45$ & $ 21.84\pm 0.13$ & $ 21.55\pm 0.14$ & $ 21.16\pm 0.21$ & $ 20.48\pm 0.13$ & $ 19.52\pm 0.01$ & $ 19.35\pm 0.01$ & $ 33.08\pm 6.38$ & $ 44.67\pm 7.17$ & $ 42.48\pm 8.86$ & $0.244$ &$0.540$ & $...$ & $ 4.78$\\
J090902.6+004737 &         6146 & $ 0           9^{\rm m} 0           9^{\rm h} 0           2\fs           42$ & $+0           0^\circ          47^\prime           39\farcs            5$ & $21.63\pm21.63$ & $ 21.56\pm 0.13$ & $ 20.92\pm 0.08$ & $ 20.48\pm 0.12$ & $ 19.89\pm 0.07$ & $...$ & $ 19.58\pm 0.01$ & $ 32.24\pm 6.37$ & $ 20.07\pm 7.18$ & $  2.26\pm 8.98$ & $...$ &$0.976$ & $0.66\pm0.22$ & $ 4.49$\\
J090928.7+002630 &         6189$^{\star}$ & $ 0           9^{\rm m} 0           9^{\rm h}           29\fs 0           3$ & $+0           0^\circ          26^\prime           31\farcs            0$ & $>22.54$ & $> 21.73$ & $> 21.05$ & $> 19.74$ & $> 19.13$ & $ 20.90\pm 0.05$ & $ 20.48\pm 0.03$ & $ 33.47\pm 6.52$ & $ 35.08\pm 7.28$ & $ 35.56\pm 8.97$ & $0.302$ &$0.430$ & $...$ & $ 4.06$\\
J090754.6+003345 &         6220$^{\star}$ & $ 0           9^{\rm m} 0           7^{\rm h}           54\fs           82$ & $+0           0^\circ          33^\prime           42\farcs            2$ & $19.71\pm19.71$ & $ 19.41\pm 0.02$ & $ 19.14\pm 0.02$ & $ 18.79\pm 0.03$ & $ 18.48\pm 0.02$ & $ 18.68\pm 0.01$ & $ 18.77\pm 0.01$ & $ 34.05\pm 6.51$ & $ 12.37\pm 7.23$ & $  8.23\pm 9.14$ & $0.518$ &$0.570$ & $0.19\pm0.04$ & $ 4.19$\\
J090946.4+003847 &         6224 & $ 0           9^{\rm m} 0           9^{\rm h}           46\fs           39$ & $+0           0^\circ          38^\prime           47\farcs            1$ & $20.48\pm20.48$ & $ 20.19\pm 0.03$ & $ 19.83\pm 0.03$ & $ 19.25\pm 0.04$ & $ 18.81\pm 0.03$ & $ 18.65\pm 0.02$ & $ 18.76\pm 0.01$ & $ 34.77\pm 6.59$ & $ 14.14\pm 7.24$ & $  4.09\pm 9.06$ & $0.998$ &$0.999$ & $0.46\pm0.12$ & $ 0.97$\\
J090832.0+002749 &         6324 & $ 0           9^{\rm m} 0           8^{\rm h}           31\fs           94$ & $+0           0^\circ          27^\prime           52\farcs            4$ & $22.04\pm22.04$ & $ 21.85\pm 0.17$ & $ 21.59\pm 0.15$ & $ 20.77\pm 0.16$ & $ 20.22\pm 0.10$ & $ 19.18\pm 0.01$ & $ 19.21\pm 0.01$ & $ 33.98\pm 6.35$ & $ 23.50\pm 7.19$ & $ 10.73\pm 9.12$ & $0.989$ &$0.972$ & $...$ & $ 2.90$\\
J090956.0+003739 &         6348 & $ 0           9^{\rm m} 0           9^{\rm h}           56\fs 0           3$ & $+0           0^\circ          37^\prime           37\farcs            2$ & $19.46\pm19.46$ & $ 19.18\pm 0.02$ & $ 18.77\pm 0.01$ & $ 18.31\pm 0.02$ & $ 17.86\pm 0.01$ & $ 18.22\pm 0.01$ & $ 18.18\pm 0.01$ & $ 33.90\pm 6.45$ & $ 12.66\pm 7.24$ & $ -3.81\pm 9.38$ & $0.999$ &$0.999$ & $0.28\pm0.06$ & $ 2.12$\\
J090840.1+001928 &         6480 & $ 0           9^{\rm m} 0           8^{\rm h}           40\fs 0           2$ & $+0           0^\circ          19^\prime           30\farcs            1$ & $19.88\pm19.88$ & $ 19.80\pm 0.03$ & $ 19.50\pm 0.02$ & $ 19.37\pm 0.05$ & $ 19.03\pm 0.03$ & $ 19.25\pm 0.01$ & $...$ & $ 33.32\pm 6.55$ & $ 25.13\pm 7.23$ & $ 27.36\pm 9.19$ & $0.995$ &$...$ & $0.31\pm0.09$ & $ 2.65$\\
J090847.1+001512 &         6792 & $ 0           9^{\rm m} 0           8^{\rm h}           47\fs           28$ & $+0           0^\circ          15^\prime           14\farcs            1$ & $18.37\pm18.37$ & $ 18.16\pm 0.01$ & $ 17.79\pm 0.01$ & $ 17.45\pm 0.01$ & $ 17.14\pm 0.01$ & $ 17.55\pm 0.01$ & $...$ & $ 33.60\pm 6.38$ & $  6.58\pm 7.14$ & $ -1.24\pm 8.87$ & $0.999$ &$...$ & $0.26\pm0.05$ & $ 2.71$\\
J090324.0+003954 &         6893 & $ 0           9^{\rm m} 0           3^{\rm h}           23\fs           95$ & $+0           0^\circ          39^\prime           52\farcs            1$ & $21.76\pm21.76$ & $ 21.35\pm 0.11$ & $ 20.98\pm 0.09$ & $ 20.48\pm 0.13$ & $ 19.86\pm 0.07$ & $ 19.17\pm 0.01$ & $ 19.32\pm 0.01$ & $ 32.93\pm 6.49$ & $ 16.27\pm 7.32$ & $ -0.91\pm 9.29$ & $0.946$ &$0.955$ & $0.49\pm0.19$ & $ 2.54$\\
J090808.2+002115 &         6962 & $ 0           9^{\rm m} 0           8^{\rm h} 0           8\fs           14$ & $+0           0^\circ          21^\prime           16\farcs            5$ & $22.38\pm22.38$ & $ 21.56\pm 0.13$ & $ 21.07\pm 0.10$ & $ 21.30\pm 0.27$ & $ 20.14\pm 0.09$ & $ 19.12\pm 0.01$ & $...$ & $ 32.57\pm 6.48$ & $ 17.35\pm 7.23$ & $ -4.52\pm 9.05$ & $0.997$ &$...$ & $...$ & $ 1.28$\\
J091308.7-005605 &         6991 & $ 0           9^{\rm m} 0           8^{\rm h} 0           8\fs           80$ & $-0           0^\circ          56^\prime 0           8\farcs            0$ & $>22.58$ & $> 21.33$ & $> 20.91$ & $> 19.72$ & $> 19.23$ & $ 20.36\pm 0.03$ & $ 20.03\pm 0.02$ & $ 30.39\pm 6.44$ & $ 38.53\pm 7.14$ & $ 27.37\pm 8.94$ & $0.987$ &$0.986$ & $...$ & $ 2.47$\\
J090915.4+003450 &         7107 & $ 0           9^{\rm m} 0           9^{\rm h}           15\fs 0           6$ & $+0           0^\circ          34^\prime           48\farcs            8$ & $20.68\pm20.68$ & $ 20.31\pm 0.04$ & $ 19.99\pm 0.03$ & $ 19.41\pm 0.04$ & $ 19.00\pm 0.03$ & $ 18.62\pm 0.01$ & $ 18.81\pm 0.01$ & $ 32.15\pm 6.41$ & $  8.28\pm 7.23$ & $  1.37\pm 9.01$ & $0.972$ &$0.976$ & $0.74\pm0.11$ & $ 5.75$\\
J090855.7+002121 &         7216 & $ 0           9^{\rm m} 0           8^{\rm h}           55\fs           64$ & $+0           0^\circ          21^\prime           23\farcs            1$ & $21.87\pm21.87$ & $ 21.44\pm 0.12$ & $ 21.04\pm 0.09$ & $ 20.82\pm 0.17$ & $ 20.38\pm 0.11$ & $ 19.34\pm 0.01$ & $ 18.88\pm 0.01$ & $ 32.34\pm 6.39$ & $ 37.62\pm 7.16$ & $ 33.56\pm 8.89$ & $0.996$ &$0.998$ & $...$ & $ 2.02$\\
J090836.5+002513 &         7826 & $ 0           9^{\rm m} 0           8^{\rm h}           36\fs           54$ & $+0           0^\circ          25^\prime 0           9\farcs            5$ & $22.11\pm22.11$ & $ 21.54\pm 0.13$ & $ 21.13\pm 0.10$ & $ 21.16\pm 0.23$ & $ 20.33\pm 0.11$ & $ 19.41\pm 0.01$ & $ 19.28\pm 0.01$ & $ 30.94\pm 6.33$ & $ 35.84\pm 7.13$ & $ 27.94\pm 9.06$ & $0.991$ &$0.992$ & $...$ & $ 3.62$\\
J090842.8+003623 &         8022 & $ 0           9^{\rm m} 0           8^{\rm h}           43\fs 0           3$ & $+0           0^\circ          36^\prime           21\farcs            7$ & $>21.85$ & $> 21.40$ & $> 20.94$ & $> 19.61$ & $> 19.20$ & $ 20.96\pm 0.06$ & $ 20.45\pm 0.03$ & $ 29.66\pm 6.41$ & $ 56.85\pm 7.21$ & $ 50.43\pm 9.18$ & $0.970$ &$0.981$ & $...$ & $ 3.07$\\
J091309.5-005417 &         9262 & $ 0           9^{\rm m} 0           8^{\rm h} 0           9\fs           85$ & $-0           0^\circ          54^\prime           19\farcs            3$ & $...$ & $...$ & $...$ & $...$ & $...$ & $ 20.42\pm 0.02$ & $ 20.04\pm 0.02$ & $ 28.12\pm 6.57$ & $ 39.70\pm 7.23$ & $ 19.64\pm 9.22$ & $0.958$ &$0.954$ & $...$ & $ 4.46$\\
J090830.0+002439 &         9290 & $ 0           9^{\rm m} 0           8^{\rm h}           29\fs           79$ & $+0           0^\circ          24^\prime           38\farcs            1$ & $>21.85$ & $> 21.40$ & $> 20.94$ & $> 19.61$ & $> 19.20$ & $ 19.93\pm 0.02$ & $ 19.72\pm 0.01$ & $ 29.27\pm 6.47$ & $ 38.29\pm 7.26$ & $ 19.79\pm 8.98$ & $0.754$ &$0.770$ & $...$ & $ 3.45$\\
J090909.4+003440 &        10110 & $ 0           9^{\rm m} 0           9^{\rm h} 0           9\fs           80$ & $+0           0^\circ          34^\prime           38\farcs            5$ & $>22.54$ & $> 21.73$ & $> 21.05$ & $> 19.74$ & $> 19.13$ & $ 21.29\pm 0.05$ & $ 20.62\pm 0.04$ & $ 27.62\pm 6.52$ & $ 37.27\pm 7.23$ & $ 27.12\pm 9.02$ & $0.558$ &$0.861$ & $...$ & $ 5.06$\\
J090820.3+002808 &        11918 & $ 0           9^{\rm m} 0           8^{\rm h}           19\fs           84$ & $+0           0^\circ          28^\prime 0           8\farcs            6$ & $22.34\pm22.34$ & $ 21.49\pm 0.13$ & $ 21.16\pm 0.10$ & $ 20.69\pm 0.15$ & $ 20.53\pm 0.13$ & $ 19.95\pm 0.03$ & $ 19.99\pm 0.03$ & $ 26.08\pm 6.45$ & $ 37.76\pm 7.28$ & $ 16.30\pm 9.11$ & $0.764$ &$0.787$ & $...$ & $ 6.92$\\
J090834.7+003635 &        15947 & $ 0           9^{\rm m} 0           8^{\rm h}           34\fs           77$ & $+0           0^\circ          36^\prime           39\farcs            1$ & $...$ & $...$ & $...$ & $...$ & $...$ & $ 20.55\pm 0.03$ & $ 20.27\pm 0.02$ & $ 22.52\pm 6.50$ & $ 37.86\pm 7.14$ & $ 32.28\pm 9.03$ & $0.731$ &$0.852$ & $...$ & $ 3.43$\\
\hline
\end{tabular}
\end{tiny}
\\Note. --
$^{\star}$~ID with color-magnitude method.
$^a$~Coordinates are the position of IRAC counterparts.
$^b$~Photometry from the VIKING survey ($Z, Y ,J , H,  K_{\rm s}$) is extracted in $2\arcsec$ diameter apertures (Sutherland et al.\ in prep).
$^c$~The reported SPIRE 350 and 500~\micron\ fluxes are measured at the postions of 250~\micron\ sources; no SNR cut is applied.
$^d$~R is reliability of the counterparts calculated separately at 3.6 and 4.5~\micron\ from the LR method (section~\ref{sec:lr}).
$^e$~Redshifts are from the SDSS catalogue for sources detected in that survey.
$^f$~The separation between the centroid of the SPIRE emission and the IRAC counterpart. 
\vspace{3mm}